\documentclass[12pt]{article}
\pdfoutput=1
\usepackage{array} 
\usepackage{amssymb,dsfont}
\usepackage{graphics,graphpap}
\usepackage{float}
\usepackage{colortbl}
\usepackage{graphicx}
\usepackage{color}
\usepackage[dvipsnames]{xcolor}
\usepackage{graphicx}% Include figure files
\usepackage{dcolumn}% Align table columns on decimal point
\usepackage{epsfig}
\usepackage{epstopdf}
\usepackage{bbm}% bold math
\usepackage{amsmath}
\usepackage{amsfonts}
\usepackage{mathtools,tikz}
\usepackage{textcomp}
\usepackage{slashed}
\usepackage{morefloats}
\usepackage{caption}
 \usepackage{subcaption}
\usepackage{setspace}
\usepackage{slashed}
\usepackage{multirow}
\usepackage{footnote}
\makesavenoteenv{table}
\usepackage{url}
\usepackage{slashed}
\usepackage{mathtools}

\usepackage{tikz}
\definecolor{dunkelgruen}{rgb}{0,0.7,0}
\definecolor{dunkelblau}{rgb}{0,0,0.7}

\usepackage{xparse}
\usetikzlibrary{matrix,backgrounds}
\pgfdeclarelayer{myback}
\pgfsetlayers{myback,background,main}

\tikzset{mycolor/.style = {line width=1bp,color=#1}}%
\tikzset{myfillcolor/.style = {draw,fill=#1}}%

\NewDocumentCommand{\highlight}{O{blue!40} m m}{%
\draw[mycolor=#1] (#2.north west)rectangle (#3.south east);
}

\NewDocumentCommand{\fhighlight}{O{blue!20} m m}{%
\draw[myfillcolor=#1] (#2.north west)rectangle (#3.south east);
}

\usepackage{xcolor}
\definecolor{red}{rgb}{1,0,0}
\definecolor{purple}{rgb}{0.5,0,0.5}
\definecolor{blue}{rgb}{0,0,1}

\usepackage{simplewick}

\newcommand{\beq}{\begin{eqnarray}}
\newcommand{\eeq}{\end{eqnarray}}

\newcommand{\bmp}{\noindent\begin{minipage}{16cm}}
\newcommand{\emp}{\end{minipage}\vskip 7mm} % 7mm untightened

\def\drawbox#1#2{\hrule height#2pt
        \hbox{\vrule width#2pt height#1pt \kern#1pt
              \vrule width#2pt}
              \hrule height#2pt}

\def\Asym#1#2{\vcenter{\vbox{\drawbox{#1}{#2}
              \kern-#2pt % line up boxes
              \drawbox{#1}{#2}}}}

\def\simge{\mathrel{%
   \rlap{\raise 0.511ex \hbox{$>$}}{\lower 0.511ex \hbox{$\sim$}}}}

\def\simle{\mathrel{
   \rlap{\raise 0.511ex \hbox{$<$}}{\lower 0.511ex \hbox{$\sim$}}}}

\def\s#1{\setbox0=\hbox{$#1$}%
\rlap{\ifdim\wd0>.7em\kern.22\wd0\else\kern.1\wd0\fi /}#1}

\definecolor{purple}{rgb}{0.5,0,0.5}

\usepackage{cite}

\begin{document}

%%%%%%%%%%%%%%%%%%%%%%%%%%%%%%%%%%%%%%%%%%%%%%%%%%%%%%%%%%%%%%%%%%%%%%%%%%%
\begin{titlepage}
\hfill {\small MPP-2017-196}\\[20mm]
\begin{center}
{\bf\Large
\boldmath{
A Comprehensive Renormalisation Group Analysis of the Littlest Seesaw Model
}
} \\[12mm]

Tanja Geib$^{\dagger}$%
\footnote{E-mail: \tt tgeib@mpp.mpg.de}~~~and~~
Stephen~F.~King$^{\star}$%
\footnote{E-mail: \texttt{king@soton.ac.uk}}
\\[-2mm]
\end{center}

\vspace*{0.5cm}

\centerline{$^{\dagger}$ \it
Max-Planck-Institut f{\"u}r Physik (Werner-Heisenberg-Institut),}
\centerline{\it
F{\"o}hringer Ring 6, 80805 M{\"u}nchen, Germany}
\vspace*{0.20cm}
\centerline{$^{\star}$ \it
School of Physics and Astronomy, University of Southampton,}
\centerline{\it
SO17 1BJ Southampton, United Kingdom }

\vspace*{1.20cm}

\begin{abstract}
{\noindent
We present a comprehensive renormalisation group analysis of the 
Littlest Seesaw model involving two right-handed neutrinos and a very constrained Dirac neutrino Yukawa coupling matrix.
We perform the first $\chi^2$ analysis of the low energy masses and mixing angles, 
in the presence of renormalisation group corrections,
for various right-handed neutrino masses and mass orderings, both with and without supersymmetry.
We find that the atmospheric angle, which is predicted to be near maximal in the absence of renormalisation group 
corrections, may receive significant corrections for some non-supersymmetric cases, bringing it into close
agreement with the current best fit value in the first octant.
By contrast, in the presence of supersymmetry, the renormalisation group corrections are relatively small,
and the prediction of a near maximal atmospheric mixing angle is maintained,
for the studied cases. Forthcoming results from T2K and NOvA will decisively test these models
at a precision comparable to the renormalisation group corrections we have calculated.
}
\end{abstract}
\end{titlepage}

%%%%%%%%%%%%%%%%%%%%%%%%%%%%%%%%%%%%%%%%%%%%%%%%%
\section{\label{sec:intro}Introduction}
%%%%%%%%%%%%%%%%%%%%%%%%%%%%%%%%%%%%%%%%%%%%%%%%%

Despite the impressive experimental progress in neutrino oscillation experiments,
\cite{nobel}, the dynamical origin of neutrino mass generation and lepton flavour mixing remains unknown~\cite{XZbook,King:2013eh}. 
Furthermore, the octant of the atmospheric angle is not determined yet, and its precise value is uncertain. While T2K prefers a close to maximal atmospheric mixing angle~\cite{Abe:2017uxa}, NOvA excludes maximal mixing at $2.6\sigma$ CL~\cite{Adamson:2017qqn}. The forthcoming results from T2K and NOvA will hopefully clarify the situation. An accurate determination of the atmospheric angle is important in order to test predictive neutrino mass and mixing models.  
The leading candidate for a theoretical explanation of neutrino mass and mixing remains 
the seesaw mechanism~\cite{Minkowski:1977sc, Yanagida:1979ss, Gell-Mann:1979ss, Glashow:1979ss, Mohapatra:1979ia}. However the seesaw mechanism
involves a large number of free parameters. 

One approach to reducing the seesaw parameters is to consider the
minimal version involving 
only two right-handed neutrinos, first proposed by one of us~\cite{King:1999mb, King:2002nf}.
In such a scheme the lightest neutrino is massless. A further simplification was considered by Frampton, Glashow and Yanagida~\cite{Frampton:2002qc}, who assumed two texture zeros in the Dirac neutrino mass matrix $M^{}_{\rm D}$ and demonstrated that both neutrino masses and the cosmological matter-antimatter asymmetry could be explained in this economical setup via the seesaw and leptogenesis mechanisms~\cite{Fukugita:1986hr}. The phenomenology of the minimal seesaw model was subsequently fully explored in the
literature~\cite{Guo:2003cc, Ibarra:2003up, Mei:2003gn, Guo:2006qa, Antusch:2011nz,
Harigaya:2012bw, Zhang:2015tea}. In particular, the normal hierarchy (NH) case in the Frampton-Glashow-Yanagida model has been shown to be already excluded by the latest neutrino oscillation data~\cite{Harigaya:2012bw, Zhang:2015tea}.

An alternative to having two texture zeros is to impose constraints on the Dirac mass matrix elements.
For example, the Littlest Seesaw (LS) model consists of two right-handed (RH) neutrino singlets 
$N^{\rm atm}_{\rm R}$ and $N^{\rm sol}_{\rm R}$
together with a tightly constrained Dirac neutrino Yukawa coupling matrix, leading to a highly predictive 
scheme~\cite{King:2013iva, Bjorkeroth:2014vha, King:2015dvf,BAU,King:2016yvg,Ballett:2016yod}. Since the mass ordering of the RH neutrinos as well as the particular choice of the Dirac neutrino Yukawa coupling matrix can vary, it turns out that there are four distinct LS cases, namely cases A, B, C and D,
as defined later.
These four cases of the LS model will be discussed in detail in the present paper.
In particular we are interested in the phenomenological viability of these four cases of the LS model
defined at the scale of some 
grand unified theory (GUT) when the parameters are run down to low energy where experiments are performed.

A first study of the renormalisation group (RG) corrections to the LS model was performed in \cite{King:2016yef}.
The purpose of the present paper is to improve on that analysis and to focus on the cases where 
the RG corrections are the most important.
It is therefore briefly reviewing the progress and limitations of the approach and results in \cite{King:2016yef}.
In \cite{King:2016yef} the authors focussed on analytically understanding the RG effects on the neutrino mixing angles for cases A and B in great detail and threshold effects were discussed due to 
two fixed RH neutrino masses, taken as $10^{12}~{\rm GeV}$ and $10^{15}~{\rm GeV}$, 
close to the scale of grand unified theories $\Lambda^{}_{\rm GUT} = 2\times 10^{16}~{\rm GeV}$  \cite{King:2016yef}. These analytical results were verified numerically. Furthermore, cases C and D were investigated numerically. 
However, the RG running of neutrino masses and lepton flavour mixing parameters were calculated at low energies, always assuming phenomenological best fit values at high energies, which was justified a posteriori by the fact that {\em in most cases} the RG corrections to the neutrino mass ratio\footnote{This is not true for the neutrino masses $m_2$ and $m_3$. Their running is significant as demonstrated in Figs. 1-4 in Ref.~\cite{King:2016yef}.} as well as the mixing angles were observed to be rather small \cite{King:2016yef}. 
Such cases with small RG corrections lead to an atmospheric mixing angle close to its maximal value, which is in some tension with the latest global fits. To account for the running of the neutrino masses, Ref.~\cite{King:2016yef} modified the Dirac neutrino Yukawa matrix by an overall factor of $1.25$ with respect to the best fit values obtained from tree-level analyses. This factor was chosen based on scaling the neutrino masses for case A to obtain appropriate values at the EW scale, and subsequently used for all four LS cases. 
In other words, the numerical analysis of Ref.~\cite{King:2016yef} chose input parameters that where extracted from a tree-level best fit, and adjusted them by an overall factor based on one specific case to include some correction for the significant running in the neutrino masses.  

There are several problems with the above approach \cite{King:2016yef}, as follows:
\begin{itemize}
\item The overall factor of $1.25$ to the Dirac neutrino Yukawa matrix implies that only the running of the neutrino masses themselves is significantly affected by the choice of input parameters, while the neutrino mixing angles are still stable. Furthermore, it assumes that keeping the ratio of the input parameters unchanged when incorporating RG effects is reasonable. Both assumptions turn out to be incorrect.
\item Having modified the Dirac neutrino Yukawa matrix based on case A, Ref.~\cite{King:2016yef} employs the same factor for cases B, C and D, although the running behaviour can change fundamentally with the LS case. 
\item 
Most importantly, as mentioned above, 
the RG running of neutrino masses and lepton flavour mixing parameters were calculated at low energies, assuming phenomenological best fit values at high energies. 
Clearly the correct approach would be to perform a complete scan of model input parameters in order to determine the optimum set of high energy input values from a global fit of the low energy parameters. This is what we will do in this paper.  As a consequence, the measure of the goodness-of-fit\footnote{
Note that the goal is to minimise the value for $\chi^2$, as defined in
e.~g.~Ref.~\cite{Bjorkeroth:2014vha}.
}
yields less than mediocre results for the input parameters used in Ref.~\cite{King:2016yef}: $\chi^2_{A,B}(\Lambda_{\rm EW})\approx 50$, and $\chi^2_{C,D}(\Lambda_{\rm EW})\approx 175$. In comparison, our complete scan here 
will reveal much improved best fit scenarios with $\chi^2_{A}(\Lambda_{\rm EW})=7.1$, $\chi^2_{B}(\Lambda_{\rm EW})=4.2$, $\chi^2_{C}(\Lambda_{\rm EW})=3.2$ and $\chi^2_{D}(\Lambda_{\rm EW})=1.5$. 
\end{itemize}

In the present paper, then, we will perform a detailed RG analysis of the LS model, including those cases where the RG corrections can become significant. As such it is no longer sufficient to fix the input parameters by fitting to the high energy masses and mixing angles.
Consequently, we perform a complete scan of model parameters for each case individually, to determine the optimum set of high energy input values from a global fit of the low energy parameters which include the effects of RG running, and to re-assess whether RG corrections might still be sufficient to obtain a realistic atmospheric mixing angle.
We shall find that the largest corrections occur in the Standard Model (SM), although we shall also perform a detailed
analysis of the Minimal Supersymmetric Standard Model (MSSM)\footnote{When we refer to the MSSM or SM we really mean the
LS models with or without supersymmetry. We shall use this rather imprecise terminology throughout the paper.} for various values of $\tan \beta$
for completeness, however, since the RG corrections there are relatively small, we relegate those results to an Appendix.
In all cases we perform a $\chi^2$ analysis of the low energy masses and mixing angles, including RG corrections
for various RH neutrino masses and mass orderings.

The layout of the remainder of the paper is as follows.
In Sec.~\ref{sec1A} we review the LS model and define the four cases A,B,C,D which we shall analyse.
In Sec.~\ref{sec:3} we discuss qualitatively the expected effects of RG corrections in the LS models. We focus on some key features that will help understand the findings in later sections, instead of aiming at a complete discussion of the RG effects.
In Sec.~\ref{sec:4} we introduce the $\chi^2$ function that we use to analyse our results.
In Sec.~\ref{sec:5} we discuss the SM results in some detail, since this is where the RG corrections can be
the largest, serving to reduce the atmospheric angle from its near maximal value at high energy to close to the 
best fit value at low energy in some cases.
Sec.~\ref{app:B} discusses the results for the RG analysis of the LS model in the MSSM.
In Sec.~\ref{sec:6} we compare the MSSM results to those of the SM, and show that 
the RG corrections in the SM are more favourable.
Sec.~\ref{sec:7} concludes the paper.
Appendix~\ref{app:A} introduces the notation needed to discuss benchmark scenarios for the LS model in the MSSM, and Appendix~\ref{app:D} displays tables with the results of all MSSM scenarios investigated.

%%%%%%%%%%%%%%%%%%%%%%%%%%%%%%%%%%%%%%%%%%%%%%%
\section{\label{sec1A} Littlest Seesaw}
%%%%%%%%%%%%%%%%%%%%%%%%%%%%%%%%%%%%%%%%%%%%%%%

The seesaw mechanism~\cite{Minkowski:1977sc, Yanagida:1979ss, Gell-Mann:1979ss, Glashow:1979ss, Mohapatra:1979ia} extends the standard model (SM) with a number of right-handed neutrino singlets $N^{}_{i{\rm R}}$
as, 
\begin{eqnarray}
-{\cal L}^{}_{\rm m} = \overline{\ell^{}_{\rm L}} Y^{}_l H E^{}_{\rm R} + \overline{\ell^{}_{\rm L}} Y^{}_\nu \tilde{H} N^{}_{\rm R} + \frac{1}{2} \overline{N^c_{\rm R}} M^{}_{\rm R} N^{}_{\rm R} + {\rm h.c.} \; ,
%     (1)
\end{eqnarray}
where $\ell^{}_{\rm L}$ and $\tilde{H} \equiv {\rm i}\sigma^{}_2 H^*$ stand respectively for the left-handed lepton and Higgs doublets, $E^{}_{\rm R}$ and $N^{}_{\rm R}$ are the right-handed charged-lepton and neutrino singlets, $Y^{}_l$ and $Y^{}_\nu$ are the charged-lepton and Dirac neutrino Yukawa coupling matrices, $M^{}_{\rm R}$ is the Majorana mass matrix of right-handed neutrino singlets. 
Physical light effective Majorana neutrino masses are generated via the seesaw mechanism, resulting in the light left-handed Majorana neutrino mass matrix
\begin{equation}
 m_\nu = - v^2 Y_{\nu} M_R^{-1} Y_{\nu}^T\,. 
 \label{seesaw}
\end{equation}

The Littlest Seesaw Model model (LS) 
extends the SM by two heavy right-handed neutrino singlets with masses $M_{atm}$ and $M_{sol}$ and imposes constrained sequential dominance (CSD) on the Dirac neutrino Yukawa couplings. 
The particular choice of structure of $Y^{A,B,C,D}_{\nu}$ and heavy mass ordering $M^{A,B,C,D}_R$
defines the type of LS, as discussed below.  All four cases predict a normal mass ordering for the light neutrinos with a massless neutrino $m_1=0$.

In the flavour basis, where the charged leptons and right-handed neutrinos are diagonal,
the {\bf Cases~A,B} are defined by the mass hierarchy $M_{atm}\ll M_{sol}$, 
and hence $\widehat{M}^{}_{\rm R} = {\rm Diag}\{ M^{}_{\rm atm}, M^{}_{\rm sol} \}$, 
and the structure of the respective Yukawa coupling matrix:
\begin{eqnarray}
{\bf Case~A}:~Y^{A}_\nu = \begin{pmatrix}
0 & b e^{{\rm i}\eta/2} \\
a & n b e^{{\rm i}\eta/2} \\
a & (n-2) b e^{{\rm i}\eta/2}
\end{pmatrix} \quad {\rm or} \quad {\bf Case~B}:~Y^{B}_\nu = \begin{pmatrix}
0 & b e^{{\rm i}\eta/2} \\
a & (n-2) b e^{{\rm i}\eta/2} \\
a & n b e^{{\rm i}\eta/2}
\end{pmatrix}
%     (2)
\label{eq:Ynu0}
\end{eqnarray}
with $a, b, \eta$ being three real parameters and $n$ an integer. These scenarios were analysed in \cite{King:2016yef} with heavy neutrino masses of $M_{\rm atm}^{} = M^{}_1 = 10^{12}~{\rm GeV}$ and $M_{\rm sol}^{} = M_2^{} = 10^{15}~{\rm GeV}$. 

Considering an alternative mass ordering of the two heavy Majorana neutrinos -- $M_{atm}\gg M_{sol}$, and consequently $\widehat{M}^{}_{\rm R} = {\rm Diag}\{ M^{}_{\rm sol}, M^{}_{\rm atm} \}$ -- we have to exchange the two columns of $Y^{}_\nu$ in Eq.~(\ref{eq:Ynu0}), namely,
\begin{eqnarray}
{\bf Case~C}: Y^{C}_\nu = \begin{pmatrix}
b e^{{\rm i}\eta/2} & 0 \\
n b e^{{\rm i}\eta/2} & a \\
(n-2) b e^{{\rm i}\eta/2} & a
\end{pmatrix} ~~~ {\rm or} ~~~ {\bf Case~D}: Y^{D}_\nu = \begin{pmatrix}
b e^{{\rm i}\eta/2} & 0 \\
(n-2) b e^{{\rm i}\eta/2} & a \\
n b e^{{\rm i}\eta/2} & a
\end{pmatrix}\,,
%     (49)
\label{eq:Ynu0I}
\end{eqnarray}
which we refer to as  {\bf Cases~C,D}. For  $M_{\rm atm}^{} = M^{}_2 = 10^{15}~{\rm GeV}$ and $M_{\rm sol}^{} = M_1^{} = 10^{12}~{\rm GeV}$, both these cases were studied in \cite{King:2016yef}.

We apply the seesaw formula in Eq.~(\ref{seesaw}), for {\bf Cases~A,B,C,D} using the Yukawa coupling matrices $Y^{A,B}_{\nu}$ in Eq.~(\ref{eq:Ynu0}) with $M^{A,B}_R=\rm{diag}(M_{atm},M_{sol})$ and  $Y^{C,D}_{\nu}$ in Eq.~(\ref{eq:Ynu0I}) with $M^{C,D}_R=\rm{diag}(M_{sol},M_{atm})$, to give (after rephasing) the light neutrino mass matrices in terms of the real parameters $m_a=a^2 v^2/M_{atm}$, $m_b=b^2 v^2/M_{sol}$ with $v=174~\rm{GeV}$:
\begin{equation}
 m^{A,C}_\nu= m_a \begin{pmatrix}
             0 & 0 & 0 \\
             0 & 1 & 1 \\
             0 & 1 & 1
            \end{pmatrix} 
       + m_b \rm{e}^{i \eta} \begin{pmatrix}
                              1 & n & (n-2) \\
                              n & n^2 & n(n-2) \\
                              (n-2) & n(n-2) & (n-2)^2
                             \end{pmatrix}\,,
\label{eq:nu_massA}
\end{equation}
\begin{equation}
 m^{B,D}_\nu= m_a \begin{pmatrix}
             0 & 0 & 0 \\
             0 & 1 & 1 \\
             0 & 1 & 1
            \end{pmatrix} 
       + m_b \rm{e}^{i \eta} \begin{pmatrix}
                              1 &  (n-2) & n \\
                               (n-2) &  (n-2)^2 &  n(n-2)  \\
                              n & n(n-2) & n^2
                             \end{pmatrix}\,.
\label{eq:nu_massB}
\end{equation}
Note the seesaw degeneracy of {\bf Cases~A,C} and {\bf Cases~B,D}, which yield the same effective
neutrino mass matrices, respectively.
Studies which ignore renormalisation group (RG) running effects do not distinguish between these degenerate cases. Of course in our RG study the degeneracy is resolved and we have to separately deal with the four physically distinct cases.

The neutrino masses and lepton flavour mixing parameters at the electroweak scale $\Lambda^{}_{\rm EW} \sim \mathcal{O}(1000~{\rm GeV})$ can be derived by diagonalising the effective neutrino mass matrix via
\begin{equation}
 U_{\nu L} m_\nu U^T_{\nu L}=\rm{diag}(m_1,m_2,m_3)\,.
\end{equation}
From a neutrino mass matrix as given in Eqs.~(\ref{eq:nu_massA}) and (\ref{eq:nu_massB}), one immediately obtains normal ordering with $m_1=0$. Furthermore, these scenarios only provide one physical Majorana phase $\sigma$.
As discussed above, we choose to start in a flavour basis, where the right-handed neutrino mass matrix $M_R$ and the charged-lepton mass matrix $M_l$ are diagonal. Consequently, the PMNS matrix is given by $U_{PMNS}=U_{\nu L}^\dagger$. We use the standard PDG parametrisation for the mixing angles, and the CP-violating phase $\delta$. Within our LS scenario, the standard PDG Majorana phase $\varphi_1$ vanishes and $-\varphi_2/2=\sigma$. 

The low-energy phenomenology in the LS model case A has been studied in detail both numerically~\cite{King:2013iva,Bjorkeroth:2014vha} and analytically~\cite{King:2015dvf}, where it has been found that the best fit to experimental data of neutrino oscillations is obtained for $n = 3$ for a particular choice of phase $\eta \approx 2\pi /3$, while for case B the preferred choice is for $n = 3$ and $\eta \approx -2\pi /3$ \cite{King:2013iva,King:2016yvg}. Due to the degeneracy of cases A,C and cases B,D at tree level, the preferred choice for $n$ and $\eta$ carries over, respectively. The prediction for the baryon number asymmetry in our Universe via leptogenesis within case A is also studied~\cite{BAU}, while a successful realisation of the flavour structure of $Y^{}_\nu$ for case B in Eq.~(\ref{eq:Ynu0}) through an $S^{}_4 \times U(1)$ flavour symmetry is recently achieved in Ref.~\cite{King:2016yvg}, where the symmetry fixes $n=3$ and $\eta = \pm 2\pi/3$.

With the parameters $n=3$ and $\eta = \pm 2\pi/3$ fixed, there are only two remaining real free Yukawa parameters in Eqs.~(\ref{eq:Ynu0}) and (\ref{eq:Ynu0I}), namely $a,b$, so the LS predictions then depend on only two real free input combinations $m^{}_a=a^2v^2/M^{}_{\rm atm}$ and $m^{}_b=b^2v^2/M^{}_{\rm sol}$, in terms of which all neutrino masses and the PMNS matrix are determined. For instance, if $m_a$ and $m_b$ are chosen to fix $m_2$ and $m_3$, then the entire PMNS mixing matrix, including phases, is determined with no free parameters. Using benchmark parameters ($m_a=26.57~{\rm meV}$, $m_b=2.684~{\rm meV}$, $n=3$, $\eta=\pm 2\pi/3$), it turns out that the LS model predicts close to maximal atmospheric mixing at the high scale, $\theta_{23}\approx 46^\circ $ for case A , or $\theta_{23}\approx 44^\circ $ for case B~\cite{King:2016yvg}, where both predictions are challenged by the latest NOvA results in the $\nu_{\mu}$ disappearance channel~\cite{nova} which indicates that $\theta_{23}=45^\circ$ is excluded at the 2.5 $\sigma$ CL, although T2K measurements in the same channel continue to prefer maximal mixing~\cite{Abe:2014ugx}.  Since no RG running is included so far, Case C and D predict the same atmospheric angles upon inserting the benchmark parameters.

%%%%%%%%%%%%%%%%%%%%%%%%%%%%%%%%%%%%%%%%%%%%%%%
\section{\label{sec:3} RGE Running in Littlest Seesaw Scenarios}
%%%%%%%%%%%%%%%%%%%%%%%%%%%%%%%%%%%%%%%%%%%%%%%
Although the best-fit input parameters in the present paper were determined by means of numerically solving the RGEs, we will briefly recap some features of the LS' RG running to facilitate comprehending the distinctive behaviour of the different cases. This qualitative discussion is based on the more thorough analytical approaches in Refs.~\cite{King:2016yef,Antusch:2005gp}. 

We switch from denoting the heavy right-handed neutrino masses by $M_{atm},\,M_{sol}$ to labeling them by $M_1, M_2$ to avoid mixing up the different cases and their opposite ordering of heavy neutrino masses. That is to say that irrespective of the case discussed, $M_2$ always denotes the higher scale and $M_1$ the lower.

For the LS, there are three different energy regimes of interest. Starting at the GUT scale, we can use the full theory's parameters and RGEs to describe the evolution down to $\mu=M_2$. At $\mu=M_2$, the heavier $N_R$ is integrated out, and the light neutrino mass matrix as well as the RGEs have to be adapted. It is important to carefully match the full theory on the effective field theory (EFT) below the seesaw scale, denoted by EFT 1. Using the modified RGEs, the parameters are further evolved down to $\mu=M_1$, where the remaining $N_R$ is integrated out, and the parameters of this intermediate EFT 1 are matched to the EFT below $M_1$, denoted by EFT 2. Once again, the light neutrino mass matrix along with the RGEs have to be determined anew. As we assume a strong mass hierarchy $M_2>>M_1$, it is important to decouple the heavy neutrinos subsequently, and describe the intermediate RG behaviour accordingly. \newline

Taking a closer look at the highest regime, we specify the LS input parameters at the GUT scale, and additionally choose the flavour basis, i.~e.~both $Y_l (\Lambda_{\rm GUT})$ and $M_R (\Lambda_{\rm GUT})$ are diagonal. For now, we are interested in the evolution of the neutrino mixing parameters, which implies narrowly watching how the mismatch between the basis, where the charged-lepton Yukawa matrix $Y_l$ is diagonal, and the one, where the light neutrino mass matrix $m_\nu$ is diagonal, unfolds. Consequently, we track the RG running of $Y_l$ and $m_\nu$. Above the seesaw threshold $\mu=M_2$,  the evolution of the flavour structure of $m_\nu$ is mainly driven by $Y_\nu Y_\nu^\dagger$. Consequently, the varying flavour structures of the Dirac neutrino Yukawa matrix need to be examined more thoroughly: 
\begin{itemize}
\item Case A: Whether we take the benchmark input parameters as stated in Sec.~\ref{sec1A} or the global-fit parameters determined in Sec.~\ref{sec:4}, there is a hierarchy $a\sim \mathcal{O} (0.04)<< b\sim \mathcal{O}(0.4)$ which allows for further simplification.
\begin{equation}
Y_\nu Y_\nu^\dagger=\begin{pmatrix}
b^2 & nb^2 & (n-2)b^2 \\
nb^2 & a^2+n^2b^2 & a^2+n(n-2)b^2 \\
(n-2)b^2 & a^2+n(n-2)b^2 & a^2+(n-2)^2 b^2
\end{pmatrix} \xrightarrow{n=3, a<<b}\begin{pmatrix}
b^2 & 3 b^2 & b^2 \\
3 b^2 &  \tikz[baseline]{\node [fill=blue!20,anchor=base,draw=black] {$9 b^2$};} & 3 b^2 \\
b^2 & 3 b^2 & b^2
\end{pmatrix} 
\end{equation}
Consequently, Ref.~\cite{King:2016yef} only considers the dominant $9 b^2$ term and thereby solves the simplified RGE for $m_\nu$ analytically. 
\item Case B: In analogy to Case A, there is a hierarchy with respect to the input parameters $a\sim \mathcal{O} (0.04)<< b\sim \mathcal{O}(0.4)$. 
\begin{equation}
Y_\nu Y_\nu^\dagger=\begin{pmatrix}
b^2 & (n-2)b^2& nb^2  \\
(n-2)b^2 & a^2+(n-2)^2b^2 & a^2+n(n-2)b^2 \\
nb^2 & a^2+n(n-2)b^2 & a^2+n^2 b^2
\end{pmatrix} \xrightarrow{n=3, a<<b}\begin{pmatrix}
b^2 & b^2 & 3 b^2 \\
b^2 & b^2 & 3 b^2 \\
3 b^2 & 3 b^2 & \tikz[baseline]{\node [fill=blue!20,anchor=base,draw=black] {$9 b^2$};}
\end{pmatrix} 
\end{equation}
Therefore, the simplified RGE of $m_\nu$, which only takes the dominant ($33$)-entry into account, can be solved analytically.
\item Case C:  Due to the opposite ordering of heavy neutrino masses, the hierarchy arising from either the benchmark or the global-fit input parameters is also reversed, namely  $a\sim \mathcal{O} (1.2)>> b\sim \mathcal{O}(0.01)$.
 \begin{equation}
Y_\nu Y_\nu^\dagger=\begin{pmatrix}
b^2 & nb^2 & (n-2)b^2 \\
nb^2 & a^2+n^2b^2 & a^2+n(n-2)b^2 \\
(n-2)b^2 & a^2+n(n-2)b^2 & a^2+(n-2)^2 b^2
\end{pmatrix} \xrightarrow{n=3, b<<a}
\begin{tikzpicture}[baseline=-\the\dimexpr\fontdimen22\textfont2\relax ]
\matrix (m)[matrix of math nodes,left delimiter=(,right delimiter=)]
{b^2 & 3 b^2 & b^2 \\
3 b^2 &  a^2 & a^2 \\
b^2 & a^2  & a^2\\
};
\begin{pgfonlayer}{myback}
\fhighlight{m-2-2}{m-3-3}
\end{pgfonlayer}
\end{tikzpicture}
\end{equation}
Even when considering only the dominant contributions arising from $a^2$, the resulting simplified RGE of $m_\nu$ cannot be solved analytically anymore due to the non-diagonal elements strongly affecting the flavour structure of $m_\nu$.
\item Case D: In analogy to Case C, there is a hierarchy to the input parameters $a\sim \mathcal{O} (1.2)>> b\sim \mathcal{O}(0.01)$. 
\begin{equation}
Y_\nu Y_\nu^\dagger=\begin{pmatrix}
b^2 & (n-2)b^2& nb^2  \\
(n-2)b^2 & a^2+(n-2)^2b^2 & a^2+n(n-2)b^2 \\
nb^2 & a^2+n(n-2)b^2 & a^2+n^2 b^2
\end{pmatrix} \xrightarrow{n=3, b<<a}
\begin{tikzpicture}[baseline=-\the\dimexpr\fontdimen22\textfont2\relax ]
\matrix (m)[matrix of math nodes,left delimiter=(,right delimiter=)]
{b^2 & b^2 & 3 b^2 \\
b^2 &  a^2 & a^2 \\
3 b^2 & a^2  & a^2\\
};
\begin{pgfonlayer}{myback}
\fhighlight{m-2-2}{m-3-3}
\end{pgfonlayer}
\end{tikzpicture}
\end{equation}
Thus, even the simplified RGE of $m_\nu$ turns out to be too involved to be solved analytically. Consequently, Case C and Case D are both investigated via an exact numerical approach in Ref.~\cite{King:2016yef}.
\end{itemize}
Note that, as apparent from the discussion below Eqs.~(\ref{eq:nu_massA}) and (\ref{eq:nu_massB}), it is $Y^A_\nu Y_\nu^{A \dagger}=Y^C_\nu Y_\nu^{C \dagger}$ and $Y^B_\nu Y_\nu^{B \dagger}=Y^D_\nu Y_\nu^{D \dagger}$. However, due to the inverted hierarchy with respect to $a,\,b$ (stemming from the inverted heavy neutrino mass ordering), different entries dominate the RG evolution of $m_\nu$, leading to different RG running behaviour. Thus, the degeneracy of the cases is resolved. This means that although (in case of starting from the same set of benchmark input parameters) the neutrino masses and mixing angles of Case A, B, C, and D at the GUT scale are all identical, the running behaviour of the mixing angles, which is mainly governed by  $Y_\nu Y_\nu^\dagger$, is quite different. Moreover, the discussion above uncovers a deeper connection among the cases $A \leftrightarrow B$ and cases $C \leftrightarrow D$ manifest in the shared respective input parameter as well as the similar/same structure of $Y_\nu Y_\nu^\dagger$ dominating the running of $m_\nu$.  \newline
Having determined $m_\nu(M_2)$ from either the analytical or numerical RG evolution, we need to diagonalise the light neutrino mass matrix. That way, we obtain not only the neutrino masses $m_{2,3}(M_2)$ but also the transformation matrix $U_\nu$. The latter in combination with the unitary transformation $U_l$, diagonalising $Y_l$, yields the PMNS matrix, and thereby the neutrino mixing parameters at the scale $\mu=M_2$.\newline
Thus, still within the high-energy regime, we focus on the charged-lepton Yukawa matrix. Since we are interested in the flavour mixing caused by the running of $Y_l$, flavour-independent terms are neglected. But besides that, the RGE for $Y_l$ can be solved analytically without further simplifications, meaning that once again $Y_\nu Y_\nu^\dagger$ drives the flavour mixing. Finally, at $\mu=M_2$, $Y_l$ is diagonalised by means of the unitary transformation $U_l$. Consequently, one would have all necessary parameters at hand to extract approximations for the mixing angles, see Ref.~\cite{King:2016yef}. \newline

Taking a closer look at the intermediate energy regime, $M_2>\mu > M_1$, we need to employ EFT 1 to describe the parameters and RG running. At the threshold $\mu=M_2$, the effective light neutrino mass matrix can be written as 
\begin{equation}
m^{(2)}_\nu=v^2 \Big(\kappa^{(2)} +\tilde{Y}_\nu M_1^{-1} \tilde{Y}^T_\nu \Big)\,,
\label{eq:EffNeutrinoMass}
\end{equation}
where $\kappa^{(2)} \propto \hat{Y}_\nu M_2^{-1} \hat{Y}^T_\nu$ stems from decoupling the heavier right-handed neutrino with mass $M_2$. The expression $\tilde{Y}_\nu$ ($\hat{Y}_\nu$) is obtained from $Y_\nu$ by removing the column corresponding to the decoupled heavy neutrino of mass $M_2$ (the right-handed neutrino of mass $M_1$). Please note that the two terms on the right-hand side of Eq.~(\ref{eq:EffNeutrinoMass}) are governed by different RGEs, leading to so-called "threshold effects". The RGEs of $\kappa^{(2)}$ and  $\tilde{Y}_\nu M_1^{-1} \tilde{Y}^T_\nu$ have different coefficients for the terms proportional to the Higgs self-coupling and gauge coupling contributions within the framework of the SM~\cite{Antusch:2005gp}. In combination with the strong mass hierarchy of the heavy right-handed neutrinos, which enforces a subsequent decoupling, the threshold effects become significant, and thereby enhance the running effects on the neutrino mixing parameters\footnote{This can be understood by assuming that if the expression $U^T \big(\kappa^{(2)} +\tilde{Y}_\nu M_1^{-1} \tilde{Y}^T_\nu \big) U$ is diagonal, then $U^T \big(x \kappa^{(2)} + \tilde{x} \tilde{Y}_\nu M_1^{-1} \tilde{Y}^T_\nu \big) U$ is only diagonal for $x=\tilde{x}$. Since this is not the case here, meaning the two terms scale differently, there is an additional "off-diagonalness".}. From the discussion in Ref.~\cite{King:2016yef}, we learn that the threshold-effect-related corrections to the neutrino mixing angles between $M_2$ and $M_1$ are dominated by an expression proportional to $\kappa^{(2)}$. 
Hence, we examine the combination $\hat{Y}_\nu M_2^{-1} \hat{Y}^T_\nu$ for the four cases:
\begin{itemize}
\item Case A:
\begin{equation}
\hat{Y}_\nu M_2^{-1} \hat{Y}^T_\nu = M_{sol}^{-1} \begin{pmatrix}
b^2 & 3 b^2 & b^2 \\
3 b^2 & 9 b^2 & 3 b^2 \\
b^2 & 3 b^2 & b^2
\end{pmatrix} 
\end{equation}
\item Case B:
\begin{equation}
\hat{Y}_\nu M_2^{-1} \hat{Y}^T_\nu = M_{sol}^{-1}\begin{pmatrix}
b^2 & b^2 & 3 b^2 \\
b^2 & b^2 & 3 b^2 \\
3 b^2 & 3 b^2 & 9 b^2
\end{pmatrix} 
\end{equation}
\item Case C and Case D: 
\begin{equation}
\hat{Y}_\nu M_2^{-1} \hat{Y}^T_\nu = M_{atm}^{-1} \begin{pmatrix} 0 & 0 & 0 \\
0 &  a^2 & a^2 \\
0 & a^2  & a^2\\
\end{pmatrix}
\end{equation}
\end{itemize}
It is evident that the different order of the heavy neutrino decoupling once again evokes distinct flavour structures. Thus, demonstrating that the connection between Case A,B and Case C,D carries on to lower energy regimes as well.
Note that, although the flavour structure of $\kappa^{(2)}$ drives the mixing parameter's running from threshold effects, its contribution comes with a suppression factor.  
Moreover, bare in mind that we only considered the threshold effects arising in EFT 1, but no further contributions from both neutrino and charged-lepton sector. These additional contributions may compete with the threshold effects in some cases, and lead to deviations from the similar features of Case A,B and Case C,D. \newline

Going below the lower threshold, $\mu<M_1$, the running effects of the mixing angles become insignificant. This is not the case for the running of the light neutrino masses, which is too complicated to describe analytically in all regimes, and therefore was not discussed above. Nevertheless, there are a few details of the neutrino matrix running that we want to briefly mention: depending on the size of the $Y_\nu$ entries, the sign of the flavour-independent contribution to the RGE of $m_\nu$ can switch; and the coefficients of the flavour-dependent contributions for the SM and MSSM differ including a sign switch in some. As a consequence, a parameter can run the opposite direction for the framework of the SM in contrast to the MSSM. This feature is most apparent for the light neutrino masses that exhibit strong overall running in opposite directions when comparing the LS in the context of the SM and in the context of the MSSM.
In order to access all parameters -- neutrino masses, mixing angles and phases -- at all scales, we turn to an exact numerical treatment using the  {\it Mathematica} package \texttt{\texttt{REAP}}~\cite{Antusch:2005gp}.\newline

There are two conclusions to be emphasised from the discussion above: 
\begin{itemize}
\item Despite yielding identical neutrino masses and mixing parameters at the GUT scale (for identical input parameters ($a$, $b$)), Case A,C and Case B,D show fundamentally different running behaviour.
\item There is an intrinsic connection between the evolution of Case $A \leftrightarrow B$ (Case $C\leftrightarrow D$) which is reflected in the parameter $b$ ($a$) dominating the running as well as $Y_\nu Y_\nu^\dagger$ being mainly diagonal (being driven by the same block matrix). This distinction between Case A,B versus Case C,D properties becomes even more evident when taking a closer look at the energy regime $M_2>\mu > M_1$. 
\end{itemize}

%%%%%%%%%%%%%%%%%%%%%%%%%%%%%%%%%%%%%%%%%%%%%%%
\section{\label{sec:4} The $\chi^2$ Function}
%%%%%%%%%%%%%%%%%%%%%%%%%%%%%%%%%%%%%%%%%%%%%%%

In the following, we fix $n=3$ and $\eta=\pm 2\pi/3$. Consequently, there are only two free real parameters remaining to predict the entire neutrino sector. 
In order to find the best-fit input parameters $m_a$ and $m_b$ while keeping $\eta=\pm 2\pi/3$ and $n=3$ fixed, we perform a global fit using the $\chi^2$ function as a measure for the goodness-of-fit~\cite{Bjorkeroth:2014vha},
\begin{equation}
 \chi^2 = \sum^{N}_{i=1} \Bigg(\frac{P_i(x)-\mu_i}{\sigma_i}\Bigg)^2\,.
 \label{eq:chi}
\end{equation}
Here, we collect our model parameters in $x=(m_a, m_b, n, \eta)$, and predict the physical values $P_i(x)$ from the Littlest Seesaw Model. The latter are compared to the $\mu_i$ that correspond to the ``data'', which we take to be the global fit values of \cite{Esteban:2016qun},
\begin{equation}
 \mu_i=\{\sin^2\theta_{12},\sin^2\theta_{13},\sin^2\theta_{23},\Delta m^2_{21}, \Delta^2_{31} (, \delta)\}\,.
 \label{eq:DataGlobal}
\end{equation}
Furthermore, $\sigma_i$ are the $1 \sigma$ deviations for each of the neutrino observables. In case the global fit distribution is Gaussian, the $1 \sigma$ uncertainty matches the standard deviation, which is the case for several of the neutrino parameters depicted in Tab.~\ref{tab:ExpVal}. However, there are a few cases where the deviations are asymmetric. To obtain conservative results, we assume the distribution surrounding the best fit to be Gaussian, and choose the smaller uncertainty, respectively. That way, we slightly overestimate the $\chi^2$ values.
\begin{table}[h!]
\setstretch{1.5}
\begin{center}
\begin{tabular}{c|c}
Parameter from \cite{Esteban:2016qun} & best-fit-values $\pm 1\sigma$ \\
\hline \hline 
$\sin^2 \theta_{12}$ & $0.306^{+ 0.012}_{-0.012}$ \\
$\sin^2 \theta_{13}$ & $0.02166^{+ 0.00075}_{- 0.00075}$ \\
$\sin^2 \theta_{23}$ & $0.441^{+0.027}_{-0.021}$ \\
\hline
$\Delta m^2_{21}$ & $\big(7.50^{+0.19}_{- 0.17}\big)~10^{-5}~\rm{eV}^2$\\
$\Delta m^2_{31}$ & $\big(2.524^{+ 0.039}_{-0.040}\big)~10^{-3}~\rm{eV}^2$\\
\hline 
$\delta $  & $-99^{\circ \,+ 51^\circ}_{\,\,\,\,- 59^\circ}$
\end{tabular}
\end{center}
\setstretch{1}
\caption{\label{tab:ExpVal} Best-fit values with $1\sigma$ uncertainty range from global fit to experimental data for neutrino parameters in case of normal ordering, taken from \cite{Esteban:2016qun}.}
\end{table}\\
Since the {\it CP}-violating phases $\delta$ and $\sigma$ are either only measured with large uncertainties or not at all, we define two different $\chi^2$ functions:
\begin{itemize}
 \item $\chi^2$ for which $N=5$, i.e., $\delta$ is {\it not} included in Eq.~(\ref{eq:DataGlobal}),
 \item $\chi^2_{\delta}$ for which $N=6$, i.e., $\delta$ is included when performing the global fit.
\end{itemize}
A $\chi^2$ function is required to have a well-defined and generally stable global minimum in order to be an appropriate  measure for the goodness-of-fit. This is the case for all CSD(n) models under the assumption that the sign of $\eta$ is fixed~\cite{Bjorkeroth:2014vha}. 
From former analyses of the LS~\cite{Bjorkeroth:2014vha,King:2016yef}, we know in which ballpark the best-fit values of $m_{a,b}$ are to be expected, respectively. That way, we can define a grid in the $(m_a,m_b)$-plane over which we scan -- meaning that we handover the respective input parameters $x=(m_a,\,m_b,\,n=3,\,\eta=\pm 2\pi/3)$ at each point of the grid to the {\it Mathematica} package \texttt{REAP}~\cite{Antusch:2005gp}. \texttt{REAP} numerically solves the RGEs and provides the neutrino parameters at the electroweak scale, i.~e.~the $P_i(x)$ in Eq.(\ref{eq:chi}). The latter are used to determine how good the fit is with respect to the input parameters $(m_a,\,m_b)$ by giving an explicit value for $\chi^2_{(\delta)}$.  In the next step, we identify the region of the global $\chi^2_{(\delta)}$ minimum, chose a finer grid for the corresponding region in the $(m_a,\,m_b)$-plane and repeat the procedure until we determine the optimum set of input values.

As we will use the {\it Mathematica} package \texttt{REAP}~\cite{Antusch:2005gp} to solve the RG equations numerically, it is important to mention that the conventions used in \texttt{REAP} slightly differ from the ones discussed in Sec.~\ref{sec:intro}. First of all, with the help of Ref.~\cite{Antusch:2005gp}, we can relate the two neutrino Yukawa matrices, which leads to $\tilde{Y}_\nu=Y_\nu^\dagger$. This needs to be taken into account when entering explicit LS scenarios into \texttt{REAP}. Secondly, note that \texttt{REAP} also uses the PDG standard parametrisation which means that the mixing angles are identical to ours, and the Majorana phase is  given by $-\varphi_2/2=\sigma$. \texttt{REAP} uses $\delta_{\texttt{REAP}} \in [0,2\pi[$ whereas we use $\delta \in [-\pi,\pi[$. Consequently, it is $\delta=\delta_{\texttt{REAP}}-2\pi$.

%%%%%%%%%%%%%%%%%%%%%%%%%%%%%%%%%%%%%%%%%%%%%%%
\section{\label{sec:5} SM Results}
%%%%%%%%%%%%%%%%%%%%%%%%%%%%%%%%%%%%%%%%%%%%%%%

We investigate the running effects on the neutrino parameters $m_2,\,m_3,\,\vartheta_{12},\,\vartheta_{13},\,\vartheta_{23},\,\delta$ and $\sigma$ numerically by means of \texttt{REAP}~\cite{Antusch:2005gp}. Our analysis involves not only the four different cases A, B, C, and D but also four settings for the heavy RH neutrino masses, namely $(M_2,\,M_1)=(10^{12},10^{10}),\,(10^{15},10^{10}),\,(10^{15},10^{12}),\,(10^{14},10^{13})$. For each case and RH mass setting, we furthermore perform vacuum stability checks which validate all scenarios under consideration. As we fixed two of the four input parameters of the LS, namely $(n,\,\eta)$, depending on the case, we minimalise $\chi^2_{(\delta)}$ with respect to the free input parameters $(m_a,\,m_b)$. From the scan of the free input parameters, we determine the optimum set of $(m_a,\,m_b)$ at the GUT scale, which are presented in Tab.~\ref{tab:SM_CaseAD_best} together with their corresponding $\chi^2_{(\delta)}$ values (obtained at the EW scale). Overall, it turns out that the values for $\chi^2_\delta$ are only slightly inferior to the ones for $\chi^2$ -- by about a few percent at most -- and both measures for the goodness-of-fit point towards the same input values $(m_a,\,m_b)$. Thus, we will refer to $\chi^2$ in the following discussion. 

\begin{table}[h!]
\centering
\setstretch{1.2}
 \begin{tabular}{c c|c|c|c|c|c}
& $M_{atm}$ [GeV] & $M_{sol}$ [GeV] & $m_a$ [meV] & $m_b$ [meV] & $\chi^2$ & $\chi^2_\delta$ \\
 \hline
 & $10^{10}$ & $10^{12}$ & 35.670 & 3.6221 & 11.778 & 11.8275 \\
  &$10^{10}$ & $10^{15}$ & 37.968 & 4.1578 & 7.16772 & 7.18596 \\
    \rowcolor{YellowGreen}
\cellcolor{white}& $10^{12}$ & $10^{15}$ & 39.505 & 4.1592 & 7.14042 & 7.15869 \\
 \multirow{-4}{*}{Case A}& $10^{13}$ & $10^{14}$ & 38.011 & 3.7985 & 10.7043 & 10.7479 \\
  \hline
& $10^{10}$ & $10^{12}$ & 35.636 & 3.6600 & 6.41862 & 6.43381 \\
 &$10^{10}$ & $10^{15}$ & 37.958 & 4.2020 & 4.40508 & 4.45905 \\
    \rowcolor{YellowGreen}
\cellcolor{white}  &$10^{12}$ & $10^{15}$ & 39.498 & 4.2031 & 4.38607 & 4.44012 \\
 \multirow{-4}{*}{Case B} &$10^{13}$ & $10^{14}$ & 37.978 & 3.8377 & 5.85644 & 5.87664 \\
  \hline
 &$10^{12}$ & $10^{10}$ & 36.950 & 3.4974 & 11.7597 & 11.8094 \\
 &$10^{15}$ & $10^{10}$ & 47.215 & 3.9735 & 3.24554 & 3.31094 \\
    \rowcolor{YellowGreen}
 \cellcolor{white}  &$10^{15}$ & $10^{12}$ & 47.226 & 4.1757 & 3.23646 & 3.30174 \\
\multirow{-4}{*}{Case C} & $10^{14}$ & $10^{13}$ & 39.029 & 3.7492 & 9.88824 & 9.93932 \\
 \hline
& $10^{12}$ & $10^{10}$ & 36.915 & 3.5340 & 6.40423 & 6.41938 \\
 & $10^{15}$ & $10^{10}$ & 47.188 & 3.9885 & 1.4981 & 1.52676 \\
    \rowcolor{YellowGreen}
 \cellcolor{white}  & $10^{15}$ & $10^{12}$ & 47.198 & 4.1913 & 1.49388 & 1.52265 \\
\multirow{-4}{*}{Case D} & $10^{14}$ & $10^{13}$ & 38.994 & 3.7843 & 5.21486 & 5.23251
 \end{tabular}
 \setstretch{1}
 \caption{\label{tab:SM_CaseAD_best} Best Fit values for SM Cases A, B, C and D with varying right-handed neutrino masses}
\end{table}

When comparing the different RH neutrino mass settings for each case, respectively, there are several observations to reflect about:
\begin{itemize}
\item The first and foremost observation is that the RH mass setting $(10^{15},10^{12})$ makes for the best fit to the global fit values given in Tab.~\ref{tab:ExpVal} for each of the LS cases individually; closely followed by the mass setting $(10^{15},10^{10})$. The scenario $(10^{14},10^{13})$ is already significantly poorer, and the goodness-of-fit further deteriorates for $(10^{12},10^{10})$. 
This shows that it is beneficial for the running effects to have $M_2$ closer to the GUT scale. In addition, the mass of $M_1$ barely -- as long as still viable for a seesaw scenario -- changes the outcome which is to say that the heavier of the RH neutrinos plays the dominant role regarding RG running behaviour and the goodness-of-fit. The detailed results for the RH mass setting $(10^{15},10^{12})$ are shown in Figs.~\ref{fig:SM_CaseA} to \ref{fig:SM_CaseD}. The results for the remaining three mass settings are displayed in Tabs.~\ref{fig:SM_AB} and \ref{fig:SM_CD}.
\item For case A the best-fit values for $m_b$ for mass settings $(10^{15},10^{12})$ and $(10^{15},10^{10})$ -- which yield nearly identical $\chi^2$'s -- are almost the same, while the $m_a$ differ notably. Furthermore, $m_b$ decreases with $M_2$. The same is true for case B.  For cases C and D, respectively, it is the best-fit values for $m_a$ that are almost identical for the comparatively good RH mass settings $(10^{15},10^{12})$ and $(10^{15},10^{10})$, and $m_b$ that does vary. Moreover, $m_a$ lowers with $M_2$. Recalling the qualitative discussion in Sec.~\ref{sec:3}, these observations can most likely be traced back to the deeper connection between Case $A \leftrightarrow B$  as well Case $C\leftrightarrow D$. For $A \leftrightarrow B$, the parameter $b \propto \sqrt{M_2 m_b}$ dominates the RG effects of the mixing angles, whereas for $C\leftrightarrow D$, the parameter $a \propto \sqrt{M_2 m_a}$ does so. This already hints towards the overall importance of the running of the mixing angles in order to predict feasible neutrino parameters at the EW scale, which we will come back to when investigating the different LS cases. This line of reasoning also explains the first observation, namely that the mass of the heavier RH neutrino impacts the goodness-of-fit predominantly.
\item Case A and B yield a nearly identical input parameter $m_a$ for each RH neutrino mass setting individually, which hints towards yet another correlation between Case A and B. The same holds true for Case C and D with slightly more deviation in $m_a$ in comparison to Case $A\leftrightarrow B$. For the input parameter $m_b$, there does not seem to be a correlation between the different LS cases. While the discussion above did feature equivalent RG behaviour of two LS cases, respectively, this observation shows a correlation with respect to the absolute value of $m_a$. The reason behind this connection, however, proves more elusive because $m_a$ is related to the lighter RH neutrino scale for Case A,B but to the heavier scale for Case C,D. Nevertheless, we will return to discussing this feature towards the end of this section. 
\end{itemize}
\begin{table}[h!]
\centering
\setstretch{1.2}
 \begin{tabular}{c|c|c|c|c}
Case & $\chi_{old}^2 (\Lambda_{\rm EW})$ & $\chi^2_{\delta\,\,old} (\Lambda_{\rm EW})$& $\chi^2 (\Lambda_{\rm EW})$ & $\chi^2_\delta (\Lambda_{\rm EW})$ \\
 \hline
A & $50.3072$ & $50.318$ & $7.14042$ & $7.15869$ \\
B & $50.3012$ & $50.3739$ & $4.38607$ & $4.44012$ \\
C & $179.711$ & $179.824$ & $3.23646$ & $3.30174$ \\
D & $172.773$ & $172.781$ & $1.49388$ & $1.52265$
 \end{tabular}
 \setstretch{1}
 \caption{\label{tab:SM_RGEcompare} $\chi^2$ values for the four cases, where the subscript $old$ denotes the input parameters used in Ref.~\cite{King:2016yef}, namely $m_a=41.5156~{\rm meV}$ and $m_b=4.19375~{\rm meV}$. In order to compare these to the results from this paper's analysis, we also include their $\chi^2$ values in the two right-handed columns of this table. Please bare in mind that the latter are based on varying input parameters $m_{a,b}$, which are specified in Tab.~\ref{tab:SM_CaseAD_best}.}
\end{table}
To emphasise the importance of performing global fits to the experimental data at the EW scale for each LS case separately, we compare the $\chi^2$ values of the modified benchmark scenarios from Ref.~\cite{King:2016yef} with the best-fit scenarios obtained from our analysis. As already mentioned in the Sec.~\ref{sec:intro}, the input values $(m_a,\,m_b)$ in Ref.~\cite{King:2016yef} are taken from a tree-level best fit, and adjusted by an overall factor of $1.25$, which was obtained from Case A and aims at including the significant running of the neutrino masses\footnote{Please note that the overall factor of $1.25$ is applied to $Y_\nu$, which translates to a factor of $1.25^2$ on the input parameters $m_{a,b}$. Furthermore, there is a typo in Ref.~\cite{King:2016yef} when quoting the parameter $m_a$. The tree-level values used are in fact $m_a=26.57~{\rm meV}$ and $m_b=2.684~{\rm meV}$.}. In contrast, our analysis scans over the model input parameters in order to determine the optimum set of high energy input values from a global fit of the low energy parameters. The $\chi^2$ values for the input parameters used in Ref.~\cite{King:2016yef} are listed in Tab.~\ref{tab:SM_RGEcompare}.  Comparing these to the $\chi^2$ values presented in Tab.~\ref{tab:SM_CaseAD_best}, there are two striking characteristics. First of all, the overall values for the goodness-of-fit improve drastically moving the $\chi^2$ values from "in tension with experimental data" to "predict experimental data nicely". Secondly, the $\chi^2$ values listed in Tab.~\ref{tab:SM_RGEcompare} suggest that Case A is most compatible with experimental data, followed closely by Case B and after a significant gap by Case D and C. It turns out that quite the opposite is true when performing global fits for  each case individually, resulting in the order: Case D yields best fit, followed by Case C, Case B and Case A. Both these features can be traced back to Ref.~\cite{King:2016yef} superficially modifying the input parameters to fit Case A. As we have already seen in the discussion above, the cases A and B are connected intrinsically while displaying detached behaviour from the also connected cases C and D, which does not only concern the running effects but also the absolute value of a suitable input parameter $m_a$. Consequently, the input parameters from Ref.~\cite{King:2016yef} work significantly better for Case A and B, but are, nevertheless, not even close to the best-fit choices due to the simplistic way of selecting them.\newline
 \begin{table}[h!]
\setstretch{1.5}
\begin{center}
\begin{tabular}{c|c}
Neutrino Parameter & best-fit-value \\
\hline \hline 
$\theta_{12}$ & $33.58^\circ$ \\
$\theta_{13}$ & $8.46^\circ$ \\
$\theta_{23}$ & $41.61^\circ$ \\
\hline
$m_2$ & $8.66~\rm{meV}$\\
$m_3$ & $50.24~\rm{meV}$\\
\hline 
$\delta $  & $-99^{\circ}$
\end{tabular}
\end{center}
\setstretch{1}
\caption{\label{tab:ExpVal_trans} Mixing angles, Dirac phase and neutrino masses in the LS as extracted from Tab.~\ref{tab:ExpVal}}
\end{table}
We choose the RH neutrino mass setting $(10^{15},10^{12})$ to further investigate the different LS cases. We are interested in understanding the features that decide how compatible with experimental data a case is and how this connects to the RG effects. \newline
From Tab.~\ref{tab:SM_CaseAD_best}, we learn that Case D is most compatible with experimental data. Moreover, Case D does not only more or less reproduce the neutrino parameters at the EW scale but does so impressively -- leading to $\chi^2=1.49$. The next best scenario is Case C with a still impressing $\chi^2=3.24$, followed closely by Case B with $\chi^2=4.39$, and with some deterioration Case A with $\chi^2=7.14$. In order to understand the underlying characteristics that make Case D most suitable and Case A least, we start by investigating the behaviour of the neutrino parameters. From the tables displayed in the upper left corner of Figs.~\ref{fig:SM_CaseA} to \ref{fig:SM_CaseD}, respectively, there are several observations to consider:
\begin{figure}[p]
\begin{minipage}[c]{7cm}
  \includegraphics[width=7cm]{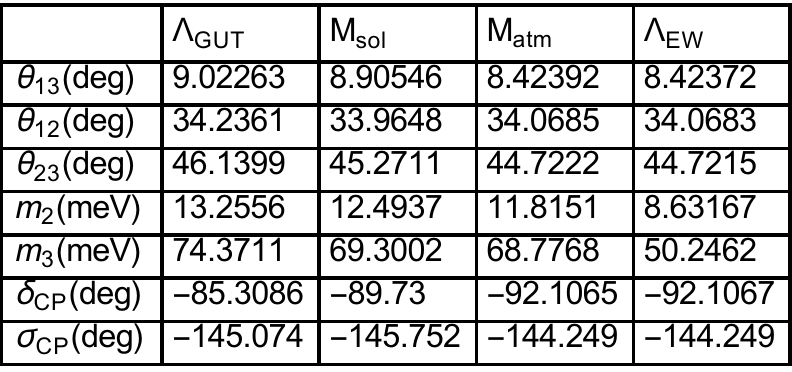}
  \end{minipage}
  \hspace*{-2cm}
  \begin{minipage}[c]{8cm}
    \includegraphics[width=7cm]{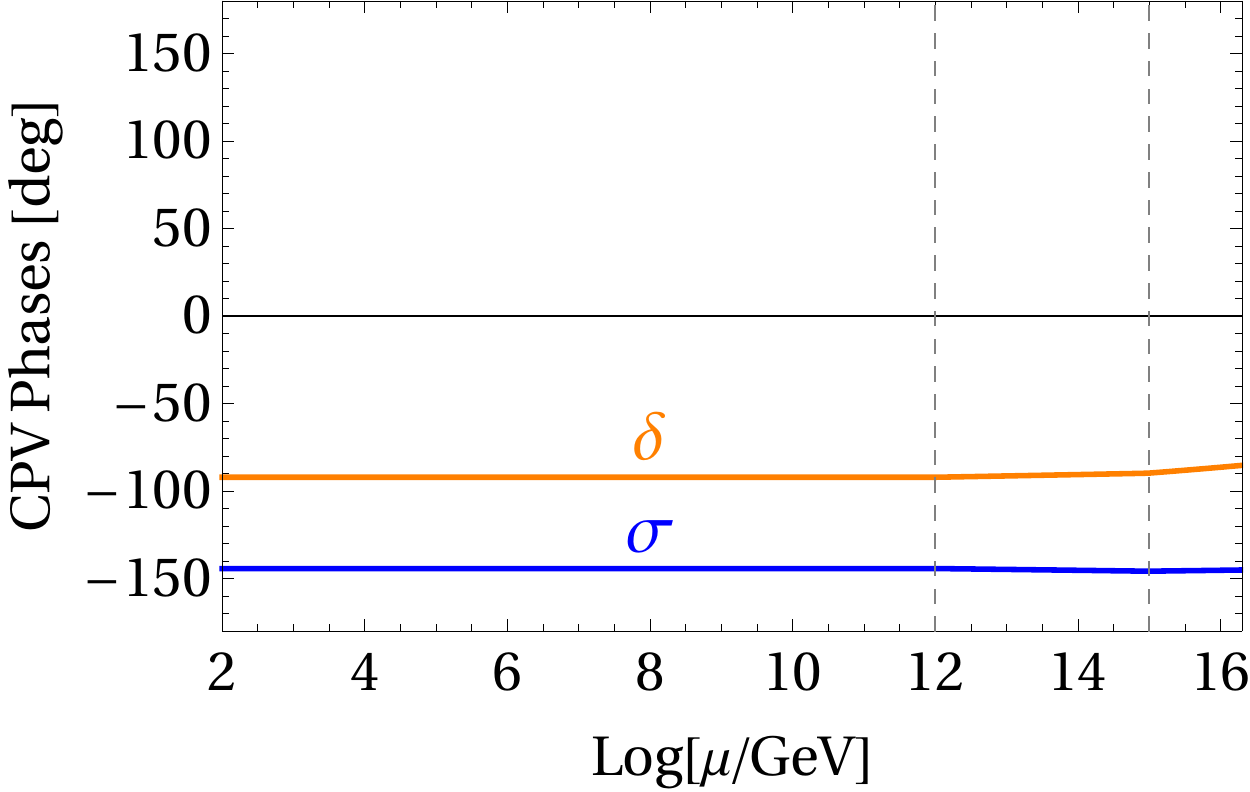}
  \end{minipage}
\begin{minipage}[c]{7cm}
  \includegraphics[width=7cm]{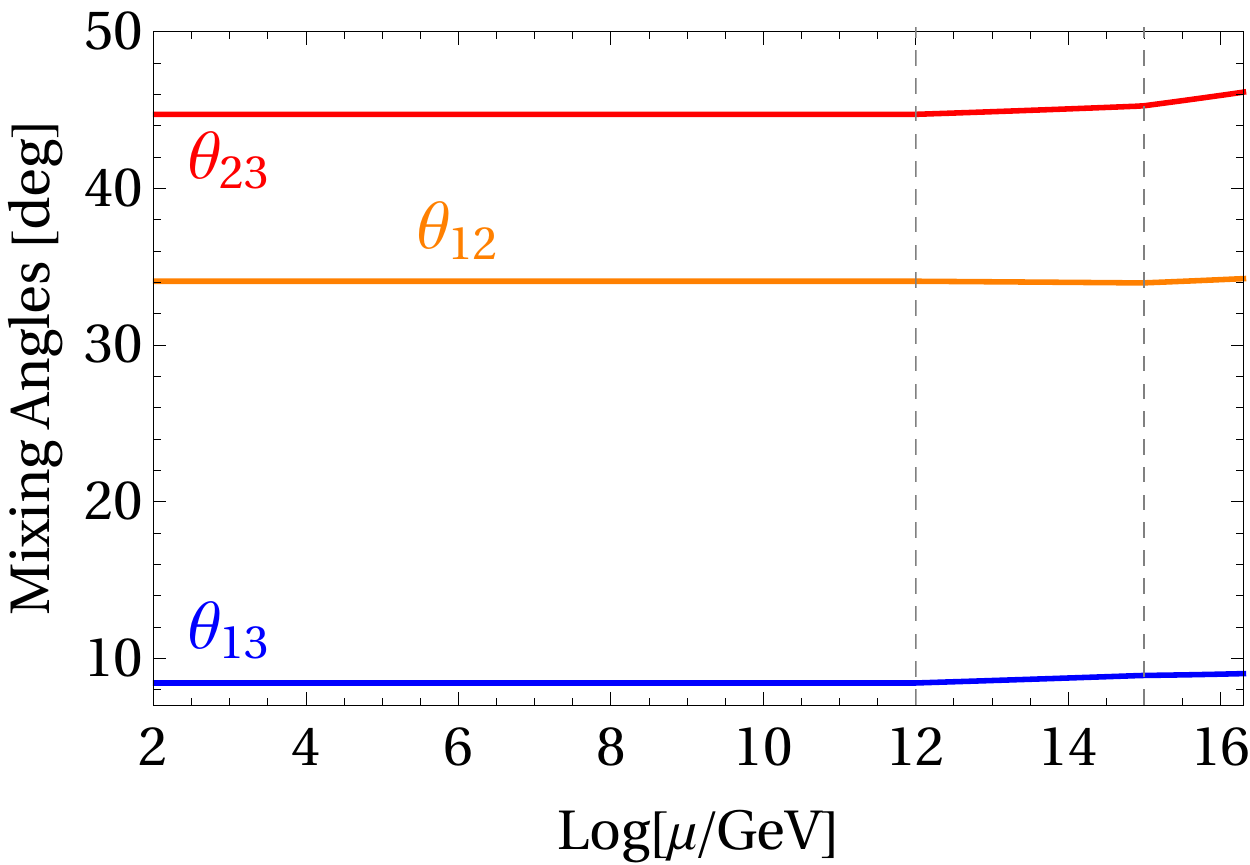}
\end{minipage}
\hspace{1cm}
\begin{minipage}[c]{7cm}
  \includegraphics[width=7cm]{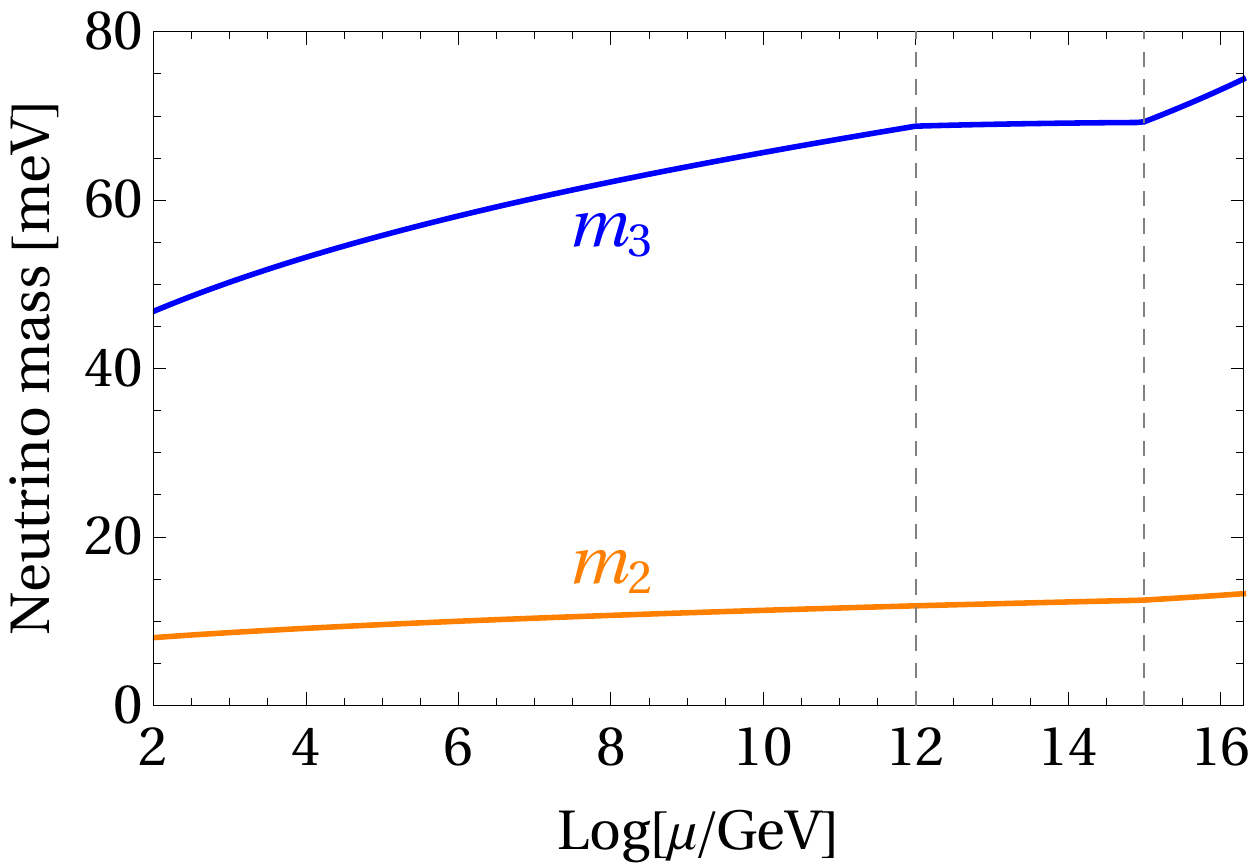}
\end{minipage}
\caption{\label{fig:SM_CaseA}Case A - SM with $M_{atm}=10^{12}~\rm{GeV}$ and $M_{sol}=10^{15}~\rm{GeV}$}
\end{figure}
\begin{figure}[p]
\begin{minipage}[c]{7cm}
  \includegraphics[width=7cm]{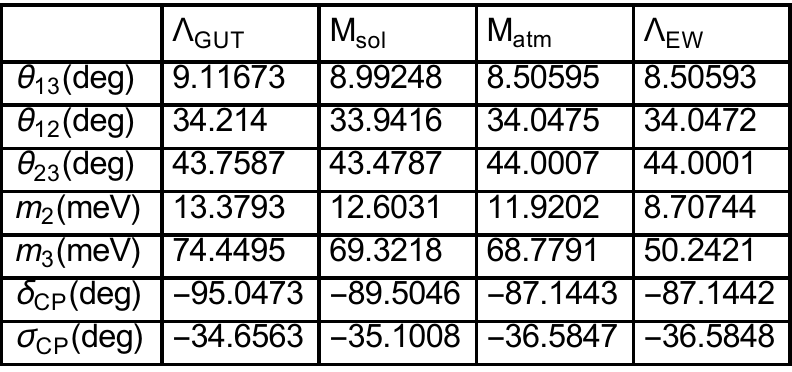}
  \end{minipage}
  \hspace*{-2cm}
  \begin{minipage}[c]{8cm}
    \includegraphics[width=7cm]{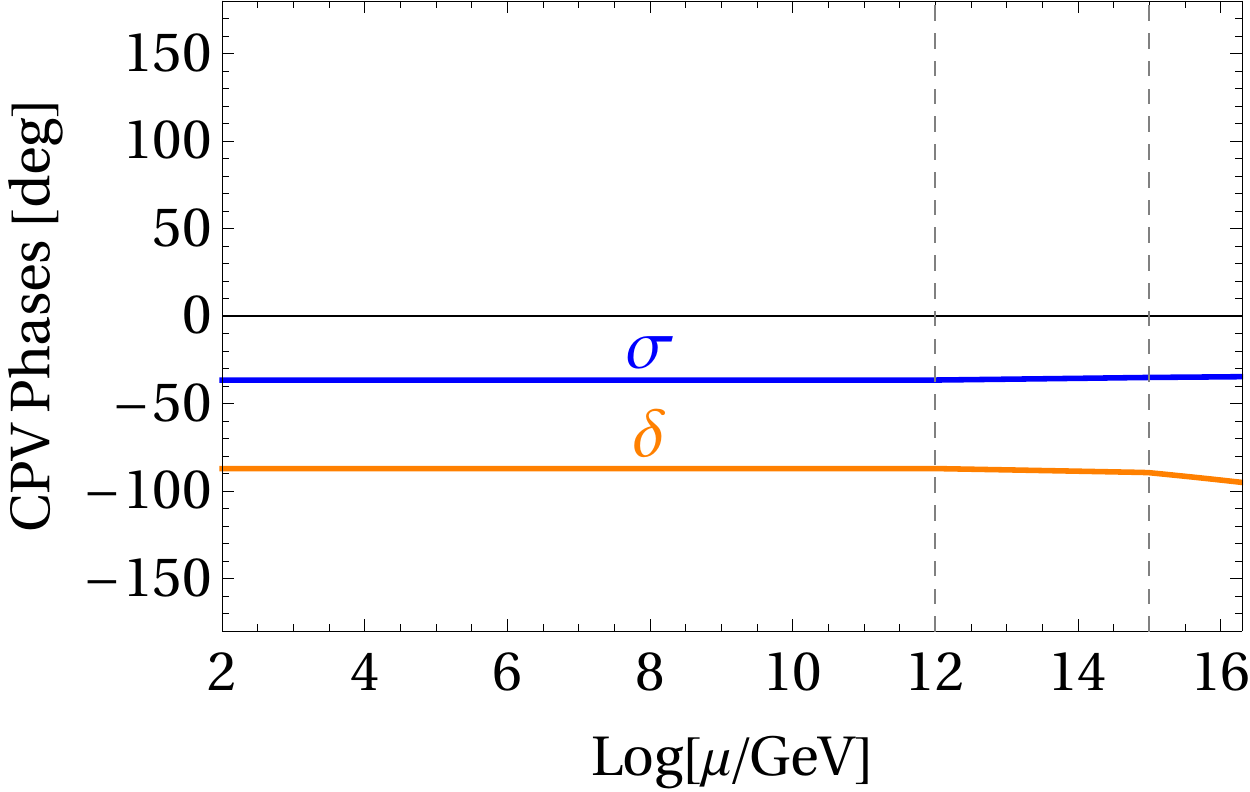}
  \end{minipage}
\begin{minipage}[c]{7cm}
  \includegraphics[width=7cm]{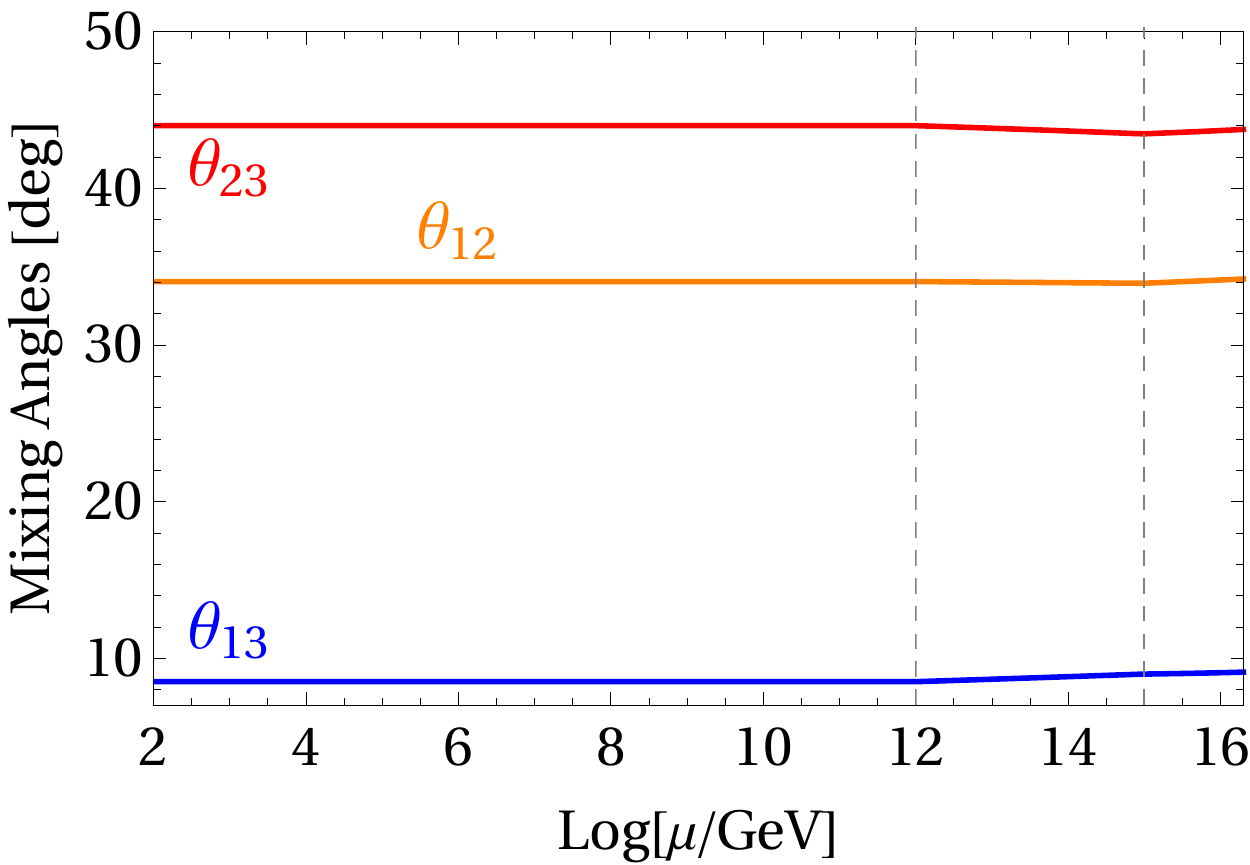}
\end{minipage}
\hspace{1cm}
\begin{minipage}[c]{7cm}
  \includegraphics[width=7cm]{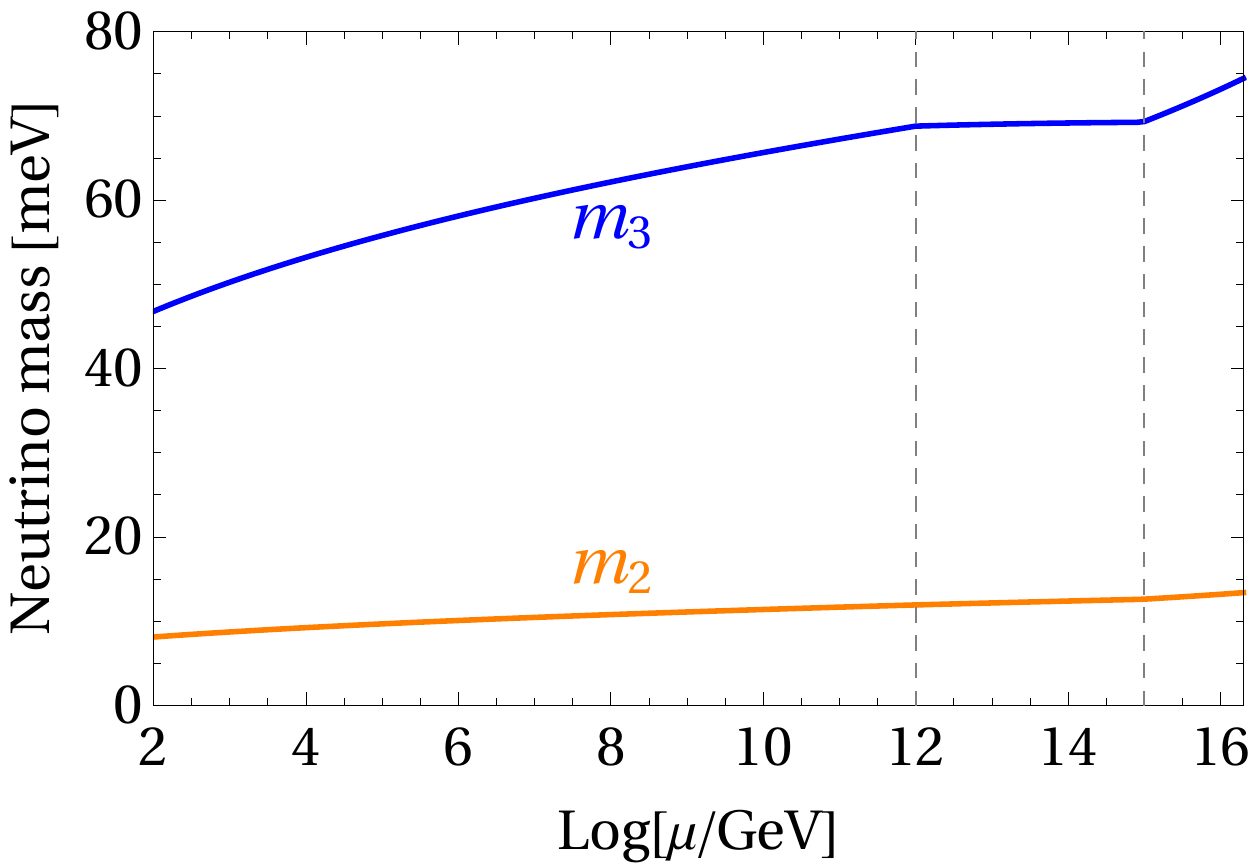}
\end{minipage}
\caption{\label{fig:SM_CaseB}Case B - SM with $M_{atm}=10^{12}~\rm{GeV}$ and $M_{sol}=10^{15}~\rm{GeV}$}
\end{figure}
\begin{figure}[p]
\begin{minipage}[c]{7cm}
  \includegraphics[width=7cm]{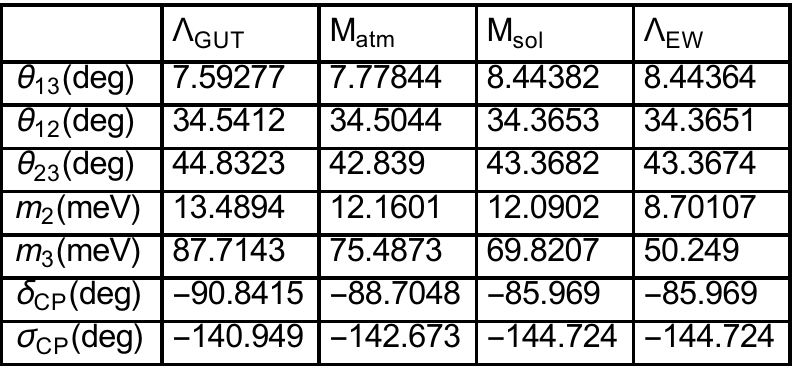}
  \end{minipage}
  \hspace*{-2cm}
  \begin{minipage}[c]{8cm}
    \includegraphics[width=7cm]{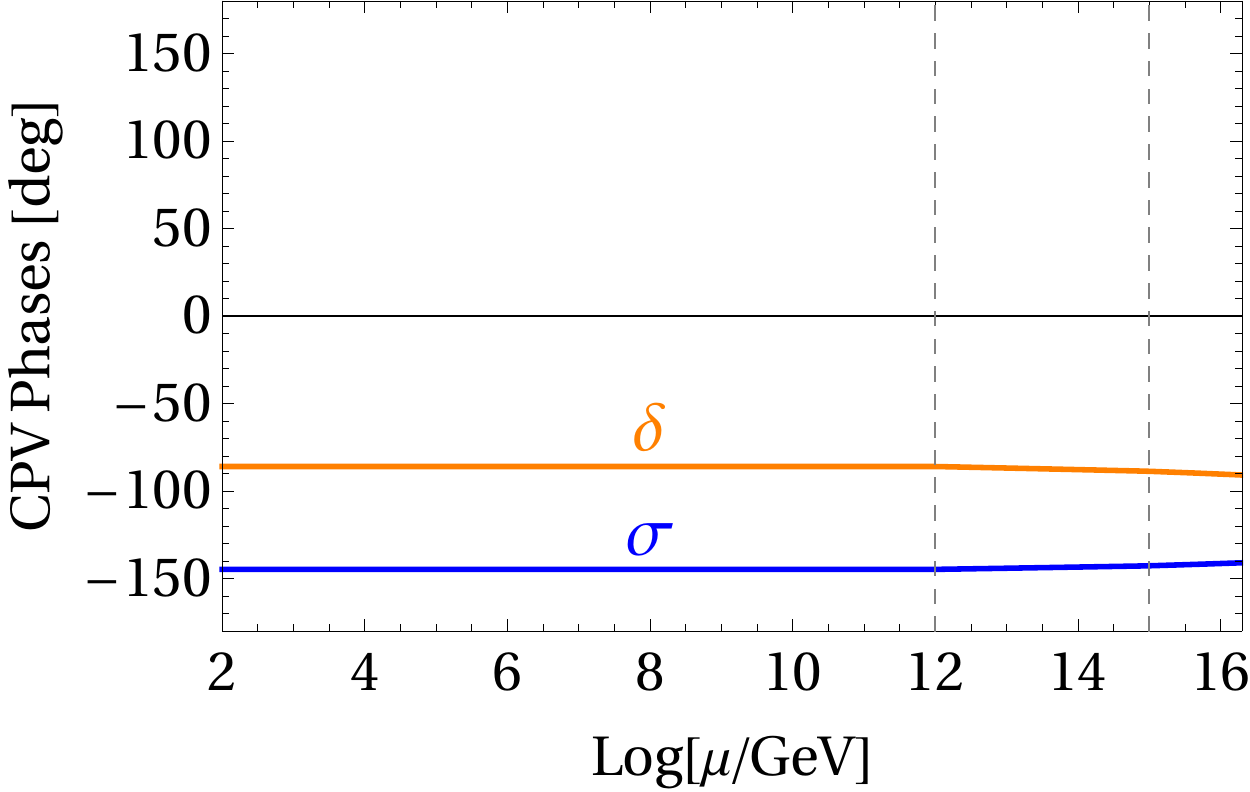}
  \end{minipage}
\begin{minipage}[c]{7cm}
  \includegraphics[width=7cm]{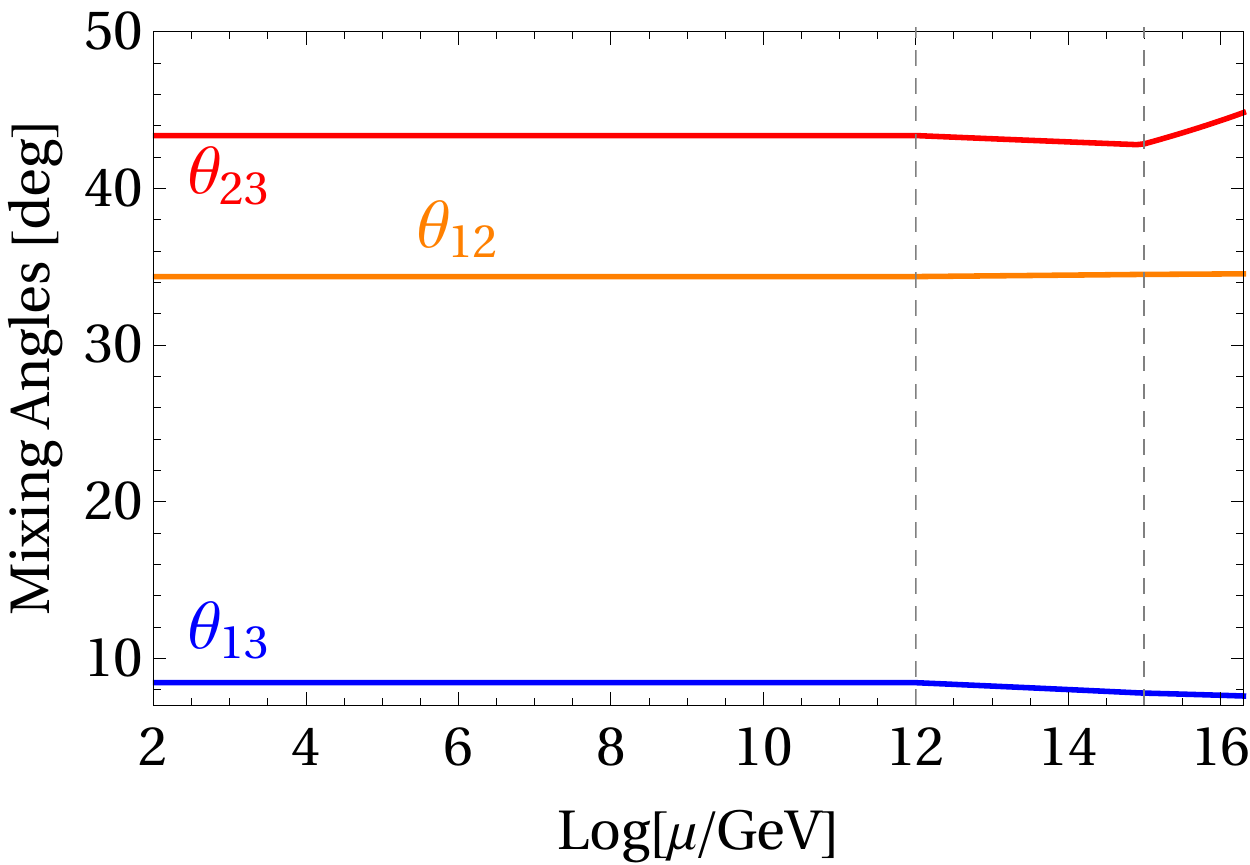}
\end{minipage}
\hspace{1cm}
\begin{minipage}[c]{7cm}
  \includegraphics[width=7cm]{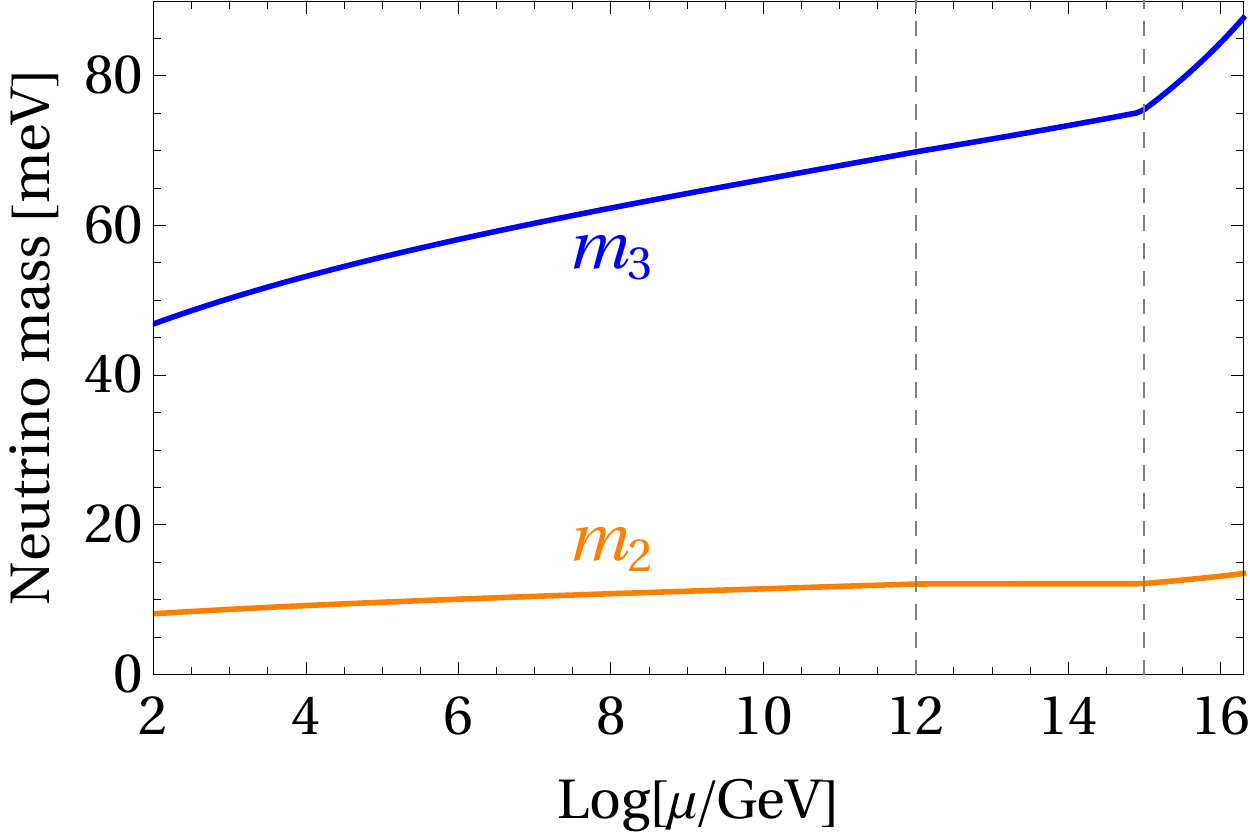}
\end{minipage}
\caption{\label{fig:SM_CaseC}Case C - SM with $M_{atm}=10^{15}~\rm{GeV}$ and $M_{sol}=10^{12}~\rm{GeV}$}
\end{figure}
\begin{figure}[p]
\begin{minipage}[c]{7cm}
  \includegraphics[width=7cm]{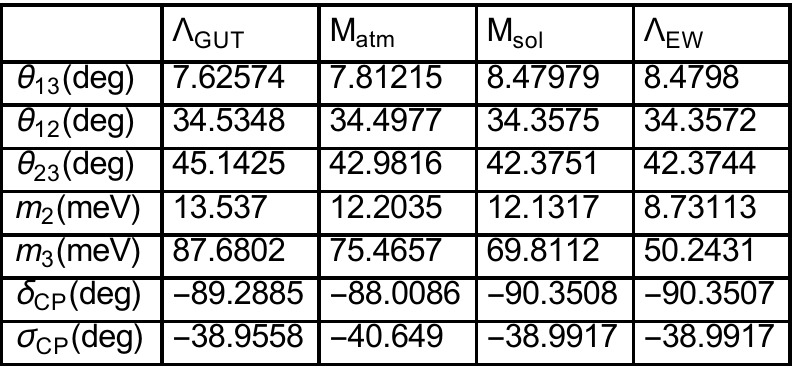}
  \end{minipage}
  \hspace*{-2cm}
  \begin{minipage}[c]{8cm}
    \includegraphics[width=7cm]{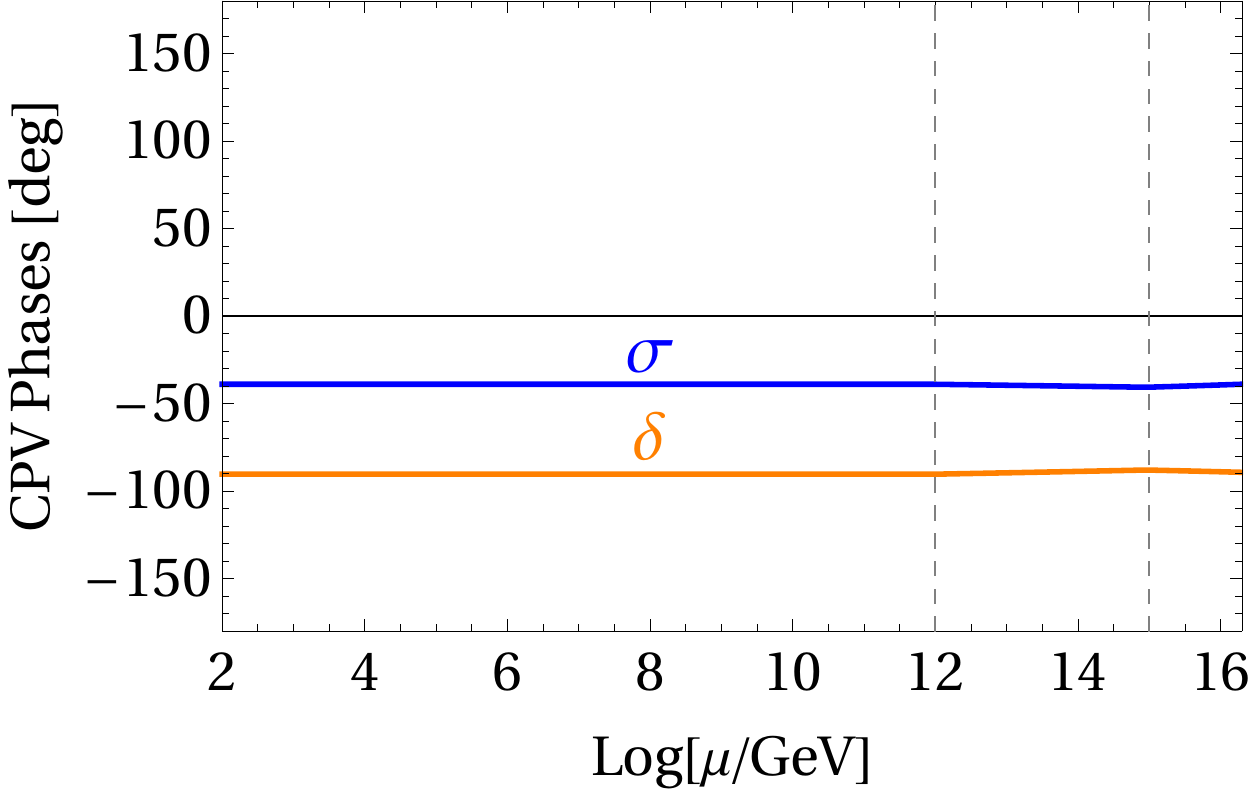}
  \end{minipage}
\begin{minipage}[c]{7cm}
  \includegraphics[width=7cm]{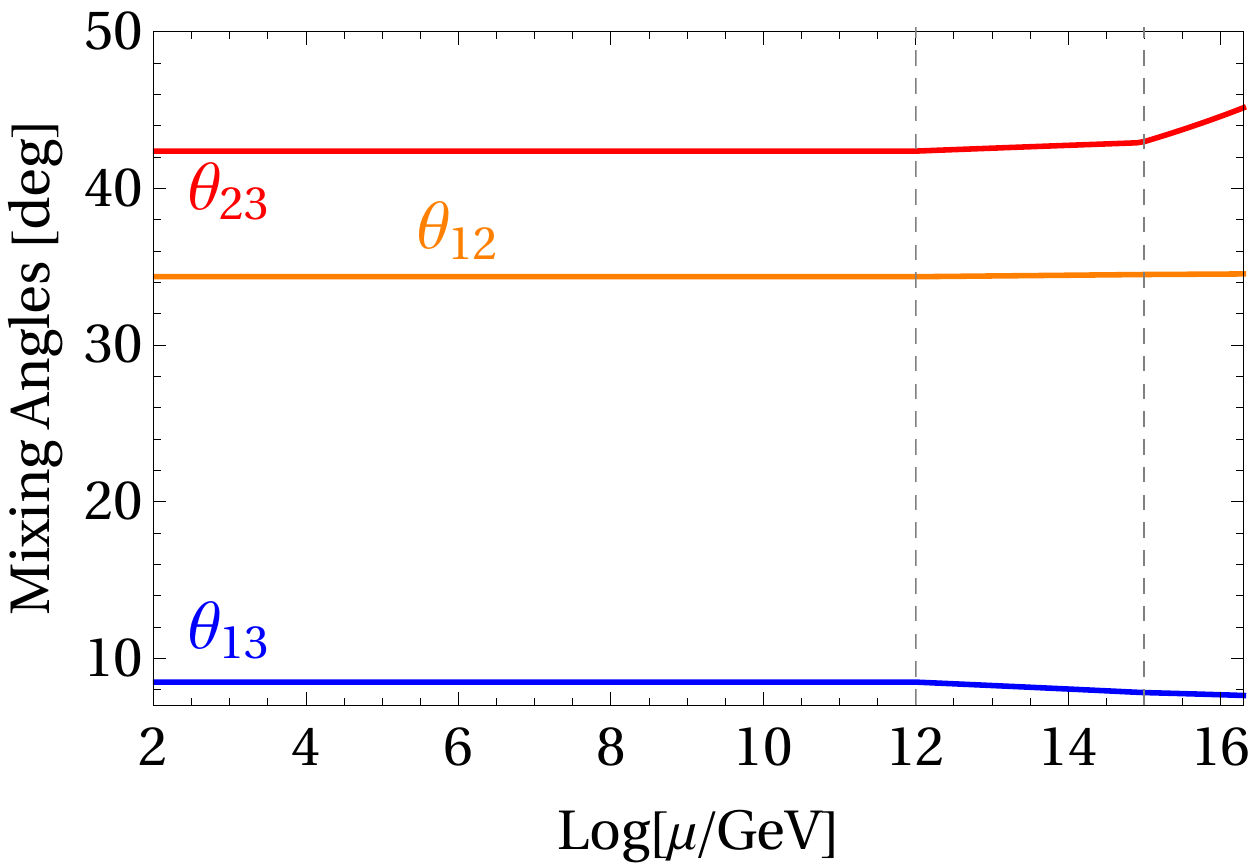}
\end{minipage}
\hspace{1cm}
\begin{minipage}[c]{7cm}
  \includegraphics[width=7cm]{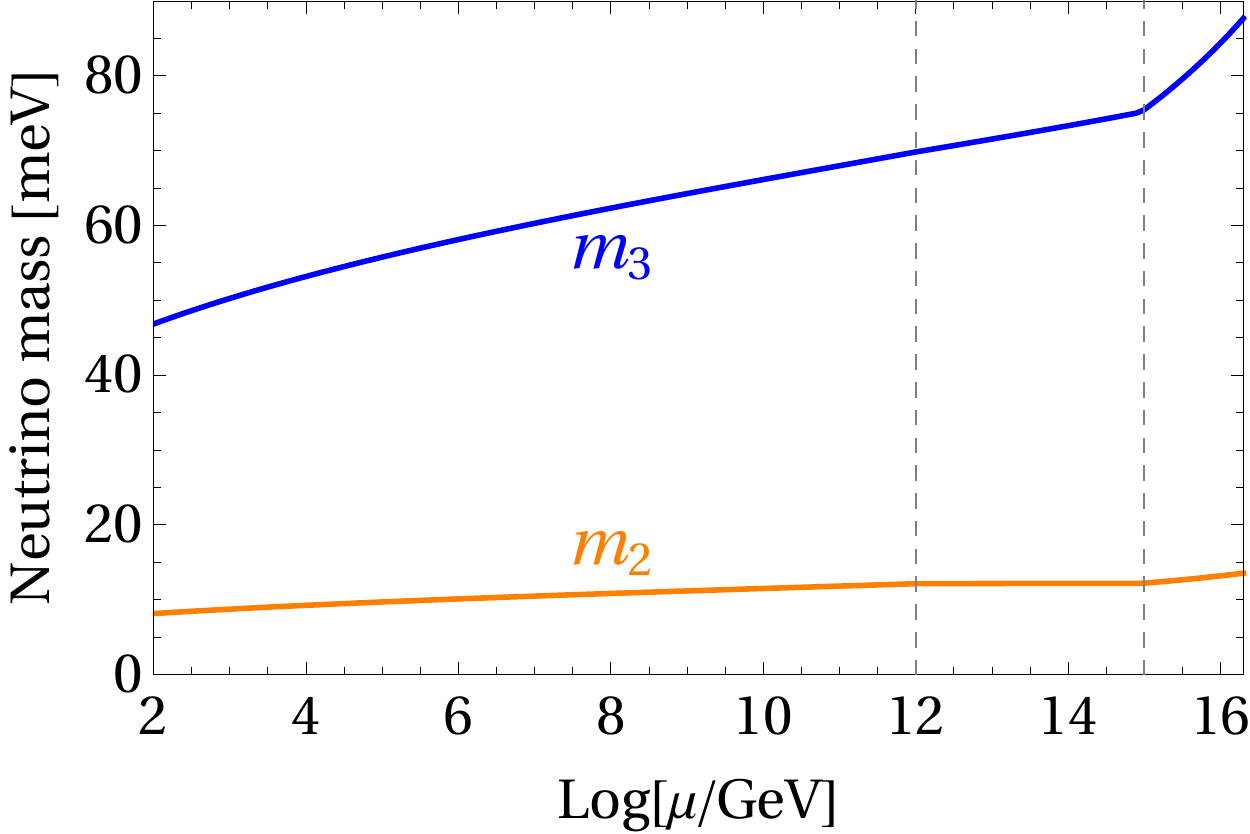}
\end{minipage}
\caption{\label{fig:SM_CaseD}Case D - SM with $M_{atm}=10^{15}~\rm{GeV}$ and $M_{sol}=10^{12}~\rm{GeV}$}
\end{figure}
\begin{itemize}
\item Starting with the mixing angle $\mathbf{\vartheta_{12}}$, we compare its experimental value to the predictions of the four different LS cases at the EW scale. Case A's best fit scenario predicts $\vartheta^A_{12}=34.07^\circ$, Case B's  $\vartheta^B_{12}=34.05^\circ$, Case C's  $\vartheta^C_{12}=34.37^\circ$, and Case D's predicts  $\vartheta^D_{12}=34.36^\circ$. First of all, we once again note the nearly identical predictions for Case A,B and Case C,D. Secondly, the measured solar angle of $\vartheta^{exp}_{12}=33.58^\circ$ -- see Tab.~\ref{tab:ExpVal_trans} -- lies below the range of predicted values with cases A and B getting closest. However, the variation among the predicted values is small. That is to say, that the $1\sigma$ deviation from the measured value gives a range of $[32.83^\circ,34.33^\circ]$, which encompassed cases A and B, and is fairly close to the values predicted for Case C and D.  Taking a closer look at the influence of the RG running effects, on display in the lower left panels of Figs.~\ref{fig:SM_CaseA} to \ref{fig:SM_CaseD}, it turns out that the overall alteration of $\vartheta_{12}$ due to the running in between the GUT and the EW scale is almost identical four all four cases. Following a decline by roughly $0.27~{\rm meV}$ with the energy scale from GUT to EW scale, the observed connection between Case A,B as well as Case C,D already occurs at the GUT scale and translates to the EW scale. 
\item Next, we analyse the mixing angle $\mathbf{\vartheta_{13}}$. From Case A, we obtain $\vartheta^A_{13}=8.42^\circ$. From Case B, the reactor angle is predicted to be $\vartheta^B_{13}=8.51^\circ$, while we obtain $\vartheta^C_{13}=8.44^\circ$ for Case C and $\vartheta^D_{13}=8.48^\circ$ for Case D. The first and somewhat unexpected observation is that for $\vartheta_{13}$, there seems to be no clear correlation between the cases from the predicted angle at the EW scale. Second of all, the measured value $\vartheta^{exp}_{13}=8.46^\circ$, see Tab.~\ref{tab:ExpVal_trans}, is right in the middle of the range of predicted angles. Including the $1\sigma$ deviations from the measured best-fit angle, one obtains a region of $[8.31^\circ,8.61^\circ]$, which covers the predicted angles from all four LS cases. The running effects for $\vartheta_{13}$ are highly case-dependent, see the lower left panel of Figs.~\ref{fig:SM_CaseA} to \ref{fig:SM_CaseD}, respectively. While the reactor angle decreases with the energy scale by roughly $0.6^\circ$ from the GUT to the EW scale for cases A and B, it increases by about $0.85^\circ$ over the same area for cases C and D. This dilutes the fact that cases A and B are indeed generating quite similar $\vartheta_{13}$ at the GUT scale, as do cases C and D. At the EW scale, all four scenarios have converged and the predicted angles do not reveal the original connection anymore.
\item For the atmospheric mixing angle $\mathbf{\vartheta_{23}}$, we obtain $\vartheta^A_{23}=44.72^\circ$ for Case A, $\vartheta^B_{23}=44.00^\circ$ for Case B, $\vartheta^C_{23}=43.37^\circ$ for Case C and $\vartheta^D_{23}=42.37^\circ$ for Case D. Firstly, as with the reactor angle, there is no apparent correlation between the four LS cases with respect to the atmospheric angle at the EW scale. Secondly, the measured value of $\vartheta^{exp}_{23}=41.61^\circ$ is below the range of predicted values. Here, however, it is Case D that is closest to the experimental best-fit value. Considering the $1\sigma$ uncertainties of the measurement, the atmospheric angle lies withing $[40.40^\circ,43.17^\circ]$, which only covers the prediction of Case D. Note that in contrast to the other two mixing angles, where either all cases where within the $1\sigma$ region or at least at close proximity, only Case C is somewhat close to the $1\sigma$ region for the atmospheric angle, whereas cases A and B are well beyond the upper margin. Furthermore, $\vartheta_{23}$ also differs from the other mixing angles in terms of its connections between the LS cases. Not only do the best-fit scenarios for the different LS cases predict quite distinct values at the EW scale, but they also exhibit no connection with respect to the values at the GUT scale and RG running behaviour. The latter manifests in Case A displaying a decrease by $0.33^\circ$ in between the GUT and the EW scale, whereas Case B has an increase by $0.24^\circ$. It is striking that Case A does not only differ from Case B in running strength but in direction. Case C, moreover, displays a decrease by $1.46^\circ$ while Case D shows an even stronger decrease of $2.77^\circ$. In combination with the already dissimilar GUT scale values, we obtain a strong preference towards Case D based on its predicted $\vartheta_{23}$. Thus, the atmospheric angle plays the decisive role with regard to the compatibility of the LS cases with experimental data.
\item When including the CP violating Dirac phase in the goodness-of-fit analysis, we also need to discuss its predicted values with respect to the measured value and the $1\sigma$ region. Since the $1\sigma$ region encompasses values within $[-158^\circ,-48^\circ]$ around a best-fit experimental value of $\delta=-99^\circ$, all Dirac phases derived from the LS cases lie within this range. Moreover, they are also within a -- relative to the $1\sigma$ region -- narrow band above the best-fit value, namely $\delta^A=-92.11^\circ$, $\delta^B=-87.14^\circ$, $\delta^C=-85.97^\circ$, and $\delta^D=-90.35^\circ$. This explains why the difference between $\chi^2$ and $\chi^2_\delta$ is negligible. The running behaviour with respect to $\delta$ differs among the four LS cases. While Case B and C have $\delta$ increasing with decreasing energy scale, Case A and D display a decreasing $\delta$. Nonetheless, the strength of the running differs with  running effects in between $1^\circ$ and $7^\circ$. So, overall, there is no hint towards a relation between any of the four LS cases in $\delta$ -- neither in the starting values at the GUT scale or the values obtained at the EW scale, nor in the total running behaviour. Since involving the Dirac phase in the global fit does not alter the results, we will focus on the other five neutrino parameters in the discussion that is to follow.
\item Turning to the neutrino masses, we start by comparing the measured value of $\mathbf{m_2}$ to the LS predictions. Case A predicts a lighter neutrino mass of $m_2^A=8.63~{\rm meV}$. For Case B we obtain $m_2^B=8.71~{\rm meV}$, for Case C $m_2^C=8.70~{\rm meV}$, and $m_2^D=8.73~{\rm meV}$ for Case D. The experimental best-fit value is given by $m_2^{exp}=8.66~{\rm meV}$ with a $1\sigma$ region of $[8.56~{\rm meV}, 8.77~{\rm meV}]$. Consequently, all four LS cases predict similarly good values well within the $1\sigma$ region. Taking a closer look at the RG running behaviour of $m_2$, see the lower right panel in Figs.~\ref{fig:SM_CaseA} to \ref{fig:SM_CaseD}, we note that the running effects are similarly strong for all four cases. In addition, all four cases display a decrease of roughly $4.7~{\rm meV}$ for $m_2$ in between the GUT and the EW scale. At the GUT scale, the $m_2$ values show no clear relation between different cases and all lie in close range to each other, which leads to the four predicted EW scale masses being equally good. 
\item For the heavier of the light neutrino masses, $m_3$, we obtain the following predictions from the four LS cases: $m_3^A=50.25~{\rm meV}$, $m_3^B=50.24~{\rm meV}$, $m_3^C=50.25~{\rm meV}$, and  $m_3^D=50.24~{\rm meV}$. These nearly identical predictions are consistent with the experimentally measured value of $m_3^{exp}=50.24~{\rm meV}$ that lies within the $1\sigma$ region given by $[49.84~{\rm meV}, 50.63{\rm meV}]$.  The running of $m_3$ between the GUT scale and the EW scale is extreme for all four LS cases, see the lower right plot in Figs.~\ref{fig:SM_CaseA} to \ref{fig:SM_CaseD}, respectively. There are, furthermore, interesting features that we want to briefly discuss. First of all, the $m_3$ values for Case A and B at the GUT scale are nearly identical, which is also true for Case C and D. Thus, once again revealing a deeper connection among the cases that gets diluted by the running. Second of all, the GUT scale values of Case C,D are significantly higher than those of Case A,B. Consequently, as the RG effects decrease the value of $m_3$ with the energy scale, the RG running in cases C and D are notably stronger. 
\end{itemize}
From the discussion of the neutrino parameters, we can summarise the following. First of all, the absolute value predicted for the parameters $\vartheta_{12}$, $\vartheta_{13}$ and $m_3$ at the GUT scale are nearly identical for Case A,B as well as for Case C,D. As opposed to this, the predictions for $\vartheta_{23}$ and $m_2$ at the GUT scale are without case induced pattern. Second of all, the RG running of $m_3$ and $\vartheta_{13}$ are similar for Case A,B and Case C,D, respectively. On top of that, $m_2$ and $\vartheta_{12}$ exhibit the same RG running behaviour for all four LS cases. The only parameter not showing any case-dependent pattern is $\vartheta_{23}$.  \newline
What can we learn from these observations and where do they come from? As we already realised when investigating the different RH neutrino mass settings, there are two additional connections between Case A and B as well as Case C and D, namely the absolute value of the input parameter $m_a$ for the best-fit scenario and the predominant dependence on either $m_b$ or $m_a$ of the RG running of the mixing angles. In order to understand the reasoning behind the above observations, we need to briefly recap some basic features of the LS and its RG running:
\begin{itemize}
\item From Ref.~\cite{King:2016yef}, we can extract the following estimates for the neutrino parameters at the GUT scale derived for Case A:
\begin{equation}
m_2\approx 3 m_b \,,\qquad m_3\approx 2m_a 
\label{eq:Neutrino_GUT}
\end{equation}
\begin{equation}
\sin \vartheta_{13} \approx \frac{\tan 2\theta}{2\sqrt{3}}\,,\,\tan \vartheta_{12} \approx \frac{1}{\sqrt{2}}\big(1-\frac{1}{4} \tan^2 2\theta \big)^{1/2}\,, \, \tan \vartheta_{23} \approx 1+\frac{2 \tan 2\theta}{\sqrt{6}} \cos \omega\,,
\label{eq:Angles_GUT}
\end{equation}
with $\tan 2\theta \approx \sqrt{6} m_b (n-1)/|m_a+m_b {\rm e}^{i\eta}(n-1)^2|$ and $\omega = {\rm arg}[m_a+m_b {\rm e}^{i\eta}(n-1)^2] -\eta$. Without running effects, these estimations also hold true for Case C. The mixing parameters for Case B, and since we do not need to consider running effects at the GUT scale also Case D, are $m_{2,3}^B=m_{2,3}^A$, $\vartheta^B_{12}=\vartheta^A_{12}$, $\vartheta^B_{13}=\vartheta^A_{13}$ and $\vartheta^B_{23}=\pi/2-\vartheta^A_{23}$.\newline
Although, we have only drawn a connections between Case A,B and Case C,D with respect to input parameter $m_a$, one has to bare in mind that the input values $m_b$ are all within a close range, namely within $[4.16~{\rm meV},\,4.20~{\rm meV}]$. A variation of only $0.04~{\rm meV}$ does not alter $\tan 2\theta$ or $\omega$ significantly. Consequently, Case A and B yield similar $\tan 2\theta$ or $\omega$. As do Case C and D. 

These estimates already answer why for similar $m_a$, as given for Case A,B and Case C,D the neutrino parameters $m_3$, $\vartheta_{12}$ and $\vartheta_{13}$ are almost identical at the GUT scale. It also explains, why the GUT scale values for parameter $m_2$ -- predominantly depending on input parameter $m_b$ -- are within a close range without exhibiting a clear case-dependent structure. And at last, it unveils why the $\vartheta_{23}$ values at the GUT scale do not show any indication of the connection between the different cases. The connection between the cases appears in the choice of $m_a$, and would suggest similar atmospheric angles for Case A and B (or analogously for Case C and D). However, due to the relation between the atmospheric angle for A and B, as given above, there is an offset of a few degrees. The same is true for Case C and D. 
\item Furthermore, from Ref.~\cite{King:2016yef}'s derivation of the mixing angles RGE running for Case A and B, we know that for $\mu>M_2$ {\it only the running of $\vartheta_{23}$} differs for Case A and B. The latter is significant as for the atmospheric angle most of the running occurs within that region. Moreover, the corrections to the GUT scale value of $\vartheta_{23}$ come with opposite signs for Case A and B, which explains why one decreases and the other increases its atmospheric angle. The running behaviour of the mixing angles in EFT 1 differs for cases A and B but is still quite similar since only the coefficients in front of a few terms are different. As the same structure is responsible for the running above $M_2$ for cases C and D, there is no sign change, which agrees with our numerical observations. On the other hand, our numerical results indicate, that for the regime $M_2>\mu >M_1$ $\vartheta_{23}$ increases for Case C but further decreases for Case D, which gives an edge to Case D regarding the global fit to data. A more in-depth investigation of this feature, however, it is beyond the scope of this work.
\end{itemize}
\begin{table}[p]
\begin{minipage}[c]{7cm}
  \includegraphics[width=7cm]{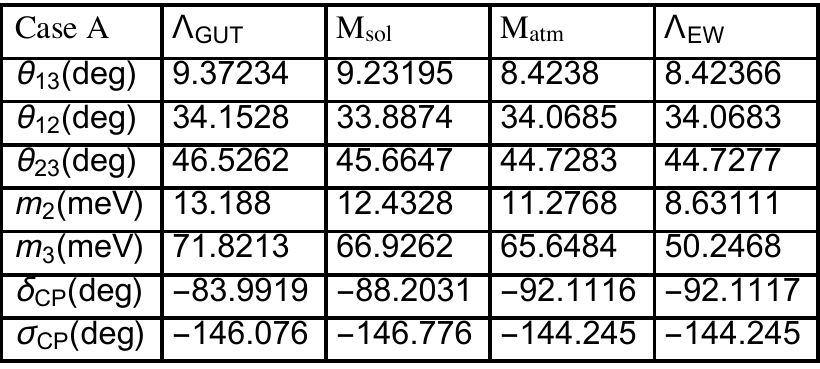}
  \end{minipage}
    \vspace*{0.2cm}
  \begin{minipage}[c]{7cm}
    \includegraphics[width=7cm]{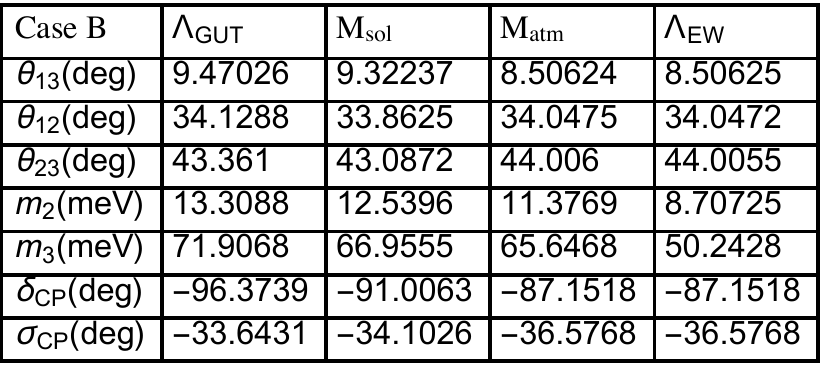}
  \end{minipage}
  \vspace*{0.2cm}
\begin{minipage}[c]{7cm}
  \includegraphics[width=7cm]{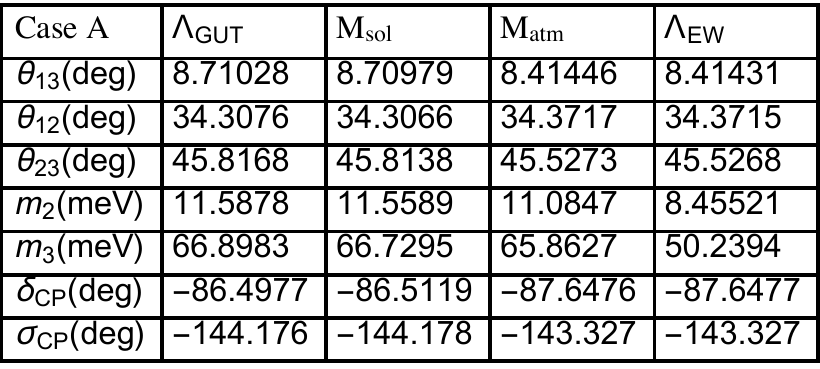}
\end{minipage}
\hspace{0.5cm}
\begin{minipage}[c]{7cm}
  \includegraphics[width=7cm]{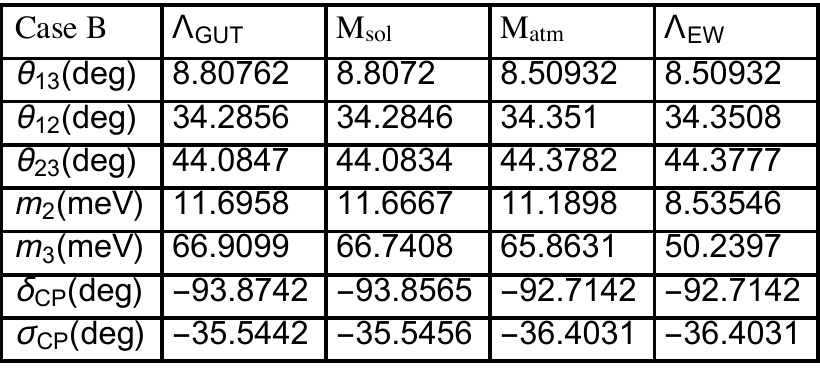}
\end{minipage}
  \vspace*{0.2cm}
\begin{minipage}[c]{7cm}
  \includegraphics[width=7cm]{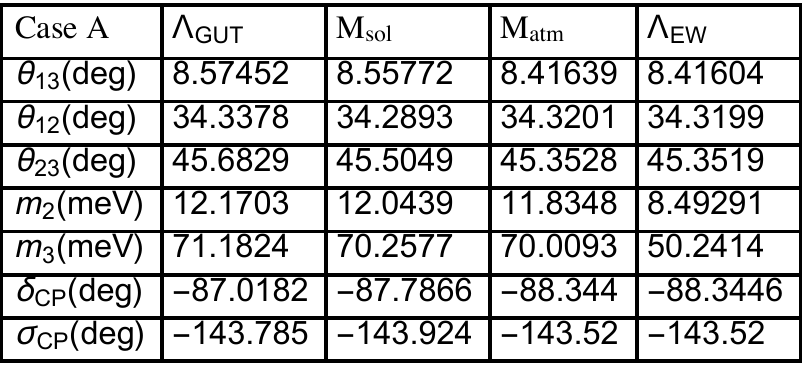}
\end{minipage}
\hspace{1.8cm}
\begin{minipage}[c]{7cm}
  \includegraphics[width=7cm]{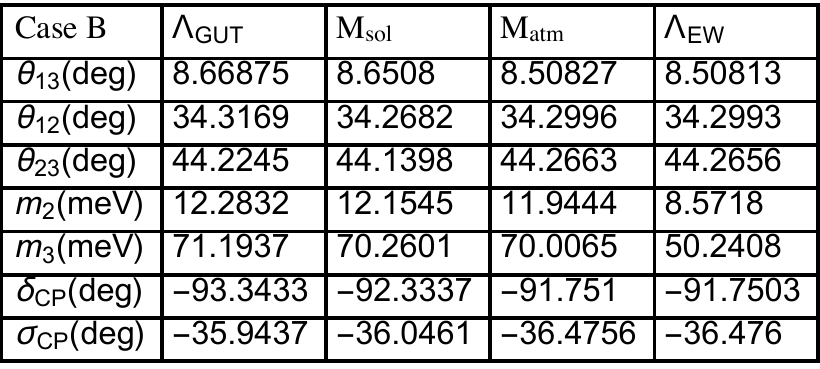}
\end{minipage}
\caption{\label{fig:SM_AB} SM cases {\it (left) } A and {\it (right) } B
with {\it (top) }$M_{atm}=10^{10}~\rm{GeV}$ and $M_{sol}=10^{15}~\rm{GeV}$ {\it (middle)} $M_{atm}=10^{10}~\rm{GeV}$ and $M_{sol}=10^{12}~\rm{GeV}$ {\it (bottom)} $M_{atm}=10^{13}~\rm{GeV}$ and $M_{sol}=10^{14}~\rm{GeV}$}
\end{table}

\begin{table}[p]
\begin{minipage}[c]{7cm}
  \includegraphics[width=7cm]{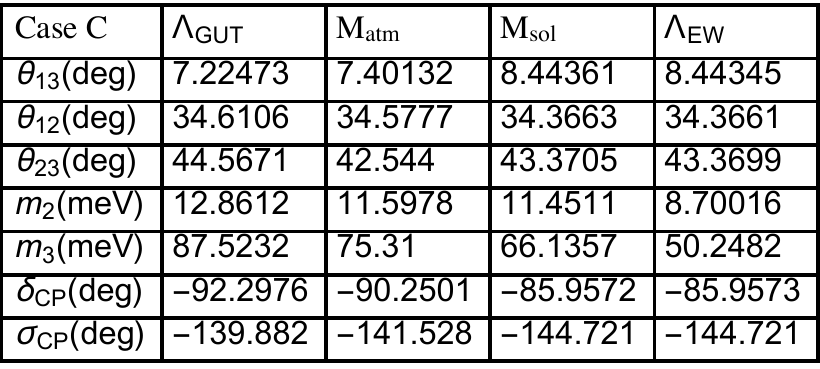}
  \end{minipage}
    \vspace*{0.2cm}
  \begin{minipage}[c]{7cm}
    \includegraphics[width=7cm]{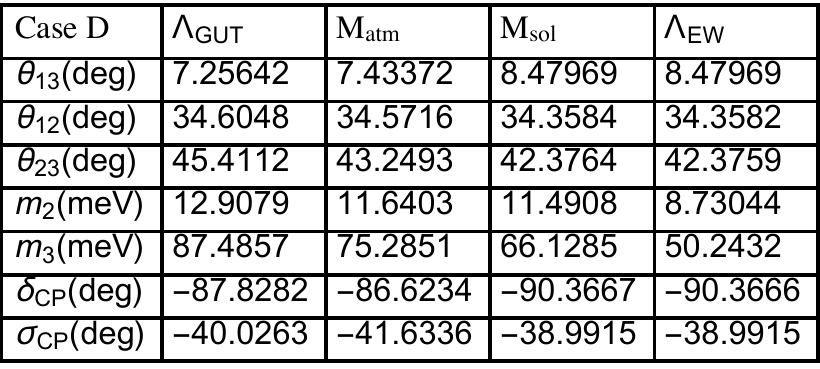}
  \end{minipage}
  \vspace*{0.2cm}
\begin{minipage}[c]{7cm}
  \includegraphics[width=7cm]{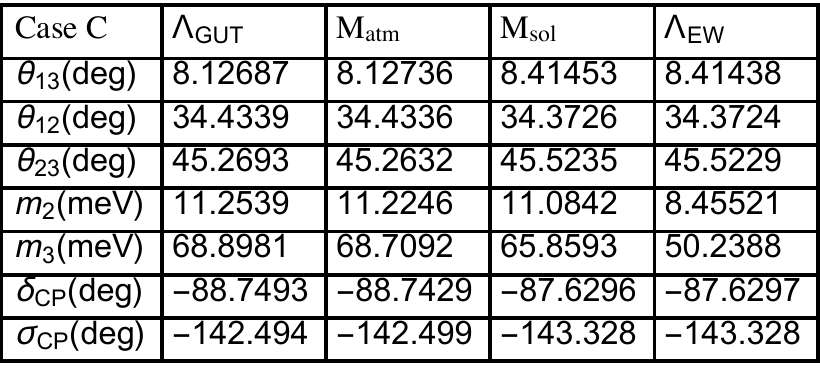}
\end{minipage}
\hspace{0.5cm}
\begin{minipage}[c]{7cm}
  \includegraphics[width=7cm]{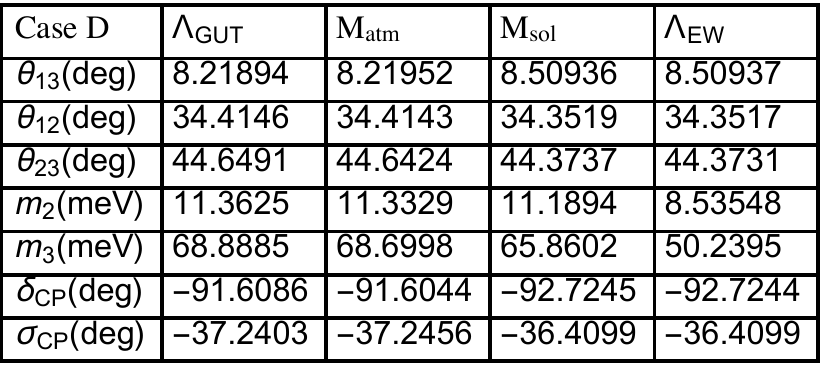}
\end{minipage}
  \vspace*{0.2cm}
\begin{minipage}[c]{7cm}
  \includegraphics[width=7cm]{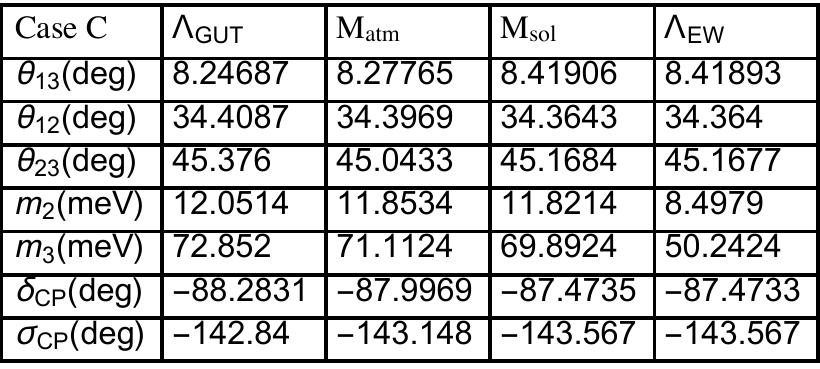}
\end{minipage}
\hspace{1.8cm}
\begin{minipage}[c]{7cm}
  \includegraphics[width=7cm]{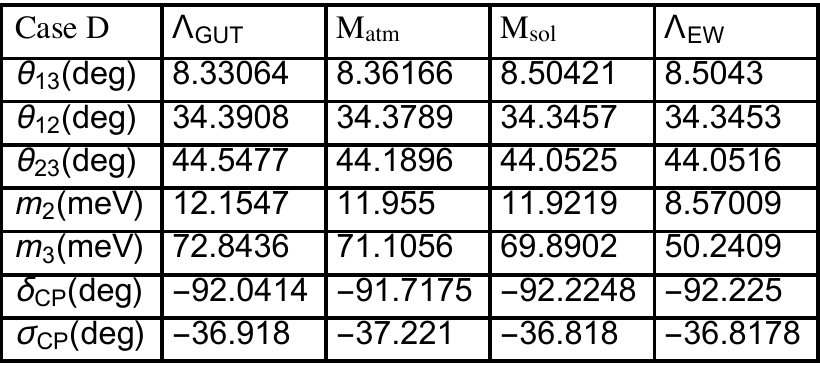}
\end{minipage}
\caption{\label{fig:SM_CD} SM cases {\it (left) } C and {\it (right) } D
with {\it (top) }$M_{atm}=10^{15}~\rm{GeV}$ and $M_{sol}=10^{10}~\rm{GeV}$ {\it (middle)} $M_{atm}=10^{12}~\rm{GeV}$ and $M_{sol}=10^{10}~\rm{GeV}$ {\it (bottom)} $M_{atm}=10^{14}~\rm{GeV}$ and $M_{sol}=10^{13}~\rm{GeV}$}
\end{table}
In summary, the connection between Case A$\leftrightarrow$ B and C $\leftrightarrow$ D stems from a combination of two features. Due to the similar running in {\it most} of the five neutrino parameters, the parameters at the GUT scale have to be similar. On top of that, we know from the estimates in Eqs.~(\ref{eq:Neutrino_GUT}) and (\ref{eq:Angles_GUT}) that similar GUT scale neutrino parameters enforce similar input parameters. Take for example the neutrino masses: from our numerical analysis, we learn that $m_2$ and $m_3$ exhibit nearly identical running for Case A and B. Since each of the neutrinos masses is directly linked to an input parameter, this already determines the suitable range of said input parameters, which is refined by means of including the mixing angles to the fit. The same can be done for Case C and D. As the running of $m_3$ is stronger in comparison to the one in Case A,B, the input parameter $m_a$ has to be higher for Case C and D, which can be observed in our results. \newline
Due to the intrinsic features of the LS cases and their connections among each other, it is possible to obtain comparably good values for $m_2$, $m_3$, $\vartheta_{12}$ and $\vartheta_{13}$ at the EW scale. For $\vartheta_{23}$, however, both the running behaviour and the relation between GUT scale value and input parameters does not follow the other neutrino parameter's connection between cases. As a consequence, the EW scale atmospheric angles show the widest spread depending on the case, and thus are most important with respect to the compatibility with experimental data. It is therefore not surprising that the hierarchy with respect to how well a scenario predicts $\vartheta_{23}$ is reflected in the goodness-of-fit values $\chi^2$. Thereby favouring Case D with a remarkable $\chi^2=1.49$ over also excellent goodness-of-fit results between $3.24$ and $7.14$ for Cases A, B and C.

%%%%%%%%%%%%%%%%%%%%%%%%%%%%%%%%%%%%%%%%%%%%%%%
\section{\label{app:B} MSSM Results}
%%%%%%%%%%%%%%%%%%%%%%%%%%%%%%%%%%%%%%%%%%%%%%%

In this section we 
examine the LS within the framework of the MSSM. 
We vary the SUSY breaking scale, considering $M_{SUSY}=1,\,3,\,10~{\rm TeV}$. For each MSSM setting with fixed $M_{SUSY}$, we furthermore investigate how $\tan \beta$ as well as the threshold effects, comprised in the parameter $\overline{\eta}_b$ and explained in Appendix~\ref{app:A}, affect the goodness-of-fit. To this end, we consider $\tan \beta=5,\,30,\,50$ and $\overline{\eta}_b=-0.6,\,0,\,0.6$. The results are collected in Tab.~\ref{tab:MSSM_CaseAB_all} and Tab.~\ref{tab:MSSM_CaseCD_all} with the corresponding predictions for neutrino masses and PMNS parameters in Figs.~\ref{fig:MSSM_CaseA_plot} to \ref{fig:MSSM_CaseD_plot} and in Appendix~\ref{app:D}, Tabs.~\ref{fig:MSSM_CaseA} to \ref{fig:MSSM_CaseD}. Note that we display detailed results for the setting with $M_{SUSY}=1~{\rm TeV}$, $\tan \beta =5$ and $\overline{\eta}_b=0.6$ in Figs.~\ref{fig:MSSM_CaseA_plot} to \ref{fig:MSSM_CaseD_plot}. We choose this MSSM setting for a more detailed representation of the neutrino parameters' running behaviour because it yields the most compatible results with experimental data for cases B, C and D. 

The MSSM results indicate the following:
\begin{itemize}
\item Independent of the SUSY breaking scale and/or $\tan \beta$, Case B yields the best fit to experimental data. The next best scenario with respect to the goodness-of-fit is Case D, which depending on the specific settings can follow Case B closely. The compatibility with experimental data deteriorates for Case A and further for Case C. How strongly the four cases vary in terms of $\chi^2$ depends on the choice of $M_{SUSY}$ and $\tan \beta$. 
\item Looking at the influence of $M_{SUSY}$ on the overall performance of a scenario, we keep $\tan \beta$ fixed and compare the goodness-of-fit measure $\chi^2$ for the three SUSY breaking scales. Performing this task for each LS case individually, we find that changing $M_{SUSY}$ barely affects the compatibility with data. There are only slight changes in $\chi^2$. We observe an increase in the absolute value of $\chi^2$ with higher $M_{SUSY}$ for $\tan \beta=5$. For $\tan \beta=30$, Case A prefers higher $M_{SUSY}$ while cases B, C and D prefer lower ones. And for $\tan \beta =50$, the goodness-of-fit increases with the SUSY breaking scale -- meaning $\chi^2$ declines. 
\item Moreover, we find that -- for each $M_{SUSY}$ and LS case -- the higher $\tan \beta$ the higher $\chi^2$, which means the poorer the overall agreement with experimental data. 
\item Just as we have ascertained for the SM, all MSSM settings yield only slightly poorer values when including the Dirac phase $\delta$ into the measure for the goodness-of-fit than they do without. Their difference is below $1~\%$ due to the comparably large uncertainty on the Dirac phase. On these grounds, we will refer to the $\chi^2$ values when further discussing the fundamental behaviour with respect to the different MSSM settings.
\item By including observations from Tabs.~\ref{tab:MSSM_CaseAB_best} to \ref{tab:MSSM_CaseCD_4}, we learn that for each LS case and setting, i.~e.~fixed SUSY breaking scale and $\tan \beta$, it is always the highest value of $\overline{\eta}_b$ under consideration that yields the best fit. How strongly the goodness-of-fit, and thereby its measure $\chi^2$, vary with $\overline{\eta}_b$ depends predominantly on $\tan \beta$. The higher $\tan \beta$, the more variation with $\overline{\eta}_b$ one observes in $\chi^2$. 
\item When taking a closer look at Tabs.~\ref{tab:MSSM_CaseAB_best} to \ref{tab:MSSM_CaseCD_4} displaying the varying threshold effects for $\tan \beta=30$, we observe unusually large values for $\chi^2$ for the threshold effects $\overline{\eta}_b=-0.6$. The latter can be explained by considering that this setting is at the border to the region where we run into trouble regarding non-perturbativity, which means that at least one of the Yukawa couplings becomes non-perturbative. 
\end{itemize}

As discussed later in Sec.~\ref{sec:6}, we know that {\it most} neutrino parameters do not only exhibit connections between cases A$\leftrightarrow$B and C$\leftrightarrow$D for the SM but also for the benchmark MSSM scenario with $M_{SUSY}=1~{\rm TeV}$, $\tan \beta =5$ and $\overline{\eta}_b=0.6$. The analogous behaviour observed among the cases is connected to their similar input parameter $m_a$, which we examine in Sec.~\ref{sec:5} for the SM. In Sec.~\ref{sec:5}, we learn that the connection for Case A$\leftrightarrow$B and C$\leftrightarrow$D originates from a combination of the similar running behaviour in most neutrino parameters, see also Sec.~\ref{sec:3}, which enforces similar starting values at the GUT scale, and the way the GUT scale parameters are linked to the two input parameters $m_a$ and $m_b$. The line of reasoning employed for the SM caries over to the MSSM -- with minor modifications, see Ref.~\cite{King:2016yef}. We, thus, expect similar $m_a$ for Case A, B and C, D, respectively, within a fixed MSSM setting, as well as an overall narrow range for $m_b$. This can indeed be observed in 
Tab.~\ref{tab:MSSM_CaseAB_all} and
Tab.~\ref{tab:MSSM_CaseCD_all}, where we give the input parameters $m_a$ and $m_b$ (in $\rm meV$). In the following, we briefly discuss how varying $M_{SUSY}$ and $\tan \beta$ influences these connections:

\begin{itemize}
\item As already discussed above, we expect the input parameter $m_a$ to reflect the connections between Case A$\leftrightarrow$B and C$\leftrightarrow$D. As well, we expect that the input parameter $m_b$ does not display any such connections but lies in a narrow region for all cases. Both projections prove to be right. How close the input parameter $m_a$ for Case A is to the one for Case B, however, depends on $\tan \beta$. The same is true for Case C and D. In other words, the higher $\tan \beta$, the further apart are the $m_a$ of the connected cases. This can be traced back to the RG running, which depends on $\tan \beta$\footnote{When switching from SM to MSSM, the vacuum expectation value $v^2$ is replaced by $v^2 \sin^2 \beta$. That way, the effective neutrino mass depends on $\tan \beta$.}. That is to say that there is -- in general -- more running for higher $\tan \beta$, and consequently, more deviation in GUT scale values depending on the case, which translates most directly to $m_a$. 
\item Fixing $\tan \beta$ to either of the three settings, one can observe an increase in both $m_a$ and $m_b$ with $M_{SUSY}$.
\item Fixing $M_{SUSY}$, on the other hand, does not yield any such clear tendency for neither $m_a$ nor $m_b$.
\item The overall range of values obtained by varying the SUSY breaking scale and $\tan \beta$ is similar for all four LS cases, namely about $1~{\rm meV}$ for $m_a$ and roughly $0.11~{\rm meV}$ for $m_b$. This means that a variation in the MSSM setting has a nearly identical impact on all four LS cases, which is further supported when taking a closer look at the relative changes in $m_a$ in between the settings studied. 
\end{itemize} 
One could in principle elaborate further on the discussion above, and also study the correlations of the LS cases on the level of neutrino parameters and that way confirm the key role of the atmospheric angle for the goodness-of-fit for all MSSM settings. This is, however, beyond the scope of this work.

\begin{table}[p]
\vspace*{-1.0cm}
\begin{subfigure}{15cm}
\hspace*{-1.0cm}
\centering
\setstretch{1.2}
 \begin{tabular}{c c|c c|c c|c c|c c}
 \multicolumn{2}{c|}{} & \multicolumn{4}{c|}{Case A} & \multicolumn{4}{c}{Case B} \\
$\tan \beta$ & $\overline{\eta}_b$ & $m_a$ [meV] & $m_b$ [meV] & $\chi^2$ & $\chi^2_\delta$ & $m_a$ [meV] & $m_b$ [meV] & $\chi^2$ & $\chi^2_\delta$ \\
 \hline
 \multirow{9}{*}{} & -0.6 & 16.115 & 1.5791 & 11.589 & 11.6373 & 16.104 & 1.5970 & 8.55635 & 8.57075 \\
 5 & 0 & 16.110 & 1.5787 & 11.5885 & 11.6368 & 16.100 & 1.5966 & 8.55518 & 8.56958 \\
  \rowcolor{YellowGreen}
&  0.6 & 16.110 & 1.5786 & 11.5885 & 11.6367 & 16.099 & 1.5965 & 8.55503 & 8.56943\\
\hline
\multirow{9}{*}{} &  -0.6 & 17.478 & 1.5123 & 47.1106 & 47.1329 & 17.355 & 1.7502 & 47.615 & 47.6264 \\
30 & 0 & 15.672 & 1.5129 & 14.2831 & 14.3278 & 15.651 & 1.5543 & 11.186 & 11.2  \\
 \rowcolor{YellowGreen}
 & 0.6 & 15.662 & 1.5128 & 14.1634 & 14.2082 & 15.641 & 1.5532 & 11.0593 & 11.0734  \\
 \hline
 \multirow{6}{*}{} 50 & 0 & 16.331 & 1.5116 & 23.8851 & 23.9203 & 16.276 & 1.6243 & 21.6018 & 21.6148 \\
 \rowcolor{YellowGreen}
& 0.6 & 16.180 & 1.5116 & 21.4738 &  21.5109 & 16.133 & 1.6081 & 18.9483 & 18.9616
 \end{tabular}
 \setstretch{1}
 \caption{\label{tab:MSSM_CaseAB_best}$M_{SUSY}=1~\rm{TeV}$}
\end{subfigure}
\begin{subfigure}{15cm}
\hspace*{-1.0cm}
\centering
\setstretch{1.2}
 \begin{tabular}{c c|c c|c c|c c|c c}
 \multicolumn{2}{c|}{} & \multicolumn{4}{c|}{Case A} & \multicolumn{4}{c}{Case B} \\
$\tan \beta$ & $\overline{\eta}_b$ & $m_a$ [meV] & $m_b$ [meV] & $\chi^2$ & $\chi^2_\delta$ & $m_a$ [meV] & $m_b$ [meV] & $\chi^2$ & $\chi^2_\delta$ \\
 \hline
\multirow{3}{*}{} & -0.6 & 16.325 & 1.5997 & 11.5855 & 11.6337 & 16.314 & 1.6178 & 8.58489 & 8.59929 \\
5 & 0 & 16.321 & 1.5993 & 11.5851 & 11.6333 & 16.310 & 1.6174 & 8.58305 & 8.59745 \\
 \rowcolor{YellowGreen}
& 0.6 & 16.320 & 1.5993 & 11.585 & 11.6332 & 16.310 & 1.6174 & 8.58295 & 8.59734 \\
 \hline
\multirow{3}{*}{} & -0.6 & 16.358 & 1.5358 & 20.2575 & 20.2957 & 16.314 & 1.6252 & 17.6666 & 17.6799 \\
30 & 0 & 15.895 & 1.5346 & 14.2506 & 14.2953 & 15.873 & 1.5763 &11.1873 & 11.2014 \\
 \rowcolor{YellowGreen}
& 0.6 & 15.886 & 1.5345 & 14.1478 & 14.1925 & 15.865 & 1.5754 & 11.0784 & 11.0924 \\
\hline
\multirow{2}{*}{} 50 & 0 & 16.527 & 1.5333 & 23.2548 & 23.2904 & 16.473 & 1.6434 & 20.9531 & 20.9661 \\
 \rowcolor{YellowGreen}
 & 0.6 & 16.402 & 1.5333 & 21.3021 & 21.3394 & 16.355 & 1.6300 & 18.8047 & 18.8179
 \end{tabular}
 \setstretch{1}
 \caption{\label{tab:MSSM_CaseAB_3}$M_{SUSY}=3~\rm{TeV}$}
\end{subfigure}
\begin{subfigure}{15cm}
\hspace*{-1.0cm}
\centering
\setstretch{1.2}
 \begin{tabular}{c c|c c|c c|c c|c c}
 \multicolumn{2}{c|}{} & \multicolumn{4}{c|}{Case A} & \multicolumn{4}{c}{Case B} \\
$\tan \beta$ & $\overline{\eta}_b$ & $m_a$ [meV] & $m_b$ [meV] & $\chi^2$ & $\chi^2_\delta$ & $m_a$ [meV] & $m_b$ [meV] & $\chi^2$ & $\chi^2_\delta$ \\
 \hline
\multirow{3}{*}{} & -0.6 & 16.563 & 1.6231 & 11.5811 & 11.6293 & 16.553 & 1.6415 & 8.61522 & 8.6296 \\
5 & 0 & 16.559 & 1.6227 & 11.5807 & 11.6289 & 16.548 & 1.6411 & 8.61422 & 8.6286 \\
 \rowcolor{YellowGreen}
& 0.6 & 16.558 & 1.6226 & 11.5807 & 11.6289 & 16.548 & 1.6410 & 8.6141 & 8.62848 \\
\hline
\multirow{3}{*}{} & -0.6 & 16.383 & 1.5592 & 17.1377 & 17.179 & 16.349 & 1.6260 & 14.3197 & 14.3334 \\
30 & 0 & 16.138 & 1.5585 & 14.2009 & 14.2456 & 16.117 & 1.6005 & 11.174 & 11.188 \\
 \rowcolor{YellowGreen}
& 0.6 & 16.130 & 1.5584 & 14.1137& 14.1585 & 16.109 & 1.5997 & 11.0816 & 11.0956 \\
 \hline
\multirow{2}{*}{} 50 & 0 & 16.744 & 1.5572 & 22.6229 & 22.659 & 16.692 & 1.6647 & 20.3071 & 20.3202 \\
 \rowcolor{YellowGreen}
 & 0.6 & 16.641 & 1.5572 & 21.0581 & 21.0956 & 16.594 & 1.6537 & 18.5858 & 18.5991
 \end{tabular}
 \setstretch{1}
 \caption{\label{tab:MSSM_CaseAB_4}$M_{SUSY}=10~\rm{TeV}$}
 \end{subfigure}
  \caption{\label{tab:MSSM_CaseAB_all} Best Fit values for Case A and B 
  with $M_{atm}=10^{12}~\rm{GeV}$ and $M_{sol}=10^{15}~\rm{GeV}$
  with {\it (top)} $M_{SUSY}=1~\rm{TeV}$ {\it (middle)} $M_{SUSY}=3~\rm{TeV}$ {\it (bottom)} $M_{SUSY}=10~\rm{TeV}$, and varying $\tan \beta$ and threshold effects denoted by $\overline{\eta}_b$.}
\end{table}

\begin{table}[p]
\vspace*{-1.0cm}
\begin{subfigure}{15cm}
\hspace*{-1cm}
\centering
\setstretch{1.2}
 \begin{tabular}{c c|c c|c c|c c|c c}
 \multicolumn{2}{c|}{} & \multicolumn{4}{c|}{Case C} & \multicolumn{4}{c}{Case D} \\
$\tan \beta$ & $\overline{\eta}_b$ & $m_a$ [meV] & $m_b$ [meV] & $\chi^2$ & $\chi^2_\delta$ & $m_a$ [meV] & $m_b$ [meV] & $\chi^2$ & $\chi^2_\delta$ \\
 \hline
\multirow{9}{*}{} & -0.6 & 16.854 & 1.6177 & 14.7629 & 14.8058 & 16.834 & 1.6377 & 9.1541 & 9.16791 \\
5 & 0 & 16.849 & 1.6173 & 14.7614 & 14.8043 & 16.830 & 1.6372 & 9.15274 & 9.16656 \\
 \rowcolor{YellowGreen}
& 0.6 & 16.849 & 1.6172 & 14.7613 & 14.8042 & 16.829 & 1.6371 & 9.15257 & 9.16639 \\
\hline
\multirow{9}{*}{} &  -0.6 & 18.324 & 1.5559 & 52.0663 & 52.0852 & 18.153 & 1.7971 & 48.4555 & 48.4663 \\
 30 & 0 & 16.357 & 1.5491 & 17.424 & 17.4638 & 16.324 & 1.5924 & 11.7735 & 11.787 \\
 \rowcolor{YellowGreen}
& 0.6 & 16.346 & 1.5490 & 17.2949 & 17.3349 & 16.313 & 1.5913 & 11.6451 & 11.6586 \\
 \hline
\multirow{6}{*}{}50 & 0 & 17.074 & 1.5507 & 27.6697 & 27.7006 & 16.994 & 1.6656 & 22.3003 & 22.3127 \\
 \rowcolor{YellowGreen}
& 0.6 & 16.910 & 1.5501 & 25.1095 & 25.1423& 16.841 & 1.6486 & 19.6235 & 19.6361
 \end{tabular}
 \setstretch{1}
 \caption{\label{tab:MSSM_CaseCD}$M_{SUSY}=1~\rm{TeV}$}
\end{subfigure}
\begin{subfigure}{15cm}
\hspace*{-1cm}
\centering
\setstretch{1.2}
 \begin{tabular}{c c|c c|c c|c c|c c}
 \multicolumn{2}{c|}{} & \multicolumn{4}{c|}{Case C} & \multicolumn{4}{c}{Case D} \\
$\tan \beta$ & $\overline{\eta}_b$ & $m_a$ [meV] & $m_b$ [meV] & $\chi^2$ & $\chi^2_\delta$ & $m_a$ [meV] & $m_b$ [meV] & $\chi^2$ & $\chi^2_\delta$ \\
 \hline
\multirow{3}{*}{} & -0.6 & 17.084 & 1.6394 & 14.8023 & 14.8451 & 17.064 & 1.6596 & 9.19151 & 9.20531 \\
5 & 0 & 17.080 & 1.6389 & 14.8011 & 14.8438 & 17.060 & 1.6592 & 9.19028 & 9.20408 \\
 \rowcolor{YellowGreen}
& 0.6 & 17.079 & 1.6389 & 14.8009 & 14.8437 & 17.059 & 1.6591 & 9.19013 & 9.20393 \\
\hline
\multirow{3}{*}{} & -0.6 & 17.103 & 1.5751 & 23.8756 & 23.9094 & 17.039 & 1.6665 & 18.343 & 18.3557 \\
30 & 0 & 16.599 & 1.5718 & 17.4356 & 17.4754 & 16.565 & 1.6155 & 11.7845 & 11.798 \\
 \rowcolor{YellowGreen}
&  0.6 & 16.589 & 1.5717 & 17.3245 & 17.3645 & 16.556 & 1.6145 & 11.674 & 11.6875 \\
\hline
 \multirow{3}{*}{} 50 & 0 & 17.288 & 1.5734 & 27.0562 & 27.0875 & 17.209 & 1.6857 & 21.6575 & 21.6699  \\
 \rowcolor{YellowGreen}
 & 0.6 & 17.151 & 1.5729 & 24.9804 & 25.0132 & 17.082 & 1.6717 & 19.4895 & 19.5021
 \end{tabular}
 \setstretch{1}
 \caption{\label{tab:MSSM_CaseCD_3} $M_{SUSY}=3~\rm{TeV}$}
\end{subfigure}
\begin{subfigure}{15cm}
\hspace*{-1cm}
\centering
\setstretch{1.2}
 \begin{tabular}{c c|c c|c c|c c|c c}
 \multicolumn{2}{c|}{} & \multicolumn{4}{c|}{Case C} & \multicolumn{4}{c}{Case D} \\
$\tan \beta$ & $\overline{\eta}_b$ & $m_a$ [meV] & $m_b$ [meV] & $\chi^2$ & $\chi^2_\delta$ & $m_a$ [meV] & $m_b$ [meV] & $\chi^2$ & $\chi^2_\delta$ \\
 \hline
\multirow{3}{*}{} & -0.6 & 17.346 & 1.6639 & 14.8467 & 14.8894 & 17.325 & 1.6845 & 9.23356 & 9.24735 \\
5 & 0 & 17.341 & 1.6635 & 14.8454 & 14.8881 & 17.321 & 1.6841 & 9.2324 & 9.24618 \\
 \rowcolor{YellowGreen}
  & 0.6 & 17.341 & 1.6635 & 14.8453 & 14.888 & 17.320 & 1.6840 & 9.23226 & 9.24604 \\
\hline
\multirow{3}{*}{} & -0.6 & 17.130 & 1.5988 & 20.5952 & 20.6317 & 17.080 & 1.6677 & 14.9701 & 14.9831 \\
30 & 0 & 16.865 & 1.5969 & 17.4333 & 17.4731 & 16.830 & 1.6409 & 11.7817 & 11.7952 \\
 \rowcolor{YellowGreen}
 & 0.6 & 16.856 & 1.5968 & 17.3391 & 17.379 & 16.822 & 1.6400 & 11.6879 & 11.7014 \\
 \hline
\multirow{2}{*}{} 50 & 0 & 17.525 & 1.5984 & 26.445 & 26.4768 & 17.449 & 1.7082 & 21.0183 & 21.0307 \\
 \rowcolor{YellowGreen}
 & 0.6 & 17.413 & 1.5979 & 24.7793 & 24.8122 & 17.343 & 1.6966 & 19.2807 & 19.2934
 \end{tabular}
 \setstretch{1}
 \caption{\label{tab:MSSM_CaseCD_4} $M_{SUSY}=10~\rm{TeV}$}
\end{subfigure}
\caption{\label{tab:MSSM_CaseCD_all} Best Fit values for Case C and D
with $M_{atm}=10^{15}~\rm{GeV}$ and $M_{sol}=10^{12}~\rm{GeV}$
with {\it (top)} $M_{SUSY}=1~\rm{TeV}$ {\it (middle)} $M_{SUSY}=3~\rm{TeV}$ {\it (bottom)} $M_{SUSY}=10~\rm{TeV}$, and varying $\tan \beta$ and threshold effects denoted by $\overline{\eta}_b$.}
\end{table}

\begin{figure}[p]
\begin{minipage}[c]{7cm}
  \includegraphics[width=7cm]{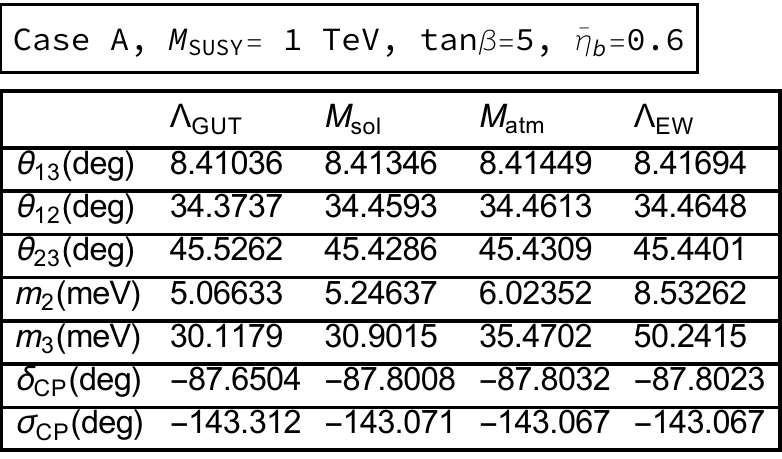}
  \end{minipage}
  \hspace*{-2cm}
  \begin{minipage}[c]{8cm}
    \includegraphics[width=7cm]{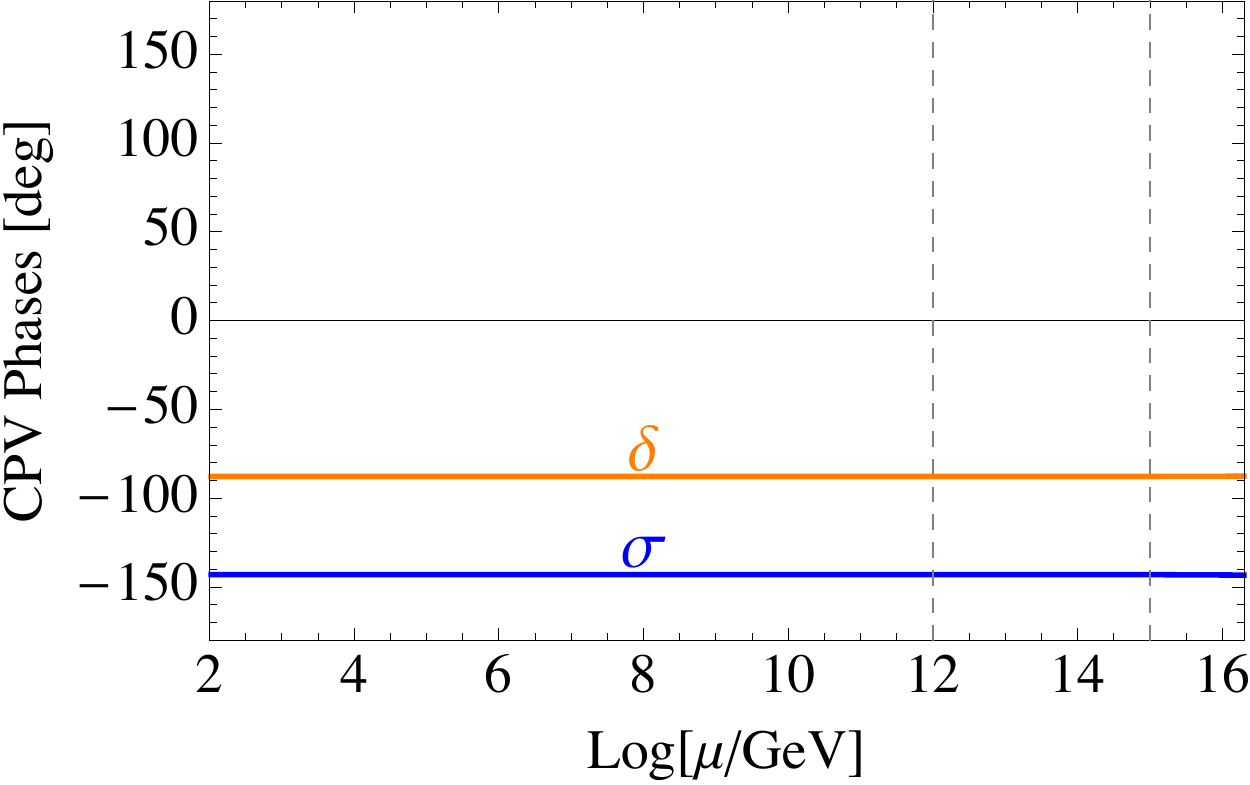}
  \end{minipage}
\begin{minipage}[c]{7cm}
  \includegraphics[width=7cm]{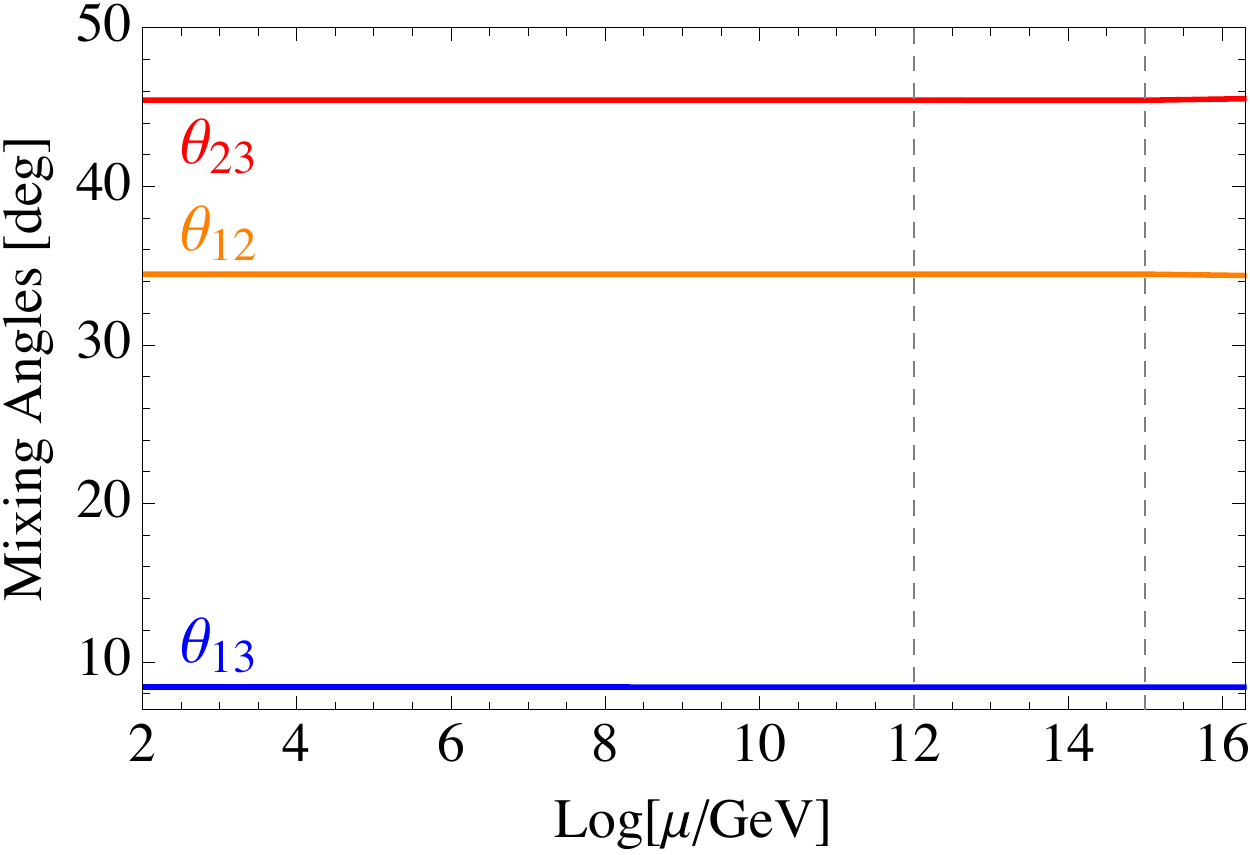}
\end{minipage}
\hspace{1cm}
\begin{minipage}[c]{7cm}
  \includegraphics[width=7cm]{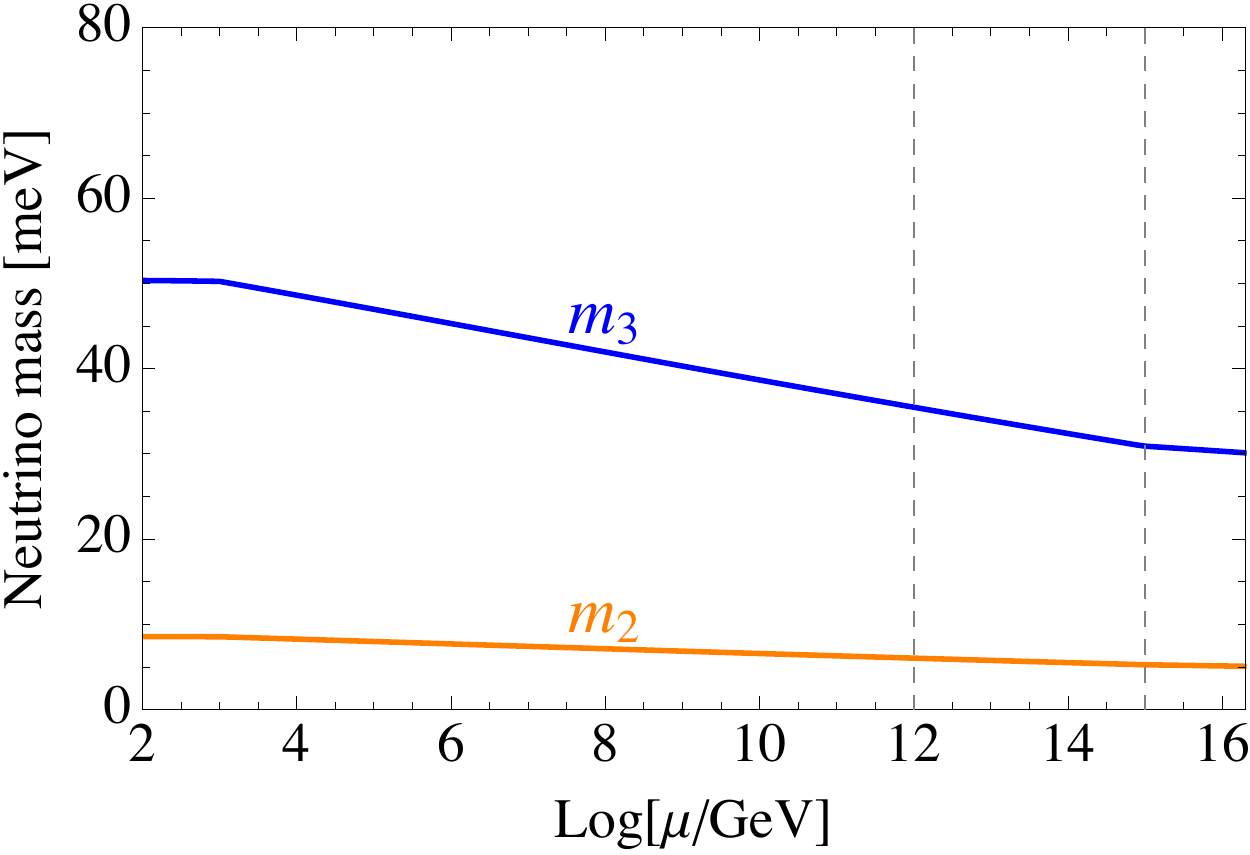}
\end{minipage}
\caption{\label{fig:MSSM_CaseA_plot}Case A - MSSM with $M_{SUSY}=1~{\rm TeV}$, $M_{atm}=10^{12}~\rm{GeV}$ and $M_{sol}=10^{15}~\rm{GeV}$}
\end{figure}
\begin{figure}[p]
\begin{minipage}[c]{7cm}
  \includegraphics[width=7cm]{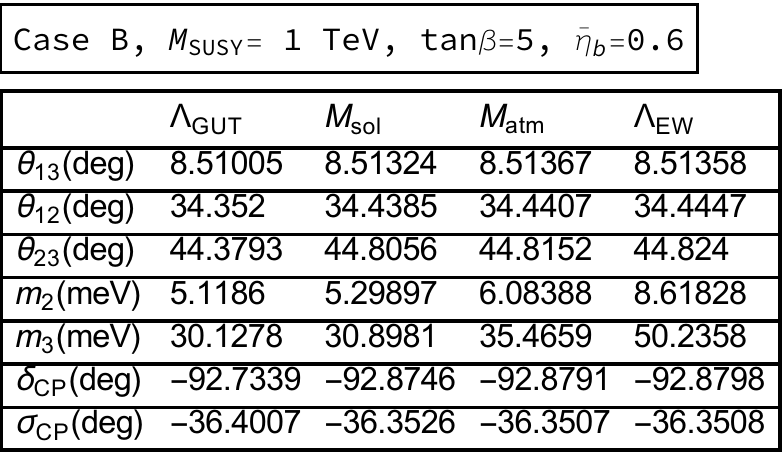}
  \end{minipage}
  \hspace*{-2cm}
  \begin{minipage}[c]{8cm}
    \includegraphics[width=7cm]{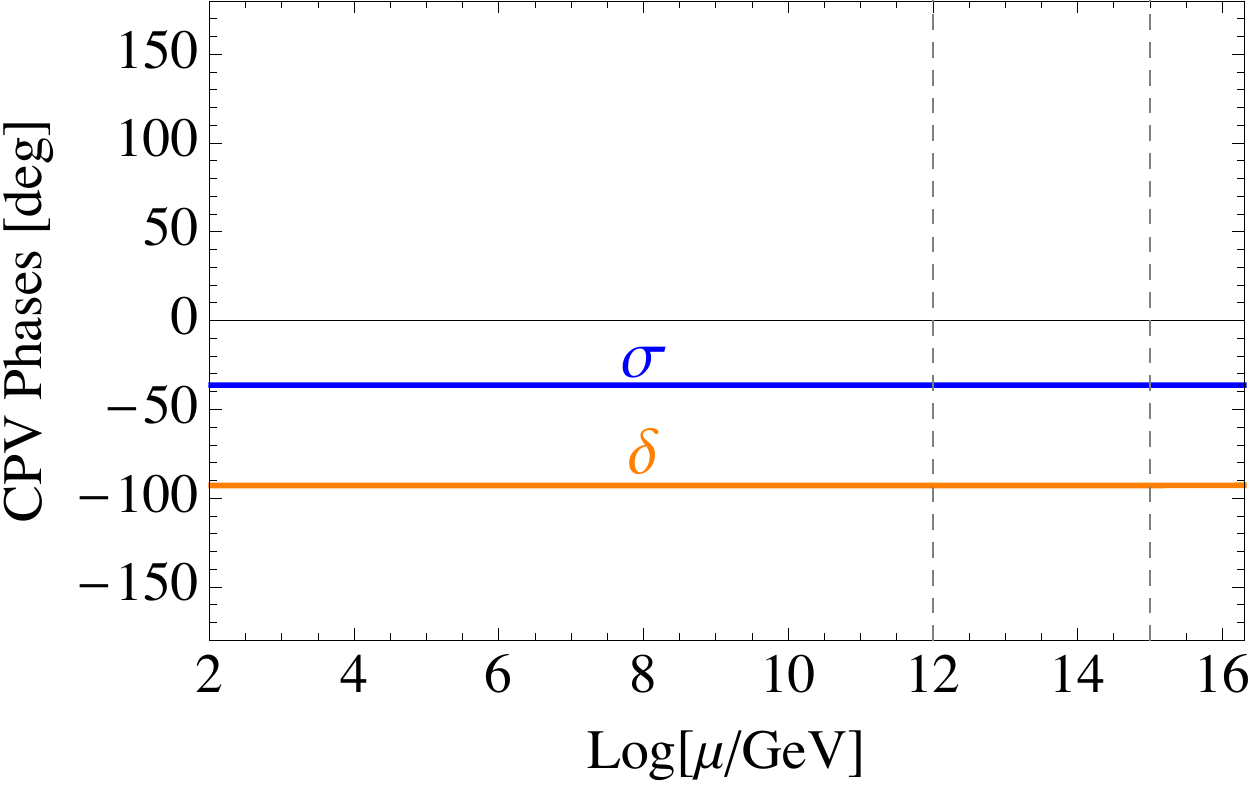}
  \end{minipage}
\begin{minipage}[c]{7cm}
  \includegraphics[width=7cm]{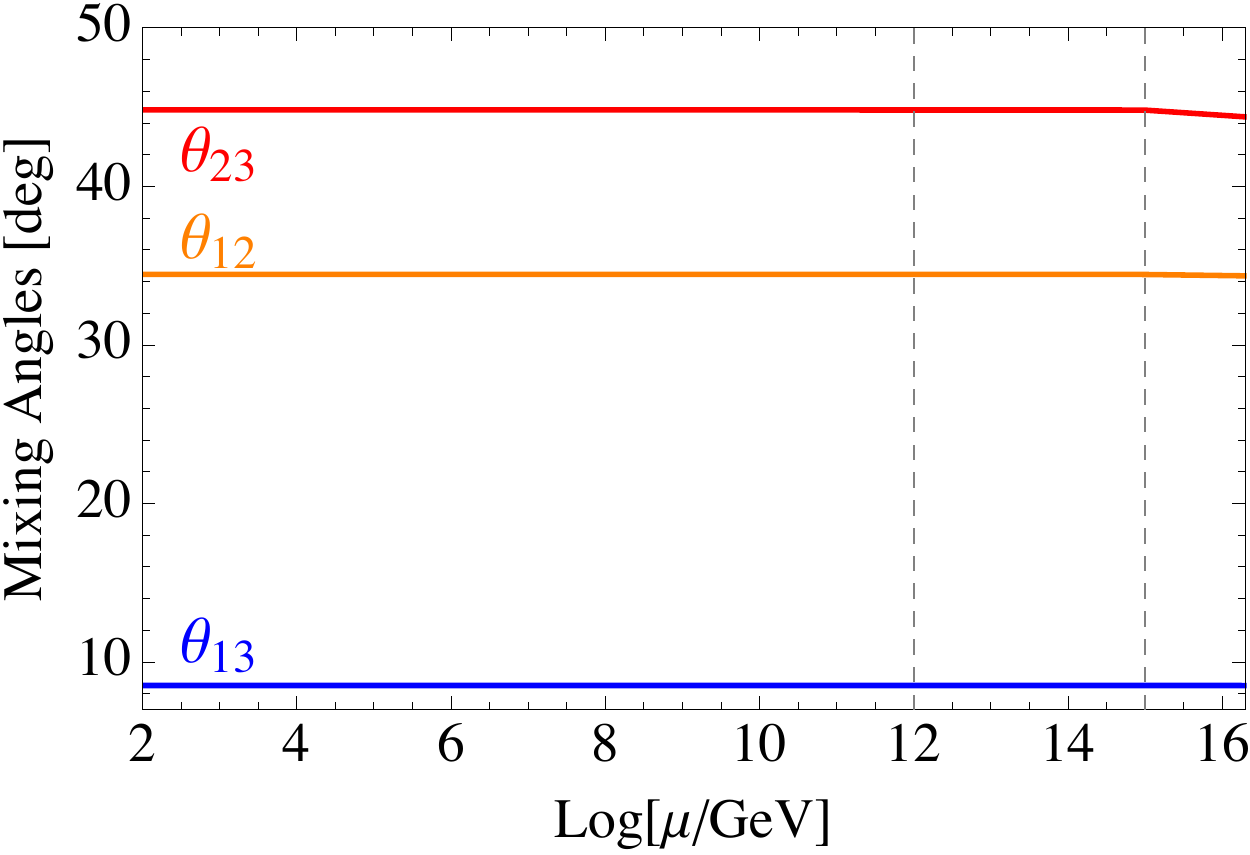}
\end{minipage}
\hspace{1cm}
\begin{minipage}[c]{7cm}
  \includegraphics[width=7cm]{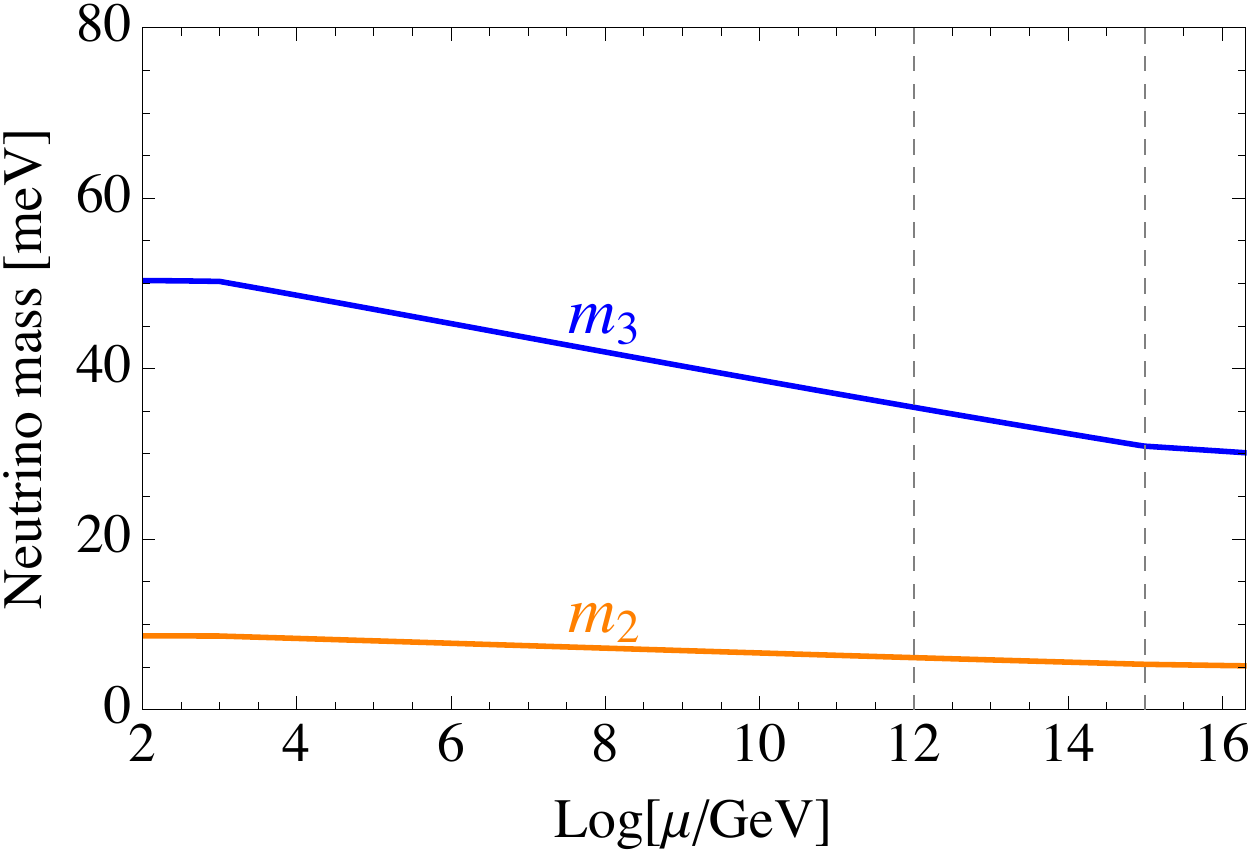}
\end{minipage}
\caption{\label{fig:MSSM_CaseB_plot}Case B - MSSM with $M_{SUSY}=1~{\rm TeV}$, $M_{atm}=10^{12}~\rm{GeV}$ and $M_{sol}=10^{15}~\rm{GeV}$}
\end{figure}
\begin{figure}[p]
\begin{minipage}[c]{7cm}
  \includegraphics[width=7cm]{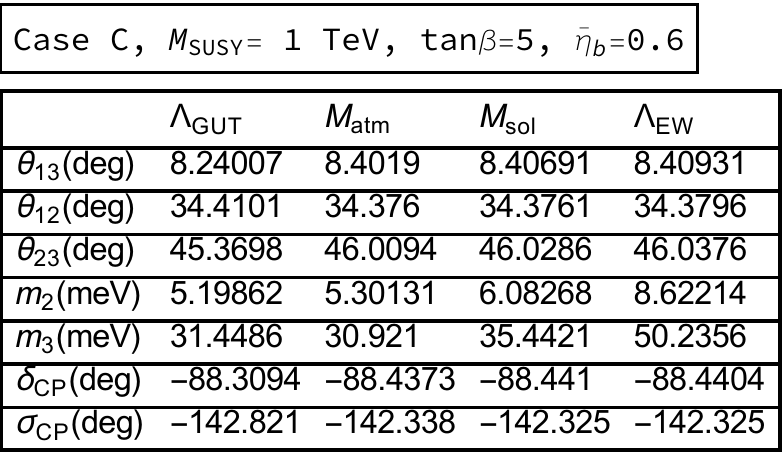}
  \end{minipage}
  \hspace*{-2cm}
  \begin{minipage}[c]{8cm}
    \includegraphics[width=7cm]{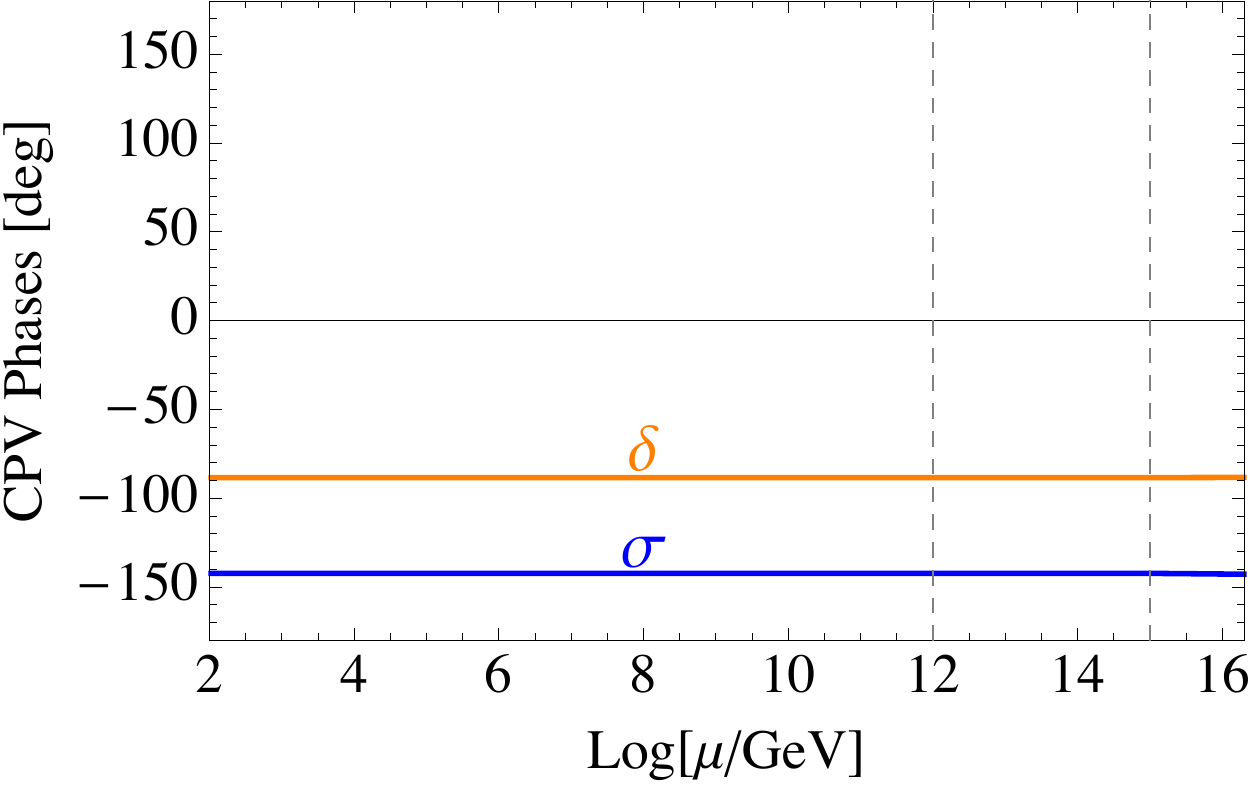}
  \end{minipage}
\begin{minipage}[c]{7cm}
  \includegraphics[width=7cm]{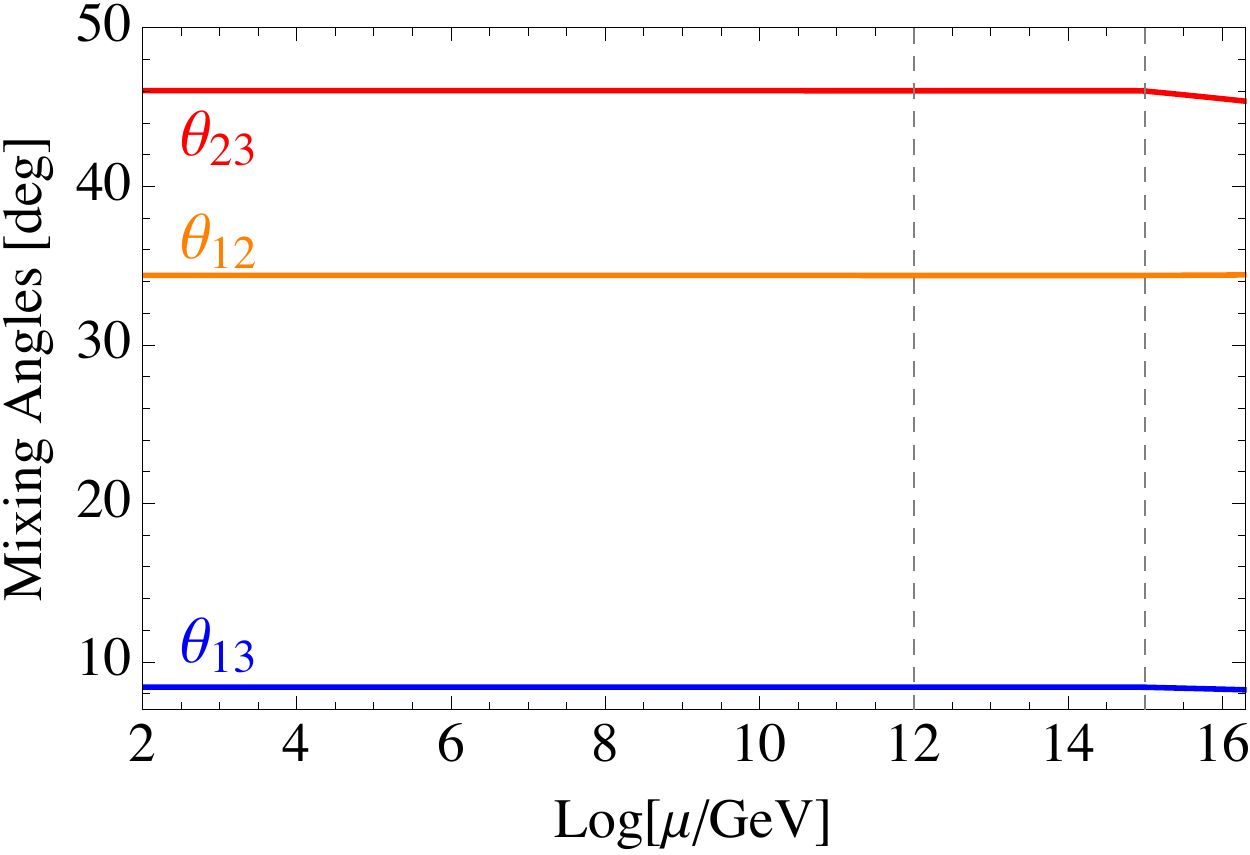}
\end{minipage}
\hspace{1cm}
\begin{minipage}[c]{7cm}
  \includegraphics[width=7cm]{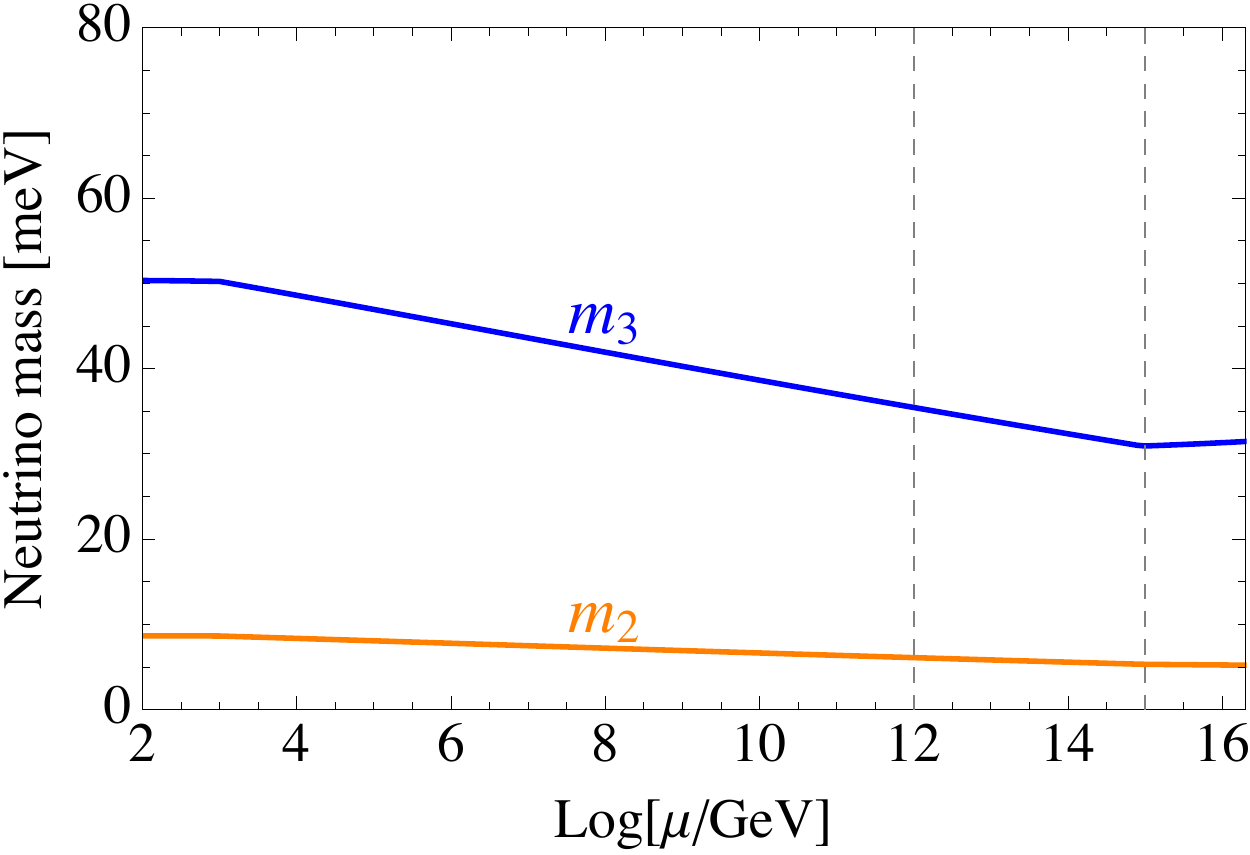}
\end{minipage}
\caption{\label{fig:MSSM_CaseC_plot}Case C - MSSM with $M_{SUSY}=1~{\rm TeV}$, $M_{atm}=10^{15}~\rm{GeV}$ and $M_{sol}=10^{12}~\rm{GeV}$}
\end{figure}
\begin{figure}[p]
\begin{minipage}[c]{7cm}
  \includegraphics[width=7cm]{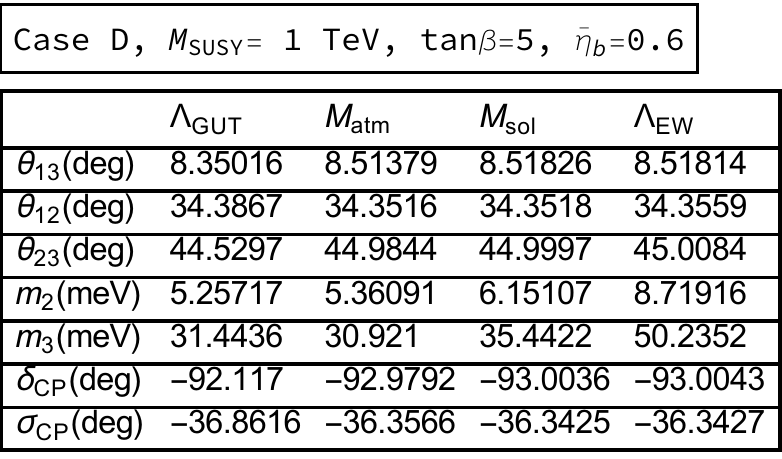}
  \end{minipage}
  \hspace*{-2cm}
  \begin{minipage}[c]{8cm}
    \includegraphics[width=7cm]{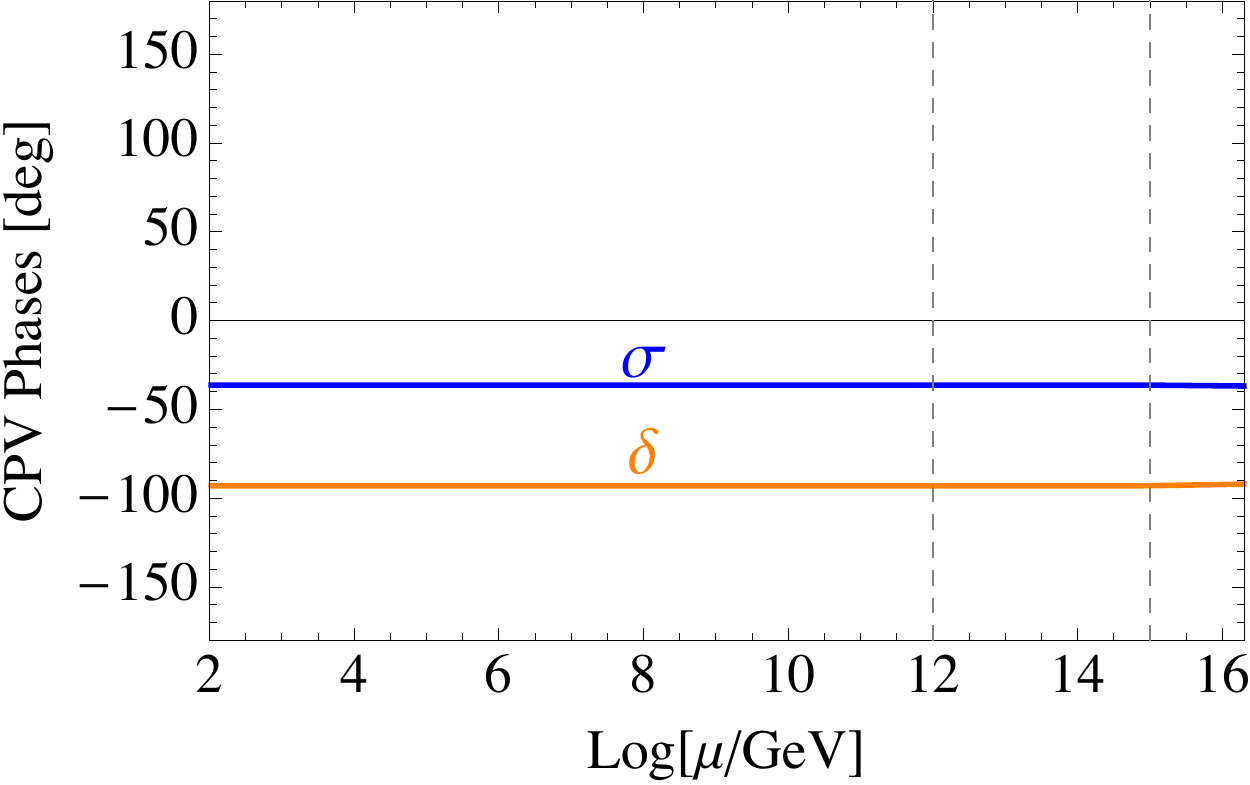}
  \end{minipage}
\begin{minipage}[c]{7cm}
  \includegraphics[width=7cm]{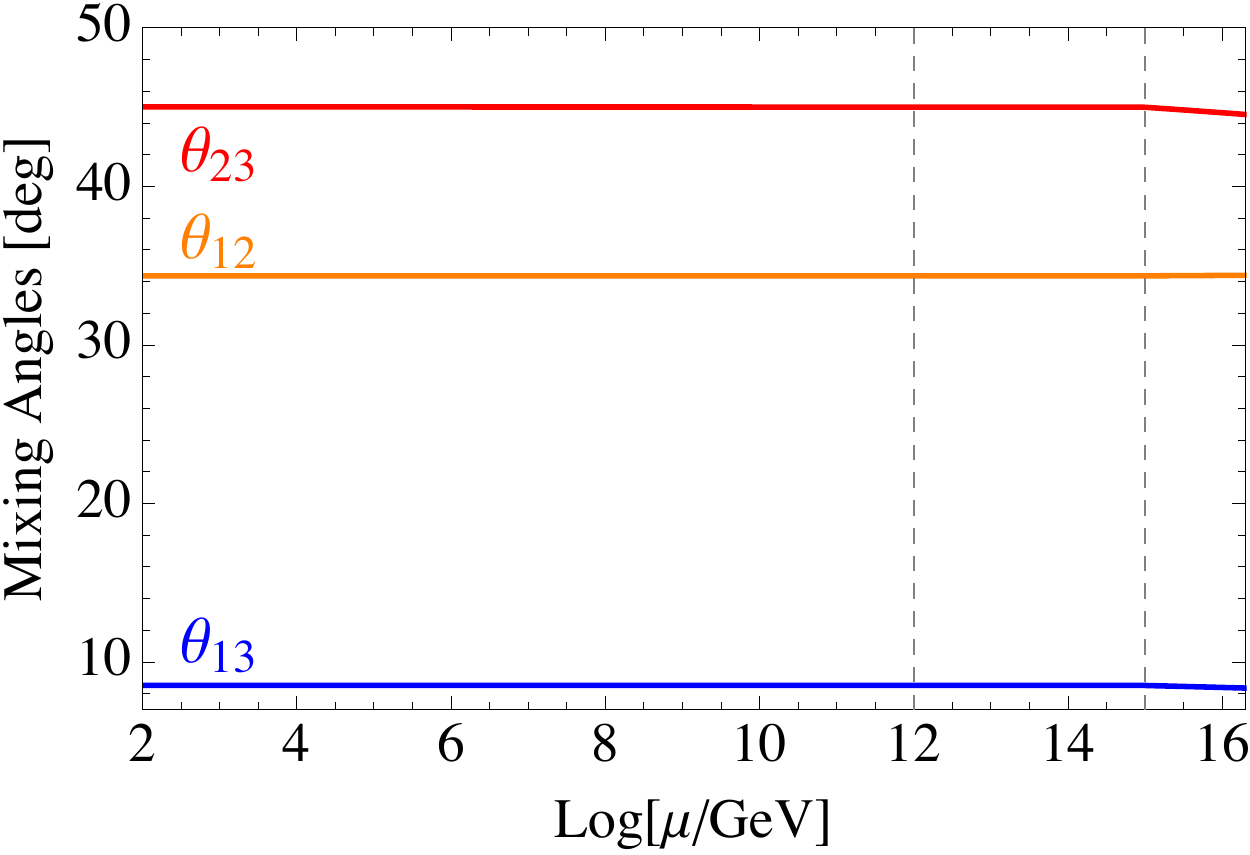}
\end{minipage}
\hspace{1cm}
\begin{minipage}[c]{7cm}
  \includegraphics[width=7cm]{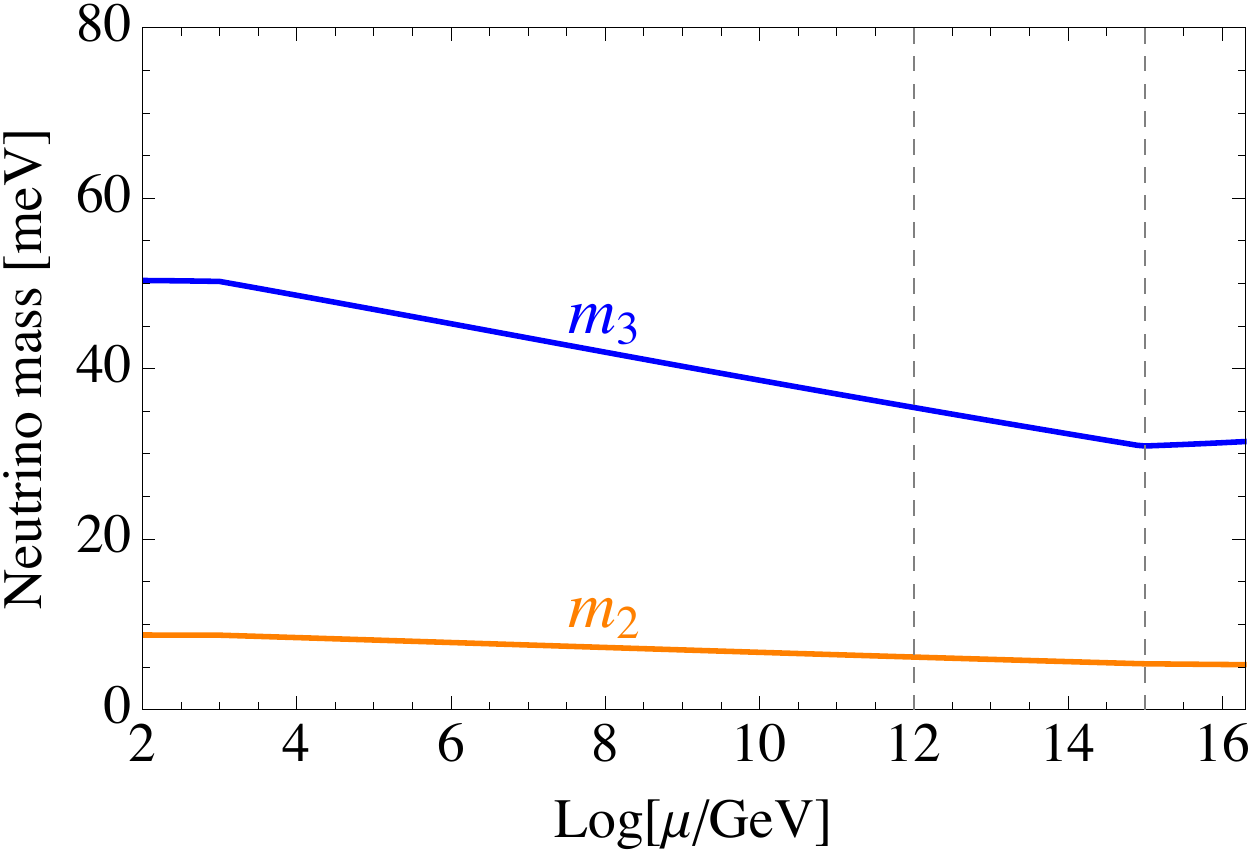}
\end{minipage}
\caption{\label{fig:MSSM_CaseD_plot}Case D - MSSM with $M_{SUSY}=1~{\rm TeV}$, $M_{atm}=10^{15}~\rm{GeV}$ and $M_{sol}=10^{12}~\rm{GeV}$}
\end{figure}

\clearpage

%%%%%%%%%%%%%%%%%%%%%%%%%%%%%%%%%%%%%%%%%%%%%%%
\section{\label{sec:6} Comparing SM and MSSM Results}
%%%%%%%%%%%%%%%%%%%%%%%%%%%%%%%%%%%%%%%%%%%%%%%

The purpose of this section is to compare the SM and MSSM behaviour. To this end, we choose one benchmark MSSM scenario with the SUSY breaking scale at $M_{SUSY}=1~{\rm TeV}$ and a threshold effect parameter of $\overline{\eta}_b=0.6$. The meaning of the latter is explained in Appendix \ref{app:A}. A more thorough discussion of the MSSM behaviour including different SUSY breaking scales and varying threshold effects can be found in the previous Sec.~\ref{app:B}. Note that we employ the RH neutrino mass setting $(10^{15},10^{12})$ GeV throughout the following analysis. \newline
In Tab.~\ref{tab:SM_MSSMcompare}, we collect the goodness-of-fit values for the SM and the benchmark MSSM scenario with varying $\tan \beta$. There are several observations worth mentioning:
\begin{itemize}
\item First and foremost, we note that the SM scenarios make for significantly better fits to the experimental data for each LS case individually. In fact, the poorest fit from the SM, namely Case A at $\chi^2=7.14$, outperforms the best for the MSSM, namely Case B with $\tan \beta =5$ at $\chi^2=8.56$.
\item While for the SM, the goodness-of-fit deteriorates from Case D via C and B to Case A, the order changes for the MSSM benchmark scenario, leading to Case B being most compatible with experimental data -- followed somewhat closely by Case D, and then by Case A and C.
\item The four LS cases of the MSSM benchmark scenario all yield a $\chi^2_\delta$ value that is only marginally poorer than the one for $\chi^2$ --by below $1~\%$. The difference between $\chi^2$ and $\chi^2_\delta$ for the SM, on the other hand, can be up to a few percent.   
\end{itemize}
To understand why the SM does yield better agreement with experimental data than the MSSM scenario as well as to understand the distinct characteristics with respect to the relative suitability of the different LS cases, we investigate and compare the behaviour of the neutrino parameters. 

\begin{table}[h!]
\centering
\setstretch{1.2}
\footnotesize{
 \begin{tabular}{c|c c c|c c c|c|c}
Case & \multicolumn{3}{c|}{$\chi_{\rm MSSM}^2 (\Lambda_{\rm EW})$} & \multicolumn{3}{c|}{$\chi^2_{\delta\,\,{\rm MSSM}} (\Lambda_{\rm EW})$} & $\chi_{\rm SM}^2 (\Lambda_{\rm EW})$ & $\chi^2_{\delta\,\,{\rm SM}} (\Lambda_{\rm EW})$ \\
& $t_{\beta} =5$ & $t_{\beta} =30$ & $t_{\beta} =50$ & $t_{\beta} =5$ & $t_{\beta} =30$ & $t_{\beta} =50$ & & \\
 \hline
A & $11.5885$ & $14.1634$ & $21.4783$ & $11.6367$ & $14.2082$ & $21.5109$ & $7.14042$ & $7.15869$ \\
B & $8.55503$ & $11.0593$ & $18.9483$ & $8.56943$ & $11.0734$ & $18.9616$ & $4.38607$ & $4.44012$ \\
C & $14.7613$ & $17.2949$ & $25.1095$ & $14.8042$ & $17.3349$ & $25.1423$ & $3.23646$ & $3.30174$ \\
D & $9.15257$ & $11.6451$ & $19.6235$ & $9.16639$ & $11.6586$ & $19.6361$ & $1.49388$ & $1.52265$
 \end{tabular}
 \setstretch{1}
 \caption{\label{tab:SM_MSSMcompare} Best fit $\chi^2$ values for the four cases for the SM as well as the MSSM with $M_{SUSY}=1~{\rm TeV}$, $\overline{\eta}_b=0.6$ and varying $t_{\beta}\equiv \tan \beta$. The corresponding $m_a$ and $m_b$ are displayed in Tabs.~\ref{tab:SM_CaseAD_best} and Tab.~\ref{tab:MSSM_CaseAB_all} and
Tab.~\ref{tab:MSSM_CaseCD_all}.
     } }
\end{table} 
As we strive to compare SM and MSSM, we focus the discussion on generic differences in the initial values (meaning at the GUT scale) and the RG running behaviour of the neutrino parameters without delving into the specifics of the MSSM. Since $\tan \beta=5$ makes for the most suitable predictions from the MSSM benchmark scenario, we use its predicted neutrino parameters when comparing to the SM.  From the upper left panels of Figs.~\ref{fig:SM_CaseA} to \ref{fig:SM_CaseD} for the SM in combination with Figs.~\ref{fig:MSSM_CaseA_plot} to \ref{fig:MSSM_CaseD_plot} for the MSSM benchmark scenario, we can condense the following characteristics with respect to the neutrino parameters:
\begin{itemize}
\item The mixing angle $\vartheta_{12}$ is predicted to be in between $[34.36^\circ,\,34.46^\circ]$ for the $\tan \beta =5$ MSSM benchmark scenario whereas it lies in between $[34.05^\circ,\,34.37^\circ]$ for the SM, both at the EW scale and depending on the LS case. The measured solar angle is $\vartheta^{exp}_{12}=33.58^\circ$ with a $1\sigma$ range of $[32.83^\circ,\,34.33^\circ]$. Consequently, the SM predictions for Case A,B are encompassed in and those for Case C,D close to the standard deviation, whereas the MSSM predictions for Case C,D lie about as close as the SM's Case C,D and the MSSM's Case A,B are further above.  Thereby, the solar angle has a bias towards the SM for cases A and B, while there is no preference when considering cases C and D.  As observed in the previous section, there is an intrinsic connection between Case A $\leftrightarrow$ B and Case C $\leftrightarrow$ D for the SM, which also appears for the MSSM benchmark scenario. That is to say, that -- in case of this MSSM scenario -- cases A and B generate quite similar values at the GUT scale, display an overall identical but minor increase based on the RG running between the GUT and the EW scale, and thus predict similar $\vartheta_{12}$ at the EW scale.  For the MSSM benchmark scenario, cases C and D behave analogously apart from a decline in the solar angle with the decrease of the energy scale and deviating absolute values at the GUT scale.
\item Analysing the predictions for the mixing angle $\vartheta_{13}$, we obtain a LS-case-dependent range of $[8.41^\circ,\,8.52^\circ]$ for the  $\tan \beta =5$ MSSM benchmark scenario at the EW scale, while the SM yields values in between $[8.42^\circ,\,8.51^\circ]$. With an experimental value of $\vartheta^{exp}_{13}=8.46^\circ$ within a $1\sigma$ range of $[8.31^\circ,\,8.61^\circ]$, both predicted ranges are centered around the measured value and fully encompassed within the $1\sigma$ region. Thus, there is no general bias towards either the SM or the MSSM scenario from the reactor angle. From the SM discussion in Sec.~\ref{sec:5}, we recall that cases A and B generate similar initial values at the GUT scale, undergo the same overall decline with the energy scale and thereby predict similar values at the EW scale. The same holds true for cases C and D, but with an increase in $\vartheta_{13}$ from the GUT to the EW scale and absolute values that differ from Case A,B at the GUT scale. Nevertheless, all four cases converge to a narrow region and predict similar reactor angles within the framework of the SM. Since the MSSM scenario displays a nearly identical range of predicted $\vartheta_{13}$, one might assume that the underlying behaviour is equivalent. This, however, does not stand up to scrutiny. From Figs.~\ref{fig:MSSM_CaseA_plot} to \ref{fig:MSSM_CaseD_plot}, we learn that the starting values at the GUT scale are spread. The RG running, on the other hand, does yet again display the connection between the cases; leading to hardly any alteration of $\vartheta_{13}$ due to running effects for cases A and B, and an increase by $0.17^\circ$ from the GUT to the EW scale for cases C and D. This allows for the EW scale values of cases A and C to be close.  The same is observed for the EW scale reactor angles of cases B and D. Since the measured mixing angle lies centered in between the different LS cases, there is no strong preference for any case to be discerned within the framework of the MSSM -- which is also true for the SM.  
\item For the atmospheric mixing angle $\vartheta_{23}$, the MSSM benchmark scenario with $\tan \beta=5$ predicts values within the range of $[44.82^\circ,\,46.04^\circ]$ at the EW scale, depending on the LS case. The atmospheric angles predicted by the LS cases within the framework of the SM are within the range of $[42.37^\circ,\,44.72^\circ]$. The measured value of $\vartheta^{exp}_{23}=41.61^\circ$ within the $1\sigma$ region of $[40.40^\circ,\,43.17^\circ]$ is below the range of predicted values in either case. Note that all atmospheric angles predicted within the framework of the SM are below the range of $\vartheta_{23}$'s derived from the MSSM scenario. Furthermore, the spread of the predicted values depending on the LS case is large in comparison to the other two mixing angles, which is true for both frameworks, SM and MSSM. The atmospheric angle also differs from the other mixing angles in terms of connections between different LS cases. For neither the SM nor the MSSM framework, there are connections for the prediction at the GUT scale, or the RG running behaviour. Consequently, the atmospheric angle plays a decisive role with respect to the compatibility of a scenario with experimental data -- and it favours the SM over the MSSM as framework for the respective LS cases. It is, therefore, not surprising that the goodness-of-fit, measured by $\chi^2$, reflects the order of how well a case and/or scenario predicts $\vartheta_{23}$. As an example of this feature take the atmospheric angles predicted by the SM's Case A, $\vartheta^{SM,A}_{23}=44.72^\circ$, and the MSSM's Case B, $\vartheta^{MSSM,B}_{23}=44.82^\circ$. The former is least suitable within the framework of the SM, whereas the latter is most compatible for the MSSM. Although, they stem from different frameworks and LS cases, their overall performance with respect to compatibility with experimental data is similar -- $\chi^2_{SM,A}=7.14$ and $\chi^2_{MSSM,B}=8.56$ -- and mirrors the ordering of their atmospheric angles. 
\item Turning to the neutrino masses, we compare the $m_2$ predictions from the SM to the ones from the MSSM benchmark scenario. From the SM, we obtain a range of $[8.63~{\rm meV},\,8.73~{\rm meV}]$. The MSSM benchmark scenario with $\tan \beta =5$ predicts lighter neutrino masses in the region of $[8.53~{\rm meV},\,8.72~{\rm meV}]$. The measured neutrino mass of $m_2^{exp}=8.66~{\rm meV}$ is embedded in the $1\sigma$ region of $[8.56~{\rm meV},\,8.77~{\rm meV}]$. Consequently, all cases but the MSSM's Case A predict values well within the $1\sigma$ region. Nevertheless, the MSSM's Case A generates a lighter neutrino mass that is in close proximity to the $1\sigma$ region. Another feature worth mentioning is the MSSM's RG running effects in distinction from the SM's RG behaviour. Within the framework of the SM, the four LS cases show no obvious connection at the GUT scale, where their absolute values are in close proximity to one another -- at roughly $m_2(\Lambda_{\rm GUT})=13.4~{\rm meV}$. Due to the RG running effects, the light neutrino mass decreases for each LS case by about the same amount, leading to equally good predictions at the EW scale. The picture is somewhat different within the framework of the MSSM. Starting from absolute values at about $m_2(\Lambda_{\rm GUT})=5.15~{\rm meV}$, the RG effects increase the light neutrino mass in between the GUT and the EW scale. Since the magnitude of the increase varies slightly, we obtain a marginally wider region of $m_2$ values at the EW scale than we do for the SM.  The opposite direction of the RG running can be traced back to the coefficients in the RGEs that differ for the SM and the MSSM, including a relative sign~\cite{Antusch:2005gp,Antusch:2003kp}. Despite the fundamental differences in terms of RG behaviour, the prediction of $m_2$ only gives a narrow edge to the SM over the MSSM for Case A. For the remaining three LS cases, there is no preference for either the SM or the MSSM from the light neutrino mass.  
\item For the neutrino mass $m_3$, the predicted values are nearly identical for any case within either the SM or the MSSM framework -- $m_3^{SM} \in [50.24~{\rm meV},\,50.25~{\rm meV}]$ and $m_3^{MSSM}=50.24~{\rm meV}$. They are also consistent with the measured value $m_3^{exp}=50.24~{\rm meV}$ which lies in the $1\sigma$ region of $[49.84~{\rm meV},\,50.63~{\rm meV}]$. Although there is no bias towards any scenario or case from the heavier of the light neutrino masses, the features leading to the EW scale value differ. As already observed for the lighter neutrino mass $m_2$, $m_3$ undergoes different alterations due to the RG effects. Recall that for the SM cases A and B start from roughly the same value at the GUT scale, as do cases C and D. The initial GUT scale values are significantly higher for the latter. All four LS cases exhibit a decrease of $m_3$ with the energy scale -- with stronger effects for Case C,D. Taking a closer look at the MSSM, we note that both Case A,B and Case C,D start from nearly identical values at the GUT scale -- with the latter being a bit higher. The RG running effects are opposite to those of the SM, meaning that $m_3$ increases from the GUT to the EW scale, which in analogy to $m_2$ is attributed to the coefficients of the RGEs~\cite{Antusch:2005gp,Antusch:2003kp}. Nevertheless, both frameworks and all four scenarios within predict the measured value perfectly, and thus give no bias regarding the goodness-of-fit.
\end{itemize}
Intriguingly, both the SM as well as the MSSM framework can generate comparably good values for the neutrino parameters $\vartheta_{13}$, $m_2$ and $m_3$, which are the parameters that have the lowest spread with respect to the LS case. Note that for $\vartheta_{13}$ and $m_3$ all four LS cases in both frameworks are within the $1\sigma$ region, and for $m_2$ there is only one outlier, namely the MSSM's Case A. The latter allows for a slight preference of the SM over the MSSM but only when considering case A. A more important distinction stems from the mixing angle $\vartheta_{12}$. First of all, $\vartheta_{12}$ has a bias towards the SM for the cases A and B while it does not display a bias for cases C and D -- giving an overall edge to the SM. Secondly, the reshuffled order with respect to how well the different LS cases do hints towards the observation that the hierarchy among the LS cases changes depending on the framework. The most decisive role with respect to compatibility with data, however, falls to the atmospheric angle $\vartheta_{23}$ once again. For $\vartheta_{23}$, there is not only the widest spread regarding the different LS cases but also the most explicit gap between the values predicted by the SM and those derived from the MSSM. In addition, the ordering of LS cases by means of how well they predict the atmospheric angle directly translates to the overall performance. It is therefore, once again, the atmospheric angle that is most significant and makes for the substantially better fits of the SM scenarios to the experimental data.

%%%%%%%%%%%%%%%%%%%%%%%%%%%%%%%%%%%%%%%%%%%%%%%
\section{\label{sec:7} Conclusions}
%%%%%%%%%%%%%%%%%%%%%%%%%%%%%%%%%%%%%%%%%%%%%%

We have performed a detailed RG analysis of the LS models, including those cases where the RG corrections can become significant. Unlike a previous analysis, where the input parameters were fixed independently of RG corrections,
we have performed a complete scan of model parameters for each case individually, to determine the optimum set of high energy input values from a global fit of the low energy parameters which include the effects of RG running.
In all cases we perform a $\chi^2$ analysis of the low energy masses and mixing angles, including RG corrections
for various RH neutrino masses and mass orderings.
We have made complete scans for each LS case individually within the framework of the SM and
the MSSM to determine the optimum set of input values $(m_a,\,m_b)$ at the GUT scale from global fits to experimental data at the EW scale. Perhaps not surprisingly, the values of $\chi^2$ that we obtain here are significantly lower than those obtained 
in the previous analysis where the input parameters were determined independently of RG corrections.

We have found that the most favourable RG corrections occur in the SM, 
rather than in the MSSM.
Amongst the three mixing angles, we find that the atmospheric angle is often the most sensitive
to RG corrections in both the SM and the MSSM, although in the latter the corrections are relatively small.
Without including RG corrections the LS predictions are in some tension with the latest global fits,
mainly because the atmospheric angle is predicted to be close to maximal.
The sensitivity of the atmospheric angle to RG corrections in the SM then allows a better fit at low energies,
corresponding to an atmospheric angle in the first octant, close to the current best fit value for a normal
hierarchy.

For the SM, we have performed the analysis with various RH neutrino masses 
and for the MSSM we investigated different SUSY breaking scales, $\tan \beta$ and threshold effects.
In the case of the SM,  it turns out that its beneficial for the running effects if the heavier of the RH neutrinos is closer to the GUT scale, with masses $(10^{15}, 10^{12})$ GeV yielding the best results. In this case we found for the SM:
$\chi^2_A=7.1$, $\chi^2_B=4.4$, $\chi^2_C=3.2$ and $\chi^2_D=1.5$, 
corresponding to exceptionally good agreement with experimental data, especially for case D.

We emphasise that the atmospheric angle plays a key role in our analysis,
and is the crucial factor in obtaining low $\chi^2$ values for a given set up.
While it is possible to obtain comparably good results for $m_2,\,m_2,\,\vartheta_{12}$ and $\vartheta_{13}$ at the EW scale for all LS cases, it is $\vartheta_{23}$ that varies most for different cases within the SM or the MSSM. 
While the SM and MSSM can generate comparably good $m_2,\,m_3$ and $\vartheta_{13}$,
and there is some preference of $\vartheta_{12}$ in favour of cases A and B of the SM,
the most decisive parameter is $\vartheta_{23}$ for which the SM predictions are significantly better.
This is partly a result of the fact that RG corrections in the MSSM are relatively small,
compared to the SM, and so the prediction of near maximal atmospheric mixing is maintained 
at low energies in the MSSM.

Forthcoming results from T2K and NOvA on the atmospheric mixing angle will test the
predictions of the LS models. The inclusion of RG corrections in a consistent way,
as done in this paper, will be crucial in confronting such theoretical models with data.

%%%%%%%%%%%%%%%%%%%%%%%%%%%%%%%%%%%%%%%%%%%%%%%%
\section*{Acknowledgements}
%%%%%%%%%%%%%%%%%%%%%%%%%%%%%%%%%%%%%%%%%%%%%%%%

We would like to thank S.~Antusch  and J.~Zhang for useful discussions. TG acknowledges partial support by the the Micron Technology Foundation, Inc.. SFK acknowledges the STFC Consolidated Grant ST/L000296/1 and the European Union's Horizon 2020 Research and Innovation programme under Marie Sk\l{}odowska-Curie grant agreements Elusives ITN No.\ 674896 and InvisiblesPlus RISE No.\ 690575. 

\clearpage

%%%%%%%%%%%%%%%%%%%%%%%%%%%%%%%%%%%%%%%%%%%%%%%%%%%%%%%%%%%%%%%%%%%%%%%%%%%%%%%%%%%%%%%
\appendix

%%%%%%%%%%%%%%%%%%%%%%%%%%%%%%%%%%%%%%%%%%%%%%%
\section{\label{app:A} Littlest Seesaw within the framework of the MSSM: Yukawa Couplings for $\mathbf{M_{SUSY}=1,\,3,\,10~\rm{\bf TeV}}$ }
%%%%%%%%%%%%%%%%%%%%%%%%%%%%%%%%%%%%%%%%%%%%%%%

Throughout this work, we use the {\it Mathematica} package \texttt{REAP}~\cite{Antusch:2005gp} to solve the RGEs numerically. It is important to employ the appropriate parameters at the GUT scale such that experimental values at low energies (e.~g.~the scale $M_Z$ of the Z boson) are reproduced correctly. To simplify matters, we stick to the common approximation that assumes only one single SUSY threshold, namely $M_{SUSY}$, at which all supersymmetric particles are integrated out. To extract the proper GUT scale values for the charged lepton and the quark Yukawa matrices, we make use of the results derived in Ref.~\cite{Antusch:2013jca}\footnote{For more information, on the framework used, the explicit low energy input values and more, please consult said reference. Note that Ref.~\cite{Antusch:2013jca} also assumes neutrino masses to be generated via the seesaw mechanism at high energies.}. We present how to calculate said values along the lines of Ref.~\cite{Antusch:2013jca} in the following. \newline
The first step is to derive the Yukawa couplings at $M_Z$ from the experimental data. The latter are handed over to \texttt{REAP}, which calculates their RG running to $M_{SUSY}$. At the SUSY breaking scale, the SM has to be matched to the MSSM. As the radiative corrections can be $\tan \beta$ enhanced, and therefore even exceed the one-loop running contributions, we must include them at the matching scale. This leads to a correction to the down-type quark as well as the charged lepton Yukawa matrix, which can be simplified to~\cite{Blazek:1995nv}
\begin{equation}
\begin{split}
Y_u^{\rm SM} & \simeq \sin \beta\,Y_u^{\rm MSSM}\,, \\
Y_d^{\rm SM} & \simeq \Big(\mathds{1}+ {\rm diag}\Big(\eta_q,\,\eta_q,\,\eta'_q+\eta_A \Big)\Big)\,Y_d^{\rm MSSM}\,\cos \beta \,,\\
Y_\ell^{\rm SM} & \simeq \Big( \mathds{1} + {\rm diag}\Big(\eta_\ell,\,\eta_\ell,\,\eta'_\ell \Big)\Big)\,Y_\ell^{\rm MSSM}\,\cos \beta\,.
\end{split}
\label{eq:Yuk_Threshold}
\end{equation}
Here, one chooses a basis where the up-type Yukawa matrix is diagonal. Note that only contributions enhanced by $\tan \beta$ are included which is accurate up to the percent-level. Furthermore, the threshold corrections to the first two generations of down-type quarks and charged leptons are assumed to be of the same size, respectively. This is a good approximation in many SUSY scenarios provided that the down and strange squark as well as the selectron and smuon are of nearly the same mass. The corrections in Eq.~(\ref{eq:Yuk_Threshold}) depend on the specific SUSY scenario under consideration, and need to be computed correspondingly. The parameters $\eta_q$ and $\eta'_q$ originate predominantly from gluino contributions in combination with some Wino and Bino loop corrections, whereas $\eta_\ell$ and $\eta'_\ell$ are caused by electroweak gauginos. The correction from $\eta_A$ is related to the trilinear soft SUSY breaking term $A_u$~\cite{Blazek:1995nv}. Note that all parameters $\eta$ contain the factor $\tan \beta$. \newline
The six parameters used in Eq.~(\ref{eq:Yuk_Threshold}) can be combined into four, namely
\begin{equation}
\begin{split}
\overline{\eta}_b &\equiv \eta'_q+\eta_A-\eta'_\ell\,, \quad \overline{\eta}_q  \equiv \eta_q-\eta'_\ell\,, \\
\overline{\eta}_\ell & \equiv \eta_\ell -\eta'_\ell\,,\quad \text{and}\quad\cos \overline{\beta} \equiv \big( 1+ \eta'_\ell \big) \cos \beta\,.
\end{split}
\end{equation}
Starting from the basis, where the SM Yukawa matrices $Y_u$ and $Y_\ell$ are diagonal at $M_{SUSY}$, the expressions for the MSSM Yukawa matrices at the SUSY breaking scale are given by~\cite{Antusch:2013jca}
\begin{equation}
\begin{split}
Y_u^{\rm MSSM} & \simeq \frac{1}{\sin \overline{\beta}}\,Y_u^{\rm SM}\,, \\
Y_d^{\rm MSSM} & \simeq {\rm diag}\Bigg(\frac{1}{1+\overline{\eta}_q},\,\frac{1}{1+\overline{\eta}_q},\,\frac{1}{1+\overline{\eta}_b} \Bigg)\,Y_d^{\rm SM}\,\frac{1}{\cos \overline{\beta}}\,, \\
Y_\ell^{\rm MSSM} & \simeq {\rm diag}\Bigg(\frac{1}{1+\overline{\eta}_\ell},\,\frac{1}{1+\overline{\eta}_\ell},\,1 \Bigg)\,Y_\ell^{\rm SM}\,\frac{1}{\cos \overline{\beta}}\,,
\end{split}
\label{eq:Yuk_Matching}
\end{equation}
with the CKM parameters fully included in the down-type quark matching condition.
As the parameters $\overline{\eta}_\ell$ and $\overline{\eta}_q$ only affect the first two generations of $Y_d$ and $Y_\ell$, which are small in comparison, their effect on the RG running can be neglected to good approximation. In other words, there are four parameters needed for the matching procedure at the SUSY breaking scale, but only two out of these, namely $\overline{\eta}_b$ and $\tan \overline{\beta}$, in order to perform the RG evolution to the GUT scale. \newline
The authors of Ref.~\cite{Antusch:2013jca} derived the GUT scale MSSM quantities for three different SUSY breaking scales, namely $M_{SUSY}=1,\,3,\,10~{\rm TeV}$, and provided them in form of data tables at~{\url{http:/particlesandcosmology.unibas.ch/RunningParameters.tar.gz}}. From these tables, one can extract the GUT scale values depending on the choice of the parameters $\overline{\eta}_\ell$,  $\overline{\eta}_q$, $\overline{\eta}_b$ and $\tan \overline{\beta}$. The proper translation between the data made available and the Yukawa couplings as well as CKM parameters we employ as input at the GUT scale is given in the captions of Figs.~1 to 3 and 5 of Ref.~\cite{Antusch:2013jca}. In order to further reduce the number of possible MSSM settings, we assume that the leptonic corrections $\eta_\ell$ and $\eta'_\ell$ can be neglected. As a consequence, it is $\overline{\eta}_\ell =0$. For $\tan \beta \geq 5$, it is $\tan \overline{\beta} =(1+\eta'_\ell)^{-1} \tan \beta \xrightarrow{\eta'_\ell=0} \tan \beta$. By this approximations only, we can extract the charged lepton Yukawa couplings, the up-type Yukawa couplings as well as the coupling of the bottom quark. In order to also extract the strange and down Yukawa couplings, we also need to specify $\overline{\eta}_q$. Since the RG running of the neutrino parameters, which is the ultimate goal of this work, depends mostly on the bottom quark's coupling, and not on the down and strange quark, we can neglect $\overline{\eta}_q$. We could have used a similar argument when setting $\overline{\eta}_\ell$ to zero as we mostly care for the effect of the $\tau$ lepton on the RG running of the neutrino parameters. As a consequence of these simplifications, we are left with the parameter $\overline{\eta}_b$ comprising the threshold effects and $\tan \beta$ when fixing the MSSM setting. Note, furthermore, that the CKM mixing angle $\theta_{12}$ and the CP violating phase $\delta$ are barely affected by threshold effects and RG running. As a consequence, we use their \texttt{REAP} default values. The CKM mixing angles $\theta_{13}$ and $\theta_{23}$, on the other hand, depend on $\overline{\eta}_q$ and $\overline{\eta}_b$. With the simplifications discussed above, we also extract their GUT scale values from the data tables in~{\url{http:/particlesandcosmology.unibas.ch/RunningParameters.tar.gz}}. Based on the data provided by the authors of Ref.~\cite{Antusch:2013jca}, we investigate MSSM scenarios with the SUSY breaking scales $M_{SUSY}=1,\,3,\,10~{\rm TeV}$. Furthermore, we choose $\tan \beta =5,\,30,\,50$ and threshold effects within the range of $\overline{\eta}_b=-0.6 \to 0.6$. For the latter, the range needs to be adapted depending on $\tan \beta$ to avoid non-perturbative Yukawa couplings. The MSSM settings investigated throughout this work are supposed to be benchmark settings that give an overview on the LS's RG behaviour within the framework of the MSSM. The corresponding initial values extracted as discussed above and handed over to \texttt{REAP} are given in Tab.~\ref{tab:Yuk_couplings}. In case one has a more specific MSSM scenario in mind and aims at a more precise analysis of its SUSY threshold corrections, there is a software extension to \texttt{\texttt{REAP}} called \texttt{SusyTc} that generates the appropriate input values from the SUSY breaking terms~\cite{Antusch:2015nwi}.
\begin{table}[h!]
\centering
\begin{subfigure}{16cm}
 \includegraphics[width=16cm]{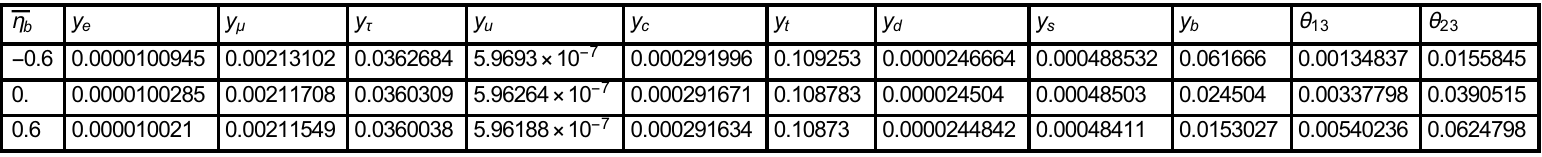}\vspace*{0.1cm}\newline
 \includegraphics[width=16cm]{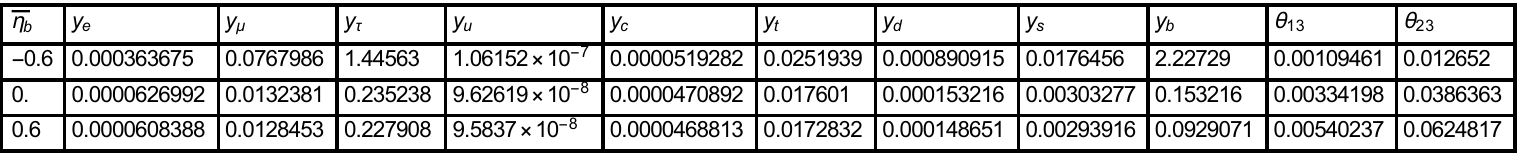}\vspace*{0.1cm}\newline
 \includegraphics[width=16cm]{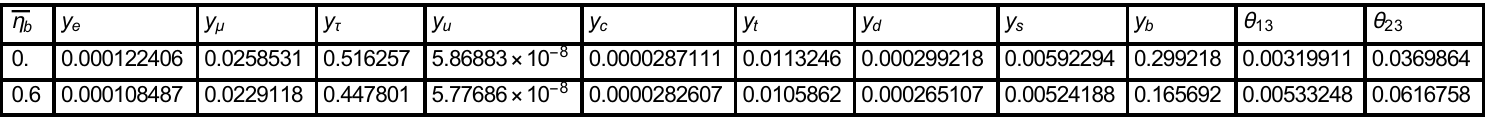}
 \caption{\label{tab:Yuk_1TeV} $M_{SUSY}=1~\rm{TeV}$ }
\end{subfigure}
\begin{subfigure}{16cm}
 \includegraphics[width=16cm]{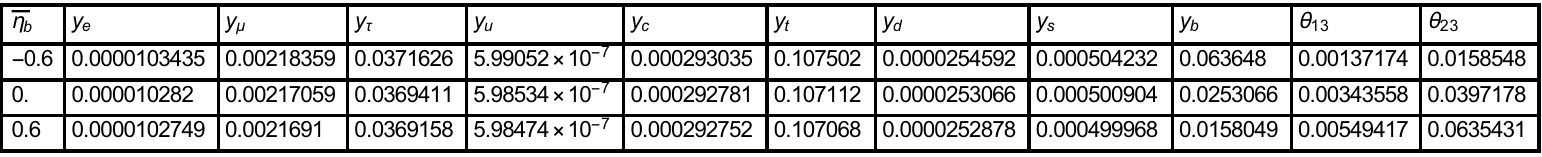}\vspace*{0.1cm}\newline
 \includegraphics[width=16cm]{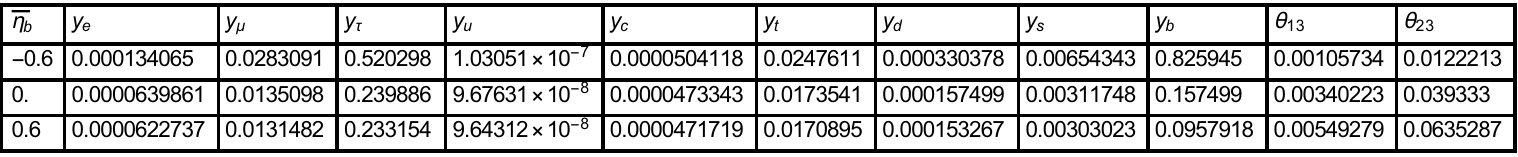}\vspace*{0.1cm}\newline
 \includegraphics[width=16cm]{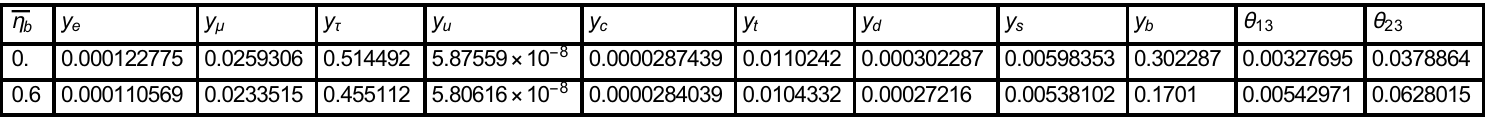}
 \caption{\label{tab:Yuk_3TeV} $M_{SUSY}=3~\rm{TeV}$}
\end{subfigure}
\begin{subfigure}{16cm}
 \includegraphics[width=16cm]{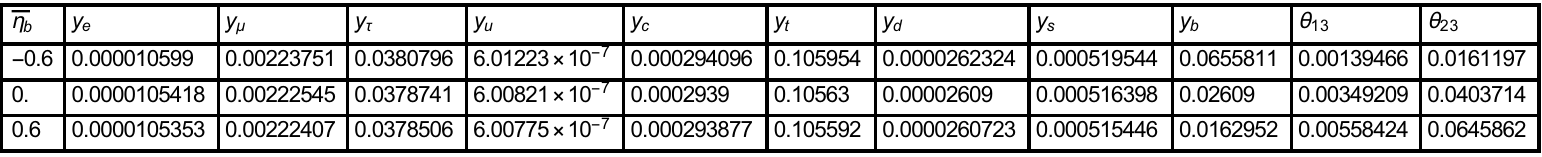}\vspace*{0.1cm}\newline
 \includegraphics[width=16cm]{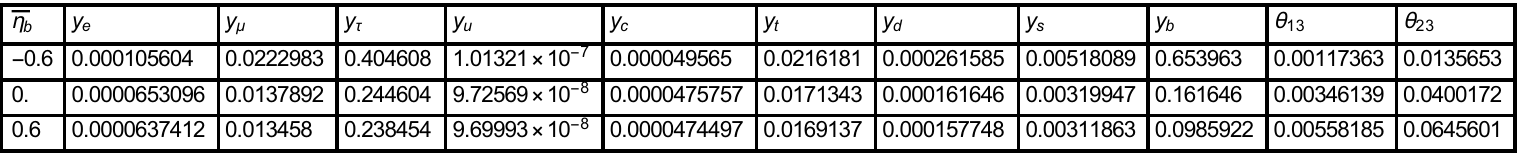}\vspace*{0.1cm}\newline
 \includegraphics[width=16cm]{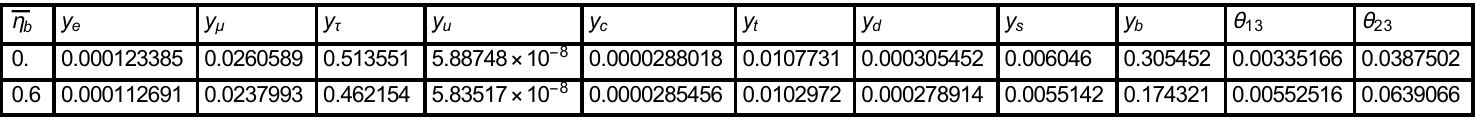}
 \caption{\label{tab:Yuk_10TeV} $M_{SUSY}=10~\rm{TeV}$}
\end{subfigure}
 \caption{\label{tab:Yuk_couplings} Yukawa couplings at the GUT scale depending on $\tan \beta$ ({\it top:} $\tan \beta=5$, {\it middle:} $\tan \beta=30$, {\it bottom:} $\tan \beta=50$), respectively, and the threshold effects represented by $\overline{\eta}_b$.}
\end{table}

\clearpage

%%%%%%%%%%%%%%%%%%%%%%%%%%%%%%%%%%%%%%%%%%%%%%%
\section{\label{app:D} MSSM Results -- Tables }
%%%%%%%%%%%%%%%%%%%%%%%%%%%%%%%%%%%%%%%%%%%%%%%

\begin{table}[h!]
\begin{tabular}{c c}
\parbox[c]{7.5cm}{
    \includegraphics[width=7cm]{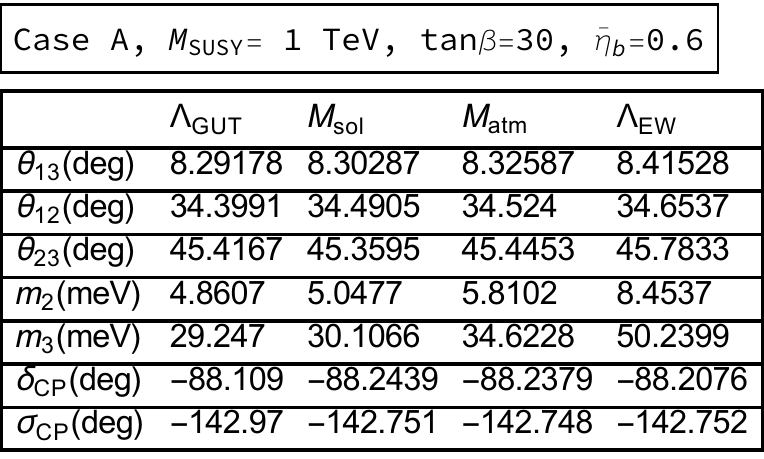}}
& \vspace*{0.3cm}
\parbox[c]{7.5cm}{
  \includegraphics[width=7cm]{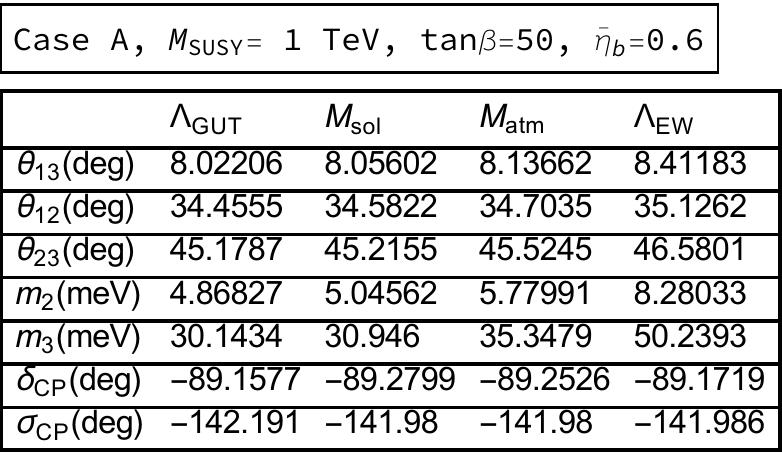}}
\\
\parbox[c]{7.5cm}{
  \includegraphics[width=7cm]{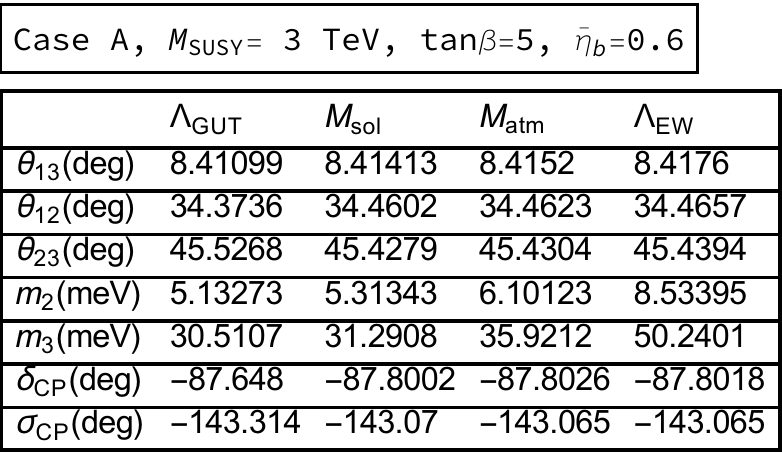}}
& \vspace*{0.3cm}
\parbox[c]{7.5cm}{
  \includegraphics[width=7cm]{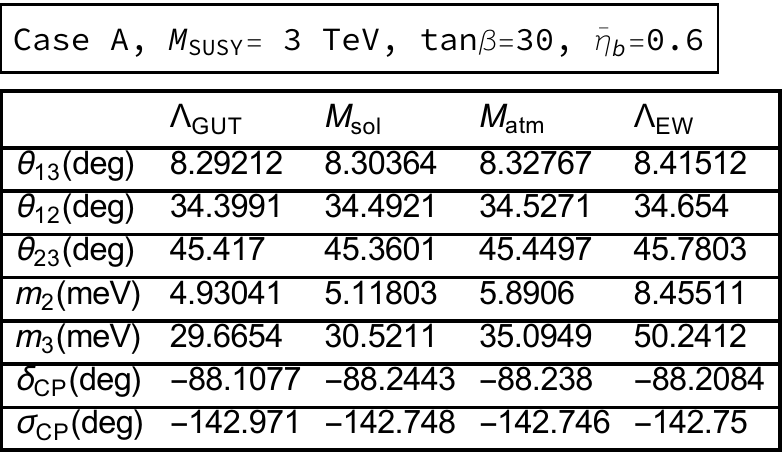}}
\\
\parbox[c]{7.5cm}{
  \includegraphics[width=7cm]{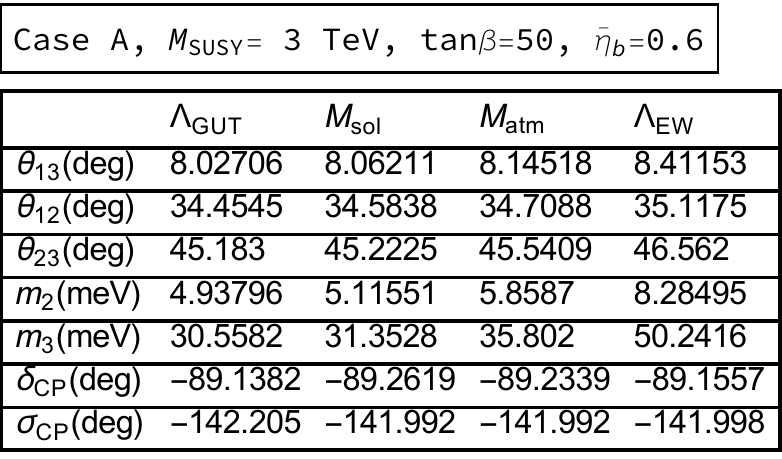}}
& \vspace*{0.3cm}
\parbox[c]{7.5cm}{
  \includegraphics[width=7cm]{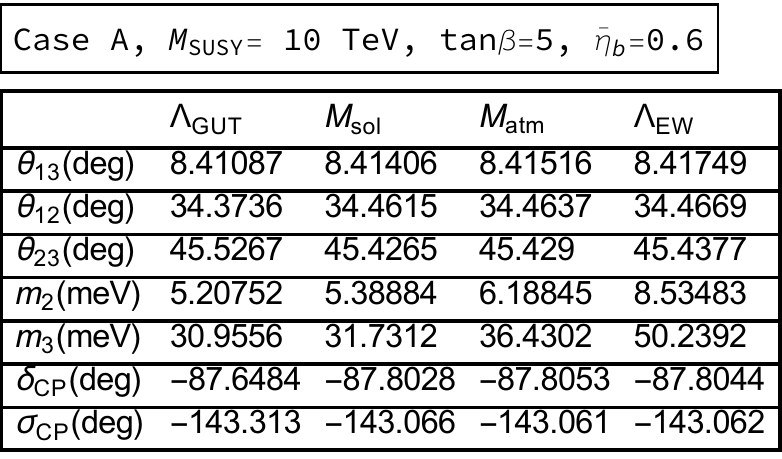}}
\\
\parbox[c]{7.5cm}{
  \includegraphics[width=7cm]{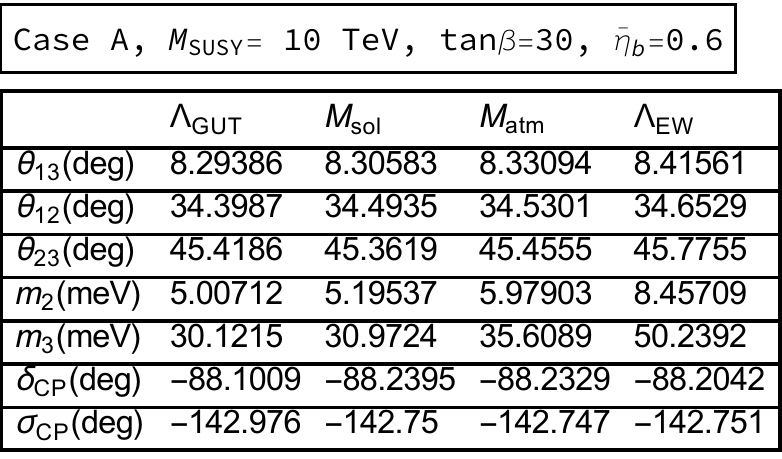}}
& \vspace*{0.3cm}
\parbox[c]{7.5cm}{
  \includegraphics[width=7cm]{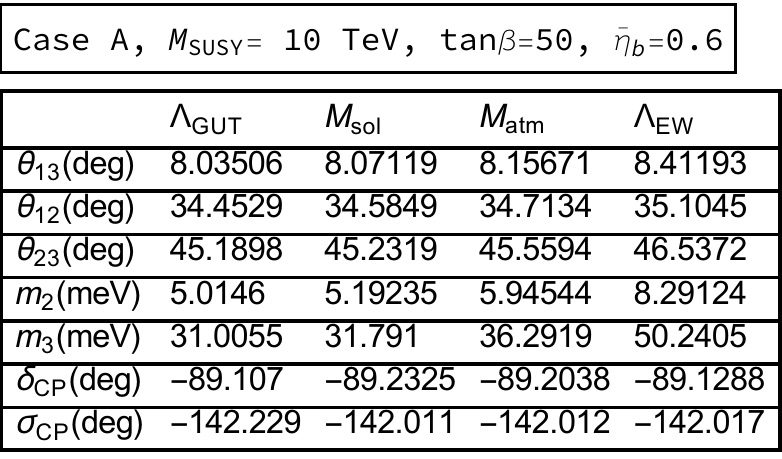}}
  \end{tabular}
  \caption{\label{fig:MSSM_CaseA}Case A - MSSM with $M_{atm}=10^{12}~\rm{GeV}$ and $M_{sol}=10^{15}~\rm{GeV}$.}
\end{table}

\begin{table}[h!]
\begin{tabular}{c c}
\parbox[c]{7.5cm}{
    \includegraphics[width=7cm]{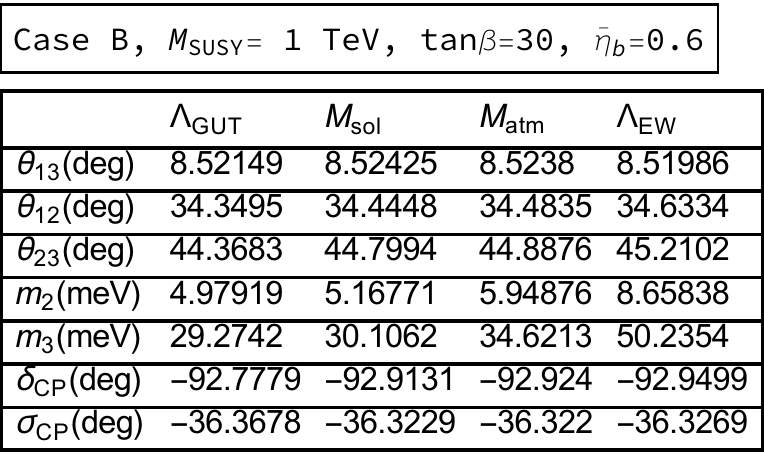}}
& \vspace*{0.3cm}
\parbox[c]{7.5cm}{
  \includegraphics[width=7cm]{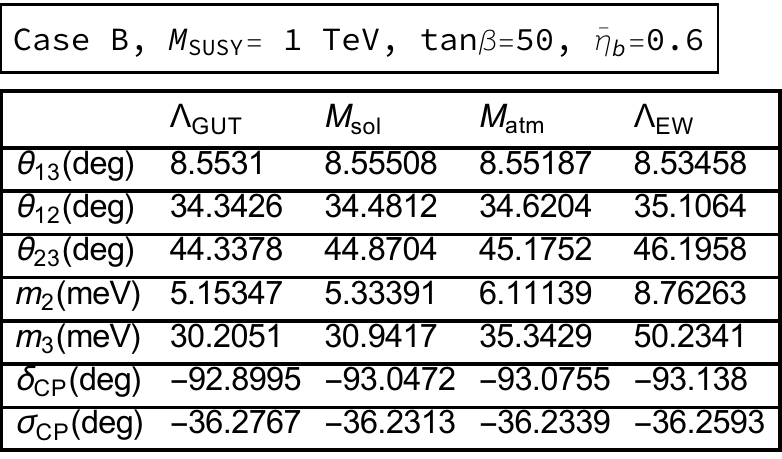}}
\\
\parbox[c]{7.5cm}{
  \includegraphics[width=7cm]{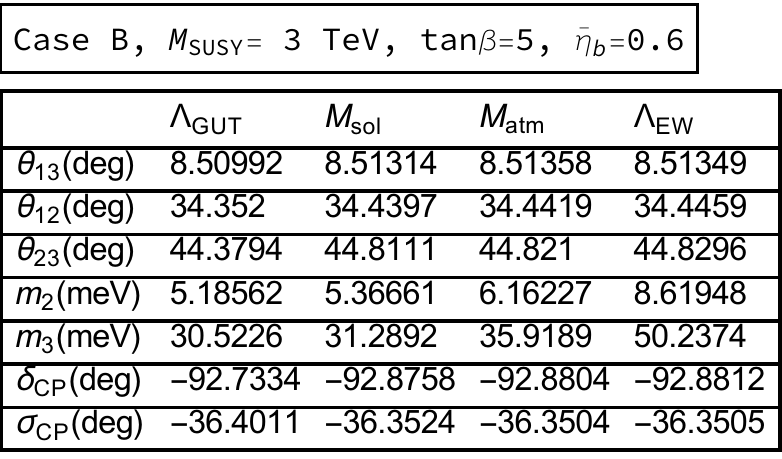}}
& \vspace*{0.3cm}
\parbox[c]{7.5cm}{
  \includegraphics[width=7cm]{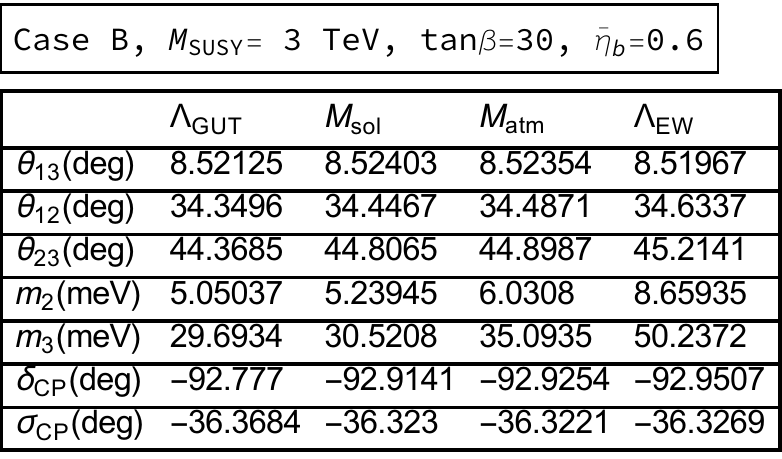}}
\\
\parbox[c]{7.5cm}{
  \includegraphics[width=7cm]{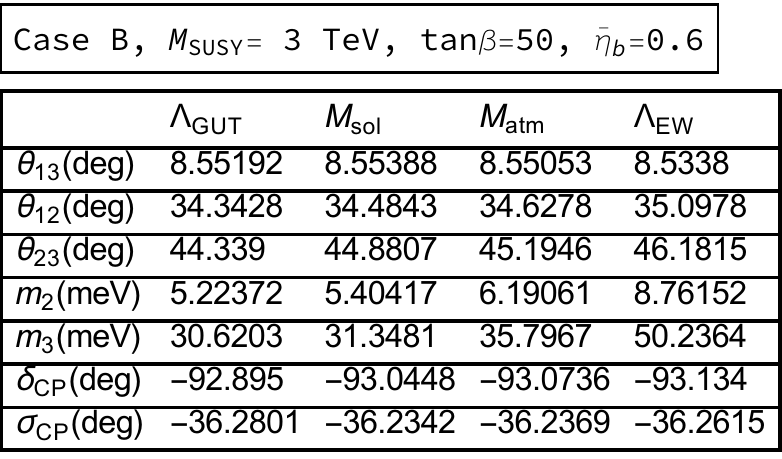}}
& \vspace*{0.3cm}
\parbox[c]{7.5cm}{
  \includegraphics[width=7cm]{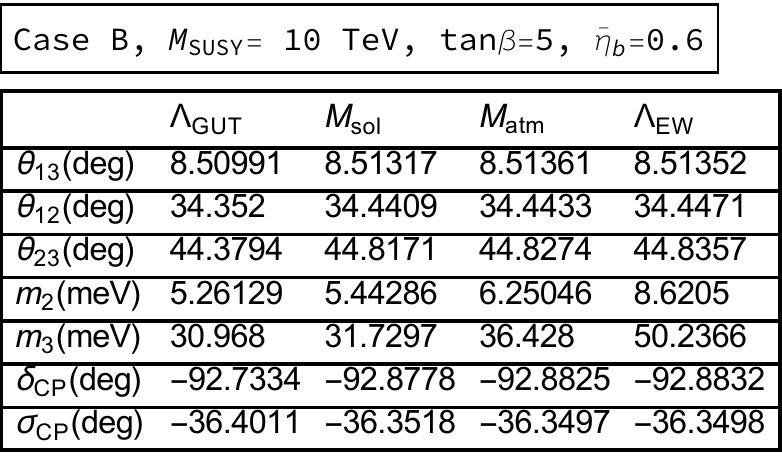}}
\\
\parbox[c]{7.5cm}{
  \includegraphics[width=7cm]{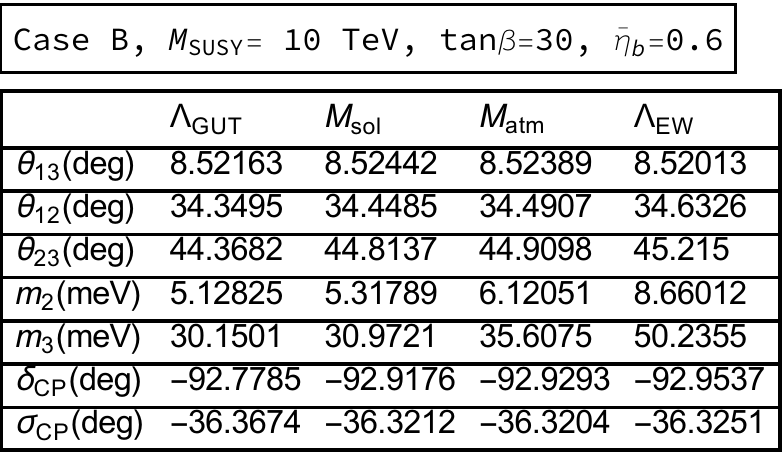}}
& \vspace*{0.3cm}
\parbox[c]{7.5cm}{
  \includegraphics[width=7cm]{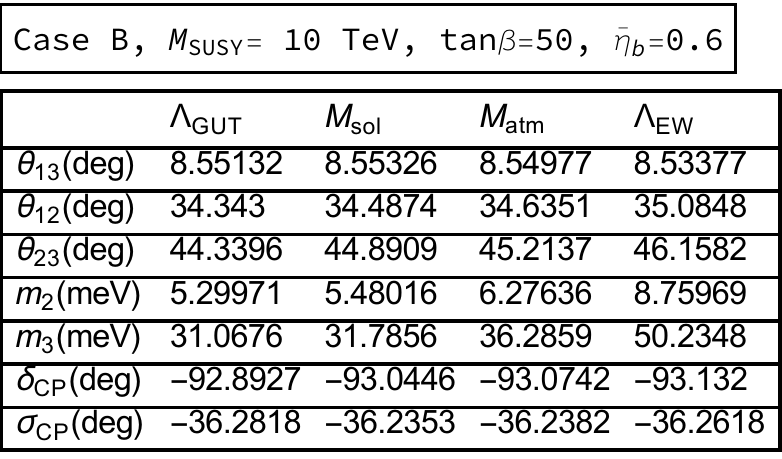}}
  \end{tabular}
  \caption{\label{fig:MSSM_CaseB}Case B - MSSM with $M_{atm}=10^{12}~\rm{GeV}$ and $M_{sol}=10^{15}~\rm{GeV}$.}
\end{table}

\begin{table}[h!]
\begin{tabular}{c c}
\parbox[c]{7.5cm}{
    \includegraphics[width=7cm]{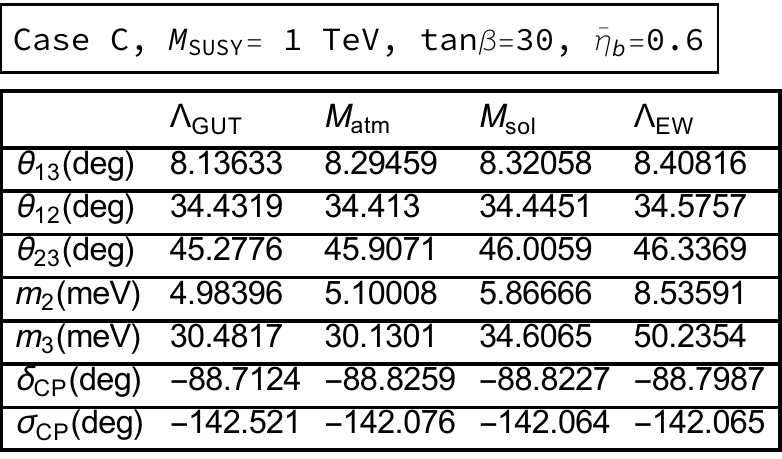}}
& \vspace*{0.3cm}
\parbox[c]{7.5cm}{
  \includegraphics[width=7cm]{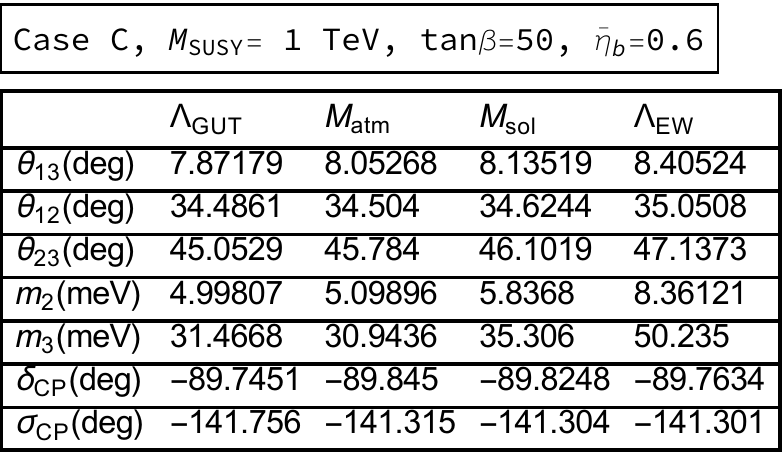}}
\\
\parbox[c]{7.5cm}{
  \includegraphics[width=7cm]{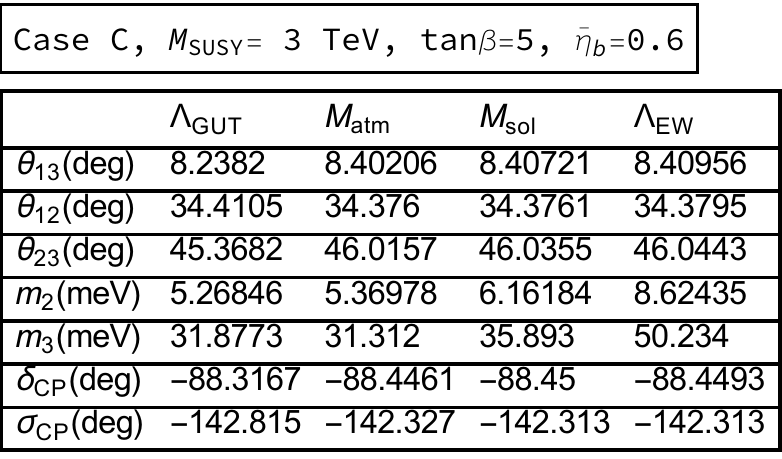}}
& \vspace*{0.3cm}
\parbox[c]{7.5cm}{
  \includegraphics[width=7cm]{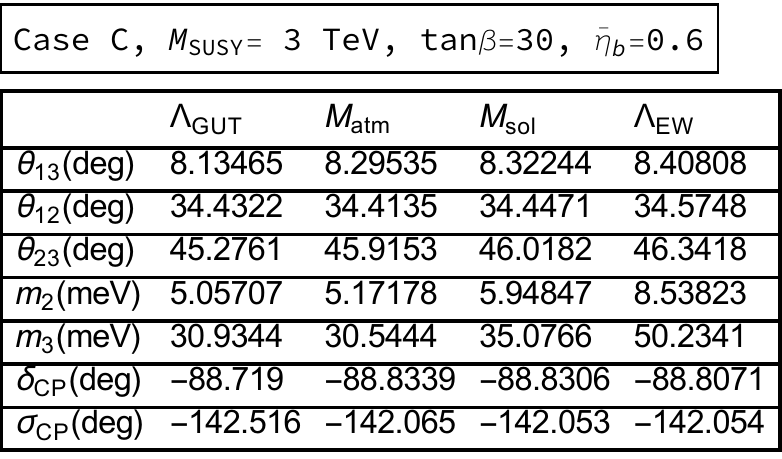}}
\\
\parbox[c]{7.5cm}{
  \includegraphics[width=7cm]{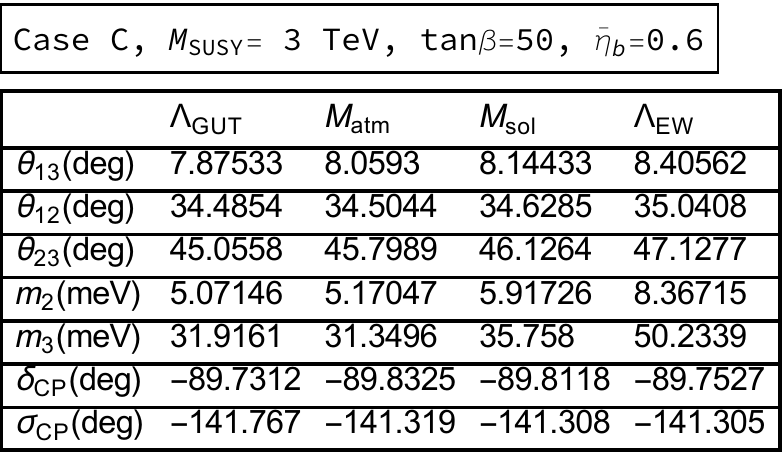}}
& \vspace*{0.3cm}
\parbox[c]{7.5cm}{
  \includegraphics[width=7cm]{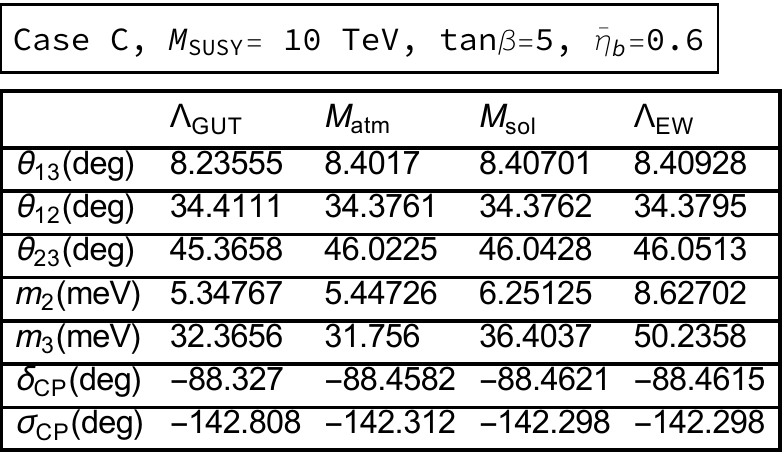}}
\\
\parbox[c]{7.5cm}{
  \includegraphics[width=7cm]{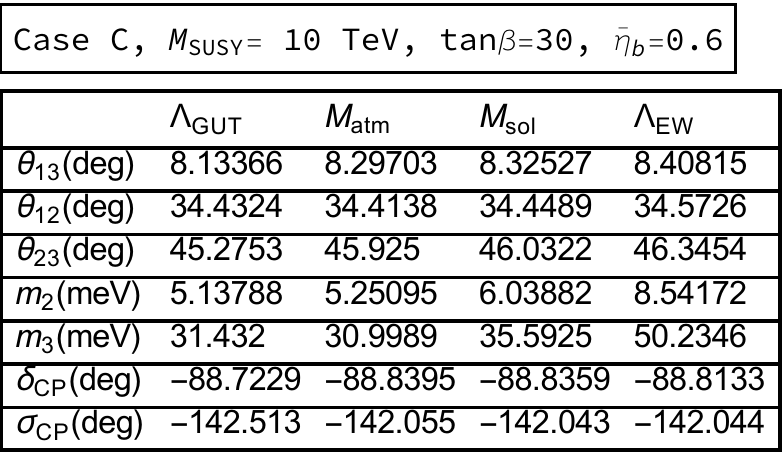}}
& \vspace*{0.3cm}
\parbox[c]{7.5cm}{
  \includegraphics[width=7cm]{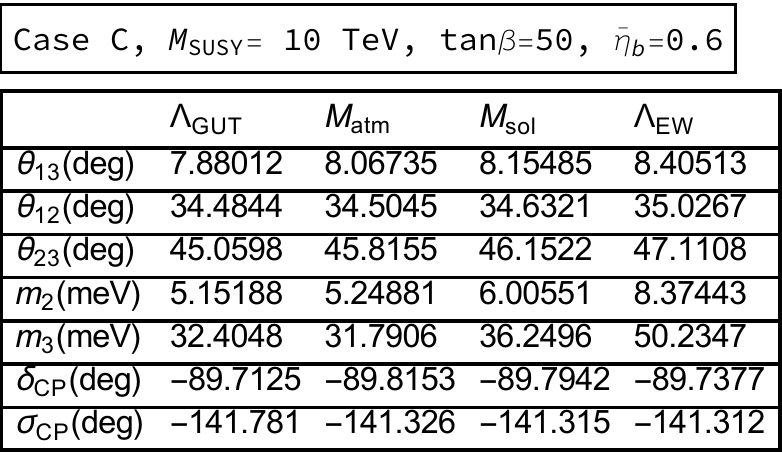}}
  \end{tabular}
  \caption{\label{fig:MSSM_CaseC}Case C - MSSM with $M_{atm}=10^{15}~\rm{GeV}$ and $M_{sol}=10^{12}~\rm{GeV}$.}
\end{table}

\begin{table}[h!]
\begin{tabular}{c c}
\parbox[c]{7.5cm}{
    \includegraphics[width=7cm]{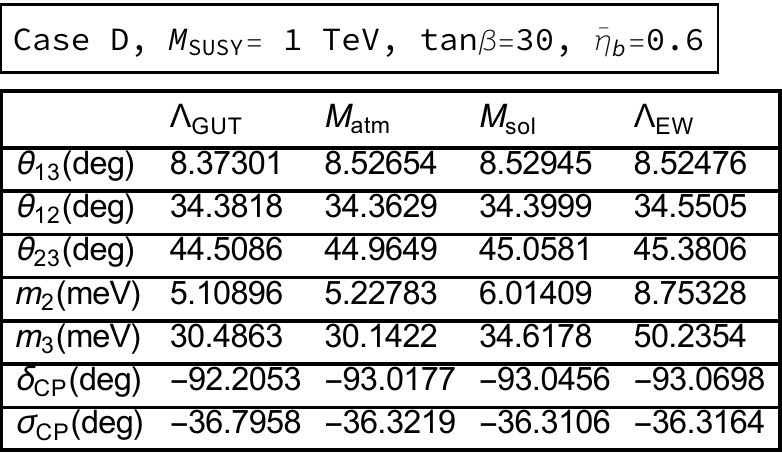}}
& \vspace*{0.3cm}
\parbox[c]{7.5cm}{
  \includegraphics[width=7cm]{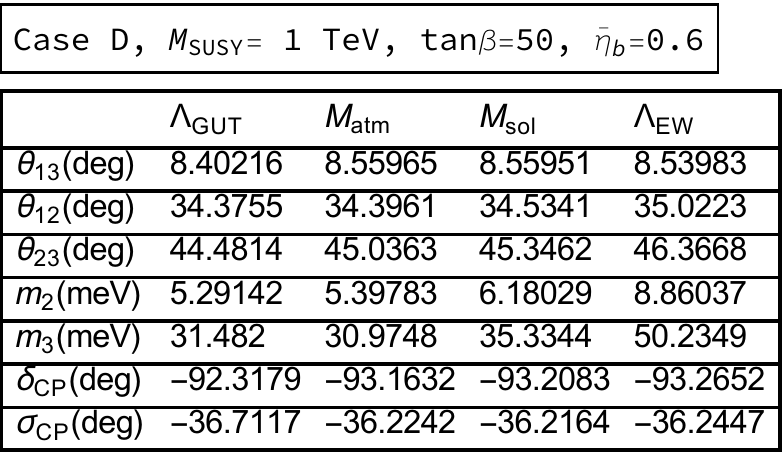}}
\\
\parbox[c]{7.5cm}{
  \includegraphics[width=7cm]{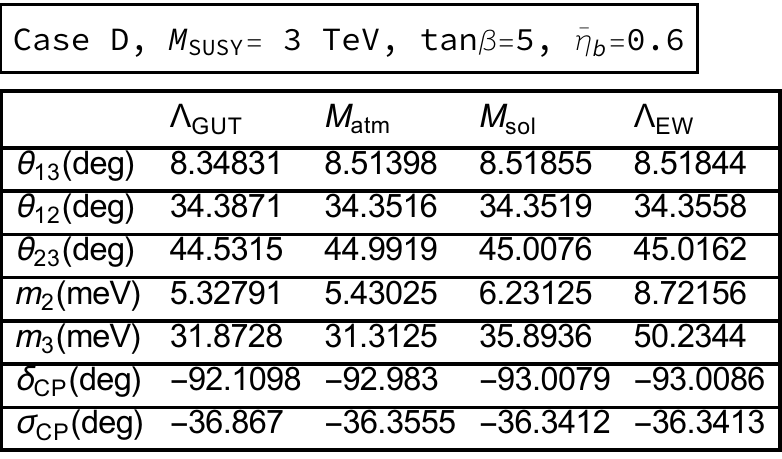}}
& \vspace*{0.3cm}
\parbox[c]{7.5cm}{
  \includegraphics[width=7cm]{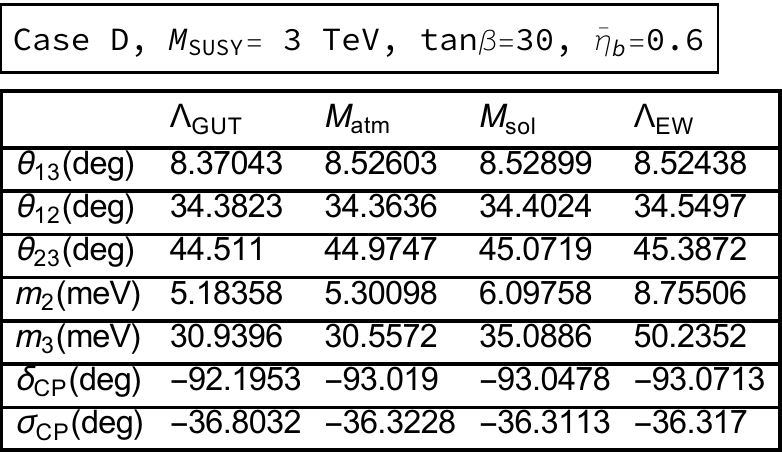}}
\\
\parbox[c]{7.5cm}{
  \includegraphics[width=7cm]{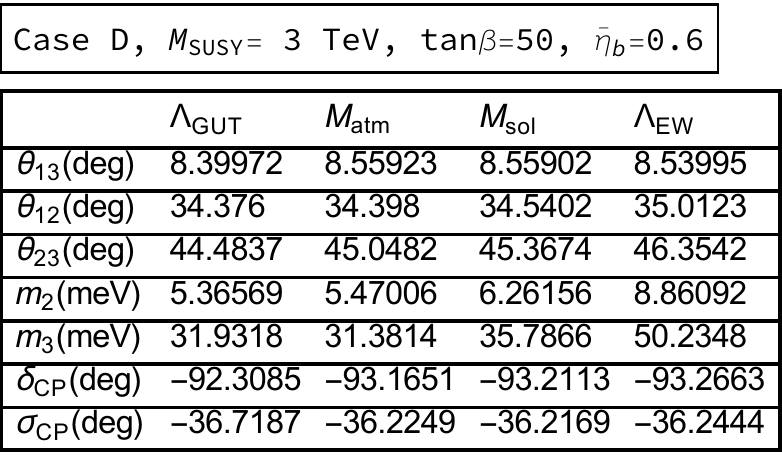}}
& \vspace*{0.3cm}
\parbox[c]{7.5cm}{
  \includegraphics[width=7cm]{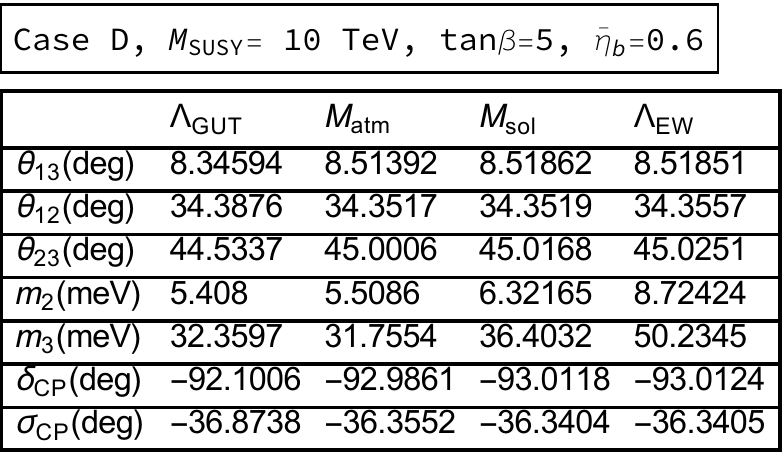}}
\\
\parbox[c]{7.5cm}{
  \includegraphics[width=7cm]{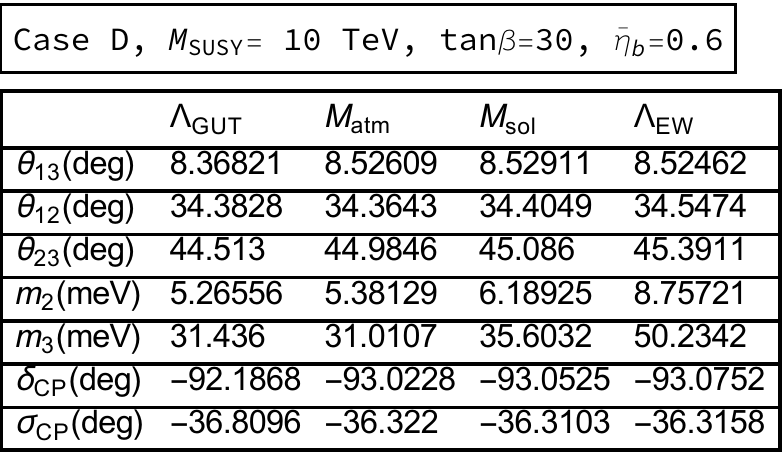}}
& \vspace*{0.3cm}
\parbox[c]{7.5cm}{
  \includegraphics[width=7cm]{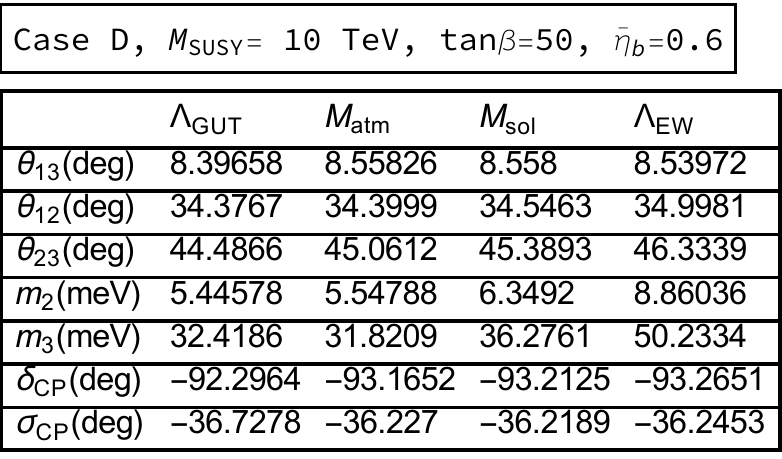}}
  \end{tabular}
  \caption{\label{fig:MSSM_CaseD}Case D - MSSM with $M_{atm}=10^{15}~\rm{GeV}$ and $M_{sol}=10^{12}~\rm{GeV}$.}
\end{table}

%=============================================================================
%\bibliographystyle{./apsrev}
%\bibliography{muon-positron}
%=============================================================================

\end{document}